\documentclass[prd,aps]{article}
\usepackage{amsmath,amsthm}
\usepackage{amssymb}
\usepackage{graphicx}
\usepackage{bm}

\DeclareMathSymbol{\C}{\mathbin}{AMSb}{"43}
\DeclareSymbolFont{AMSb}{U}{msb}{m}{n}

\makeatletter
\@addtoreset{equation}{section}
\makeatother

\newtheorem{Lemma}{Lemma}[section]

\def\be{\begin{eqnarray}}
\def\ee{\end{eqnarray}}

\newcommand{\ca}{\mathcal A}

\newcommand{\cf}{\mathcal F}
\newcommand{\cg}{\mathcal G}

\newcommand{\ci}{\mathcal I}

\newcommand{\ck}{\mathcal K}

\newcommand{\cv}{\mathcal V}

  \newcommand{\Fn}{\mathfrak{N}}
\newcommand{\fo}{\mathfrak{o}}  
  \newcommand{\Fp}{\mathfrak{P}}

\newcommand{\fs}{\mathfrak{s}}  
  
\newcommand{\fu}{\mathfrak{u}}

\renewcommand{\a}{\alpha}
\renewcommand{\b}{\beta}
\newcommand{\g}{\gamma}

\newcommand{\eps}{\epsilon}

\newcommand{\sig}{\sigma}

\renewcommand{\l}{\lambda}

\renewcommand{\O}{\Omega}
\renewcommand{\t}{\tau}

\newcommand{\rmd}{\mathrm d}

\newcommand{\lt}{\left}
\newcommand{\rt}{\right}

\newcommand{\lag}{\left\langle}
\newcommand{\rag}{\right\rangle}

\newcommand{\tr}{\mathrm{tr}}
\newcommand{\half}{\frac{1}{2}}
\newcommand{\bbc}{\mathbb{C}}

\title{\bf Commuting Simplicity and Closure Constraints\\
for 4D Spin Foam Models}

\author{Muxin Han$^{1,2}$ \ \  Thomas Thiemann$^{3,4}$\\
\\
{\small 1. Max-Planck-Institut f\"ur Gravitationsphysik, Am M\"uhlenberg 1, 14476 Potsdam-Golm,
Germany}\\
{\small 2. Centre de Physique Th\'eorique, CNRS UMR7332, Aix-Marseille Universit\'e and Universit\'e de Toulon, 13288 Marseille, France}\\
{\small 3. Institut f\"ur Theoretische Physik III, Universit\"{a}t Erlangen-N\"{u}rnberg, Staudtstra{\ss}e
7, 91058 Erlangen, Germany}\\
{\small 4. Perimeter Institute for Theoretical Physics, 31 Caroline Street N, Waterloo, ON N2L 2Y5,
Canada}}

\begin{document}

\maketitle

\begin{abstract}
Spin Foam Models are supposed to be discretised path integrals for quantum gravity constructed from
the Plebanski-Holst action. The reason for there being several models currently
under consideration is that no consensus has been reached for how to implement the simplicity
constraints.

Indeed, none of these models strictly follows from the original path integral with commuting
B fields, rather, by some non standard manipulations one always ends up with non commuting
B fields and the simplicity constraints become in fact anomalous which is the source for
there being several inequivalent strategies to circumvent the associated problems.

In this article, we construct a new Euclidian Spin Foam Model which is
constructed by standard methods from the Plebanski-Holst path integral
with commuting B fields discretised on a 4D simplicial complex. The resulting
model differs from the current ones in several aspects, one of them being
that the closure constraint needs special care. Only when dropping the closure constraint by hand and only in the large spin limit can the vertex amplitudes of this model be related to those of the
$\text{FK}_\gamma$ Model but even then the face and edge amplitude differ.

Interestingly, a non-commutative deformation of the $B^{IJ}$ variables leads
from our new model to the Barrett-Crane Model in the case of $\gamma=\infty$.
\end{abstract}

\newpage

\tableofcontents

\newpage

\section{Introduction}\label{introduction}

Loop Quantum Gravity (LQG) is an attempt to make a background independent, non-perturbative
quantization of 4-dimensional General Relativity (GR) -- for reviews, see \cite{book,rev,sfrevs}. It is
inspired by the formulation of GR as a dynamical theory of connections \cite{connection}. Starting
from this formulation, the kinematics of LQG is well-studied and results in a successful kinematical
framework (see the corresponding chapters in the books \cite{book}), which is also unique in a
certain sense \cite{unique}. However, the framework of the dynamics in LQG is still largely open so
far. There are two main approaches to the dynamics of LQG, they are (1) the Operator formalism of
LQG, which follows the spirit of Dirac quantization of constrained dynamical system, and performs
a canonical quantization of GR \cite{QSD,master}; (2) the Path integral formulation of LQG, which
is currently understood in terms of the Spin-foam Models (SFMs) \cite{sfrevs,BC,EPRL,FK,QSF}. The
relation between these two approaches is well-understood in the case of 3-dimensional gravity
\cite{perez}, while for 4-dimensional gravity, the situation is much more complicated and there are
some attempts \cite{links} for relating these two approaches.

The present article is concerned with the following issue in the framework of spin-foam models. The
current spin-foam models are mostly inspired by the 4-dimensional Plebanski formulation of GR
\cite{plebanski} (Plebanski-Holst formulation by including the Barbero-Immirzi parameter $\g$),
whose action reads
\be
S_{\text{PH}}[A,B,\varphi]:=\int \lt(B+\frac{1}{\g}*B\rt)^{IJ}\wedge F_{IJ}+ \frac{1}{4}\int\rmd^4x\
\varphi^{\a\b\g\delta}B^{IJ}_{\a\b}B^{KL}_{\g\delta}\eps_{IJKL}
\ee
where $B$ is a so(4)-valued 2-form field, $F:=\rmd A+A\wedge A$ is the curvature of the
so(4)-connection field $A$ and $\varphi^{\a
\b\g\delta}=\varphi^{[\a\b][\g\delta]}$ is a densitized tensor, symmetrized under interchanging $[\a
\b]$ and $[\g\delta]$, and traceless $\eps_{\a\b\g\delta}\varphi^{\a\b\g\delta}=0$.
For the illustrative purposes of this article, we consider only Euclidean GR in the present article,
however, the lessons learnt will extend also to the Lorentzian theory.
One can show
that the equations of motion implied by the Plebanski-Holst action are equivalent to the Einstein
equations of GR. Moreover, if we consider formally the following path integral partition function of
the Plebanski-Holst action and perform the integral of $\varphi^{\a\b\g\delta}$
\be
Z:=\int [DA\ DB\ D\varphi]\ e^{iS_{\text{PH}}[A,B,\varphi]}=\int [DA\ DB]\ \delta\lt(\eps_{IJJKL}B^{IJ}_
{\a\b}B^{KL}_{\g\delta}-\cv\eps_{\a\b\g\delta}/4!\rt)\ e^{i\int \lt(B+\frac{1}{\g}*B\rt)^{IJ}\wedge F_{IJ}}
\label{PL}
\ee
we obtain the partition function of BF theory \cite{bf} whose paths are, however, constrained by 20
\emph{Simplicity Constraint} equations
\be
\eps_{IJJKL}B^{IJ}_{\a\b}B^{KL}_{\g\delta}-\frac{1}{4!}\cv\eps_{\a\b\g\delta}
\ee

The point of this formulation is of course that the path integral of BF theory has been formulated as a concrete spin-foam model (subject to the divergence issue, see the corresponding chapters in \cite{book}) and thus the idea is to rely on those results
and to implement the simplicity constraints properly into the partition function of BF theory.
We remark that even for Euclidian gravity, the partition function (\ref{PL}) is unlikely to be derived
from the canonical formulation because of the presence of second class constraints which affect the choice
of the measure in (\ref{PL}), see the first and third reference in \cite{links} for a detailed discussion.
Since in current spin foam models the proper choice of measure is also regarded as a nontrivial problem and as we want to draw attention to a different issue for the current spin foam models, we also will not deal with the measure issue in this article and leave this for future research.

The partition function of BF theory, after
discretization on a 4-dimensional simplicial complex $\ck$ and its dual complex $\ck^*$, can be
expressed as a sum over certain spin-foam amplitudes. Here a spin-foam amplitude is obtained
by (1) assigning an SO(4) unitary irreducible representation to each triangle $f$ of $\ck$ (we label
the representation by a pair $(j_f^+,j_f^-)$ for each triangle); (2) assigning a 4-valent SO(4)
intertwiner to each tetrahedron $t$ of $\ck$ (we label the intertwiner by a pair $(i^+_t,i^-_t)$ for
each tetrahedron). Then the partition function of BF theory can be written as
\be
Z_{\text{BF}}(\ck)=\sum_{\{j_f^\pm\}_f}\sum_{\{i^\pm_t\}_t}\prod_f\dim(j_f^+)
\dim(j_f^-)\prod_\sig\big
\{15j\big\}_{\text{SO(4)}}\lt(j_f^\pm,i_t^\pm\rt)
\label{BF}
\ee
where the $15j$-symbol is the 4-simplex/vertex amplitude corresponding to the 4-simplex $\sig$.
The partition function $Z_{\text{BF}}$ turns out to be formally independent of the triangulation $\ck$.
Clearly, as shown explicitly in Eq.(\ref{PL}), in order to obtain the partition function for quantum gravity
as a sum of spin-foam amplitudes, one has to impose the simplicity constraint in the BF theory measure.
When doing that, the resulting partition function is no longer triangulation independent\footnote{As it should
not be because GR is not a TQFT in the classical level. Triangulation independence is understood as a feature in the quantization of classical TQFT, which should not be expected in the quantization of gravity.} and thus one should in fact consider all possible discretizations
and not only simplicial ones. This is also necessary in order to make contact with the canonical
LQG Hilbert space which contains all possible graphs and not only 4-valent ones. This has been recently
emphasised in \cite{KKL,DHR} and the current spin foam models already have been generalised in that respect.
We believe our model also to be generalisable but will not deal with this aspect in the present work as this
would draw attention away from our main point.

Essentially, the very method of imposing the simplicity constraint {\it defines} the
corresponding candidate spin-foam model for
quantum gravity which why its proper implementation deserves so much attention. Currently the three most
studied spin-foam models for quantum gravity
(in Plebanski or Plebanski-Holst formulation) are the Barrett-Crane Model \cite{BC}, the EPRL
Model \cite{EPRL}, and $\text{FK}_\g$ Model \cite{FK}. These three, a priori, different models are
defined by three different ways to impose simplicity constraint on the measure of the BF partition function
$Z_{\text
{BF}}$. We will review these different methods of imposing the simplicity constraint briefly in what
follows.

First of all,
in the context of the discretized path integral, the simplicity constraint also takes a discretized
expression. For each triangle $f$ we define an so(4) Lie algebra element $B_f$ which corresponds to the
integral of the two form $B$ over the triangle $f$. Then in terms of the $B_f$ for each 4-simplex $\sig$
the discretised simplicity constraints read
\be
&&\epsilon_{IJKL}B^{IJ}_{f}B^{KL}_{f'}=0,\ \ \ \ \ f,f'\text{ belong to the same tetrahedron $t$}\label
{3simple}\\
&&\epsilon_{IJKL}B^{IJ}_{f_1}B^{KL}_{f'_1}=\epsilon_{IJKL}B^{IJ}_{f_2}B^{KL}_{f'_2},\ \ \ \ \ f_i,f'_i
\text{ belong to the two different tetrahedrons in $\sig$}\label{4simple}
\ee
The Barrett-Crane Model, the EPRL Model, and the $\text{FK}_\g$ Model all explicitly impose the
first type of simplicity
constraint Eq.(\ref{3simple}), called tetrahedron constraint, in some way to the spin-foam partition
function of BF
theory. On the other hand,  all of them replace the second type of simplicity constraint, called
4-simplex constraint Eq.(\ref{4simple}) by the so called \emph{Closure
Constraint}
\be
\sum_{f\subset t}B^{IJ}_f=0\ \ \ \ \ \text{for each tetrahedron $t$.} \label{closure}
\ee
It is not difficult to see that the closure constraints together with the tetrahedron constraints
imply the 4-simplex constraints but not vice versa. Thus, sprictly speaking, imposing the closure
constraint constrains the BF measure more than the classical theory would precribe. It is unknown
and also beyond the scope of the present paper whether this replacement is harmless or is in conflict
with the classical theory. In this paper, as we are merely interested in comparing the standard way
of imposing the simplicity constraints (commuting B fields) with the non standard methods
defining the BC, EPRL and FK models (non commuting B fields), we proceed as in those other spin foam
models and also replace the 4-simplex constraint by the closure constraint. To distinguish
these two different types of constraints, in what follows we use the terminology ``simplicity constraint''
for Eq.(\ref{3simple}) and ``closure constraint'' for Eq. (\ref{closure}).
Notice that the BC Model, EPRL Model, and $\text{FK}_\g$ Model argue that the closure
constraint is ``automatically'' implemented in their spin-foam amplitude. We will come back to
this argument in a moment. Because of that argument, in none of these models the closure constraint is
further analysed. The proper implemementation of the simplicty and closure constraints is one of the
most active research areas in the spin foam model community and there are many issues that
yet have to be understood \cite{Alexandrov}.

For both the Barrett-Crane Model and EPRL Model, the strategy for imposing the simplicity constraint
is the following: In order to take advantage of the knowledge of BF spin-foam Model, one formally takes
the delta distribution on the B variables out of the integral over B by a standard trick known from
ordinary quantum field theories: One (formally) just has to replace $B$ by $\delta/\delta F$ because
the integrand of the B integral is of the form $\exp(iF\cdot B)$. Due to the
discretization upon which $F$ is replaced by a holonomy around a face of the dual triangulation and
B by an integral over a triangle of the triangulation,
$\delta/\delta F$ can be rewritten in terms of the right invariant vector fields $X$ on the copy of $SO(4)$
corresponding to the given holonomy with holonomy dependent coefficients. One now argues that these
coefficients can be replaced by their chromatic evaluation (setting the holonomy equal to unity) because
the integration over $B$ leads to $\delta(F)$ enforcing the measure on the space of connections to be
supported on flat ones. Clearly, this argument is not obviously water tight because
$\delta(\delta^2/\delta F^2)\cdot \delta(F)$ may not be supported at $F=0$. In fact it should not be
if we are interested in gravity rather than BF theory. See the chapter on spin-foams in the second
reference of \cite{book} for more details. In any case, this way of proceeding now leads to replacing
the commutative derivations $\delta/\delta F$ by the non commutative right invariant vector fields $X$.

An alternative argument that has been given is the following: The kinetical boundary Hilbert
space of the spin foam path integral
should be the canonical LQG Hilbert space (restricted to the 4-valent boundary graph of the given simplicial
triangulation) and here the $B$ field would be quantised as $\delta/\delta A$ where $A$ is the underlying
connection. On functions of holonomies this again becomes a right invariant vector field labelled
by the triangles dual (in the 3D sense) to the corresponding boundary edges which in turn correspond to the
faces of the dual triangulation dual (in the 4D sense) to those triangles. The physical boundary Hilbert
space should therefore be the kernel of that quantised boundary simplicity constraints.
In order to write the corresponding spin foam model, one has to define the projector on that physical
Hilbert space. To do this properly, one should canonically quantize Plebanski -- Holst gravity, identify
all the first and second class constraints and define the projector via Dirac bracket and group averaging
which then leads to a spin foam path integral.
How complicated this becomes if one really performs all the necessary steps is outlined in
\cite{links}. However, this is not what is done in \cite{EPRL}. The first observation is that since the
spin foam path integral naturally involves SO(4), the kinematical boundary Hilbert space is naturally
also in terms of SO(4) spin network functions. One now studies the restrictions
that the simplicity constraints impose on the spins and intertwiners of the boundary SO(4) Hilbert space
spin network functions. The detailed structure of these restrictions suggests a natural one to one map
with spin network states in the canonical SU(2) Hilbert space. Finally, using locality arguments, one
conjectures that these restrictions should not only hold on the boundary but also in the bulk of the
BF SO(4) spin foam model. See \cite{Rovelli} for a particularly simple and clear exposition of this procedure.
It has recently been criticised in \cite{Alexandrov} on the ground that the BF symplectic structure and
the LQG symplectic structure have wrongly been identified in the afore mentioned identification map.

In any case, whether or not the map is the correct correpondence, the simplicity constraints were
again quantised as non commuting (anomalous) constraints. If one understands the kernel in the strong
operator topology then one obtains the BC model, if one understands it
in the weak operator topology (Gupta -- Bleuler procedure) one obtains the EPRL model.
Because of the anomaly, imposing the constraint operators strongly apparently makes the Barrett-Crane
Model lose some important information about non-degenerate quantum geometry \cite{BCtrouble}.
Imposing the constraints weakly is less restrictive and thus may lead to a better behaved model.
More in detail, first of all the quadratic
expression of the simplicity constraint Eq.(\ref{3simple}) is replaced by a linearized expression. It
is given by asking that
for each tetrahedron $t$, there exists a unit vector $u_t^I$, such that
\be
*B_f^{IJ}u_{t,I}=0\label{LSC}
\ee
The equivalence of the linearized simplicity constraint Eq.(\ref{LSC}) with original simplicity
constraint Eq.(\ref{3simple}) will be reviewed in Section \ref{implementation} (in the gravitational
sector of the solution). In the original construction of EPRL spin-foam model in \cite{EPRL}, the unit
vector $u_t^I$ is gauge fixed to be $\delta^{0,I}$, and a ``Master constraint'' $M_f:=
\sum_jC_f^jC_f^j$ is defined (to replace the cross-diagonal part of the simplicity constraint Eq.(\ref
{3simple})), where $C_f^j:=*B_f^{0j}$ from Eq.(\ref{LSC}). The corresponding ``Master constraint
operator'' is defined by replacing $B_f^{IJ}$ by right invariant derivatives. This Master constraint
solves the problem of non-commutativity/anomaly of the quantum simplicity to a certain extent,
because a single Master constraint replaces all the cross-diagonal components of Eq.(\ref
{3simple}). Moreover the diagonal part of Eq.(\ref{3simple}) and this Master constraint operator
restrict the Hilbert space spanned by the 4-valent SO(4) spin-networks to its subspace, which can be
identified with 4-valent SU(2) spin-networks and thus can be imbedded into the kinematical Hilbert space of
LQG. For each of these SU(2) spin-networks, the SU(2) unitary irreducible representations labelled
by $k\in \frac{1}{2}\mathbb{N}$ has the following relation with the original SO(4) representations
on all the boundary edges dual to the boundary triangles
\be
j^\pm=\frac{|1\pm\g|}{2}k\label{jk}
\ee
Here the Barbero-Immirzi parameter $\g$ can only take discrete values, i.e.
\be
\text{If}\ |\g|>1:&& \g=\frac{j^+_{f}+j_{f}^-}{j^+_{f}-j_{f}^-}\nonumber\\
\text{If}\ |\g|<1:&& \g=\frac{j^+_{f}-j_{f}^-}{j^+_{f}+j_{f}^-}
\ee
More importantly, the recent results in \cite{dingyou,DHR} show that the boundary Hilbert space used in
the EPRL Model solves the linear version of simplicity constraint Eq.(\ref{LSC}) (and the closure
constraint Eq.(\ref{closure})) weakly, i.e. the matrix elements (with respect to the
boundary SO(4) Hilbert space) of the constraint operators vanish on
the space of solutions
\be
\lag f,\hat{C}f'\rag=0, \ \ \ \ \ \text{for all $f,f'$ in the Hilbert space of solutions.}
\ee
in contrast to the strong implementation of the constraints in the Barrett-Crane Model. Finally the
(Euclidean) EPRL spin-foam partition function is expressed by
\be
Z_{\text{EPRL}}(\ck)=\sum_{\{k_f\}_f}\sum_{\{i_t\}_t}\prod_f\dim(k_f)
\prod_\sig\sum_{i_t^\pm}\big\{15j\big\}_{\text{SO(4)}}\lt(j_f^\pm,i_t^\pm\rt)\prod_{\overrightarrow
{(\sig,t)}}f^{i_t}_{i^+_t,i^-_t}\lt(j^\pm_f,k_f\rt)
\ee
where for each spin-foam amplitude, an SU(2) unitary irreducible representation $k_f$ is assigned
to each triangle $f$, satisfying the relation Eq.(\ref{jk}), and an SU(2) 4-valent intertwiner $i_t$ is
assigned to each tetrahedron $t$. Here
\be
\sum_{i_t^\pm}\big\{15j\big\}_{\text{SO(4)}}\lt(j_f^\pm,i_t^\pm\rt)\prod_{\overrightarrow{(\sig,t)}}f^
{i_t}_{i^+_t,i^-_t}\lt(j^\pm_f,k_f\rt)
\ee
is the 4-simplex/vertex amplitude for the EPRL Model, where $f^{i_t}_{i^+_t,i^-_t}$ are a fusion
coefficients defined in \cite{EPRL}.

The $\text{FK}_\g$ Model follows a different strategy to impose the simplicity constraint, namely by using the
coherent states for SU(2) group \cite{LS,Perelomov}. Given a unitary irreducible representation
space $V^j$ of SU(2), the coherent state is defined by
\be
\left|j,n\rag:=n\left|j,j\rag=\sum_{m=-j}^j\left|j,m\rag\pi^j_{mj}(n)\ \ \ \ \ n\in\text{SU(2)}
\ee
We then immediately have the resolution of identity on $V^j$
\be
1_j=\dim(j)\int_{SU(2)}\rmd n\ \left|j,n\rag\lag j,n\right|\label{ri}
\ee
This coherent state has a certain geometrical interpretation, which can be seen by computing the
expectation value of the su(2) generator ($\sig_i$ are Pauli matrices)
\be \label{interpretation}
\lag j,n\right|\hat{X}\left|j,n\rag=\lag j,n\right|\hat{J}^i\left|j,n\rag\sig_i=jn\sig_3n^{-1}\label{expect}
\ee
If we identify the Lie algebra su(2) with $\mathbb{R}^3$, we can see that the coherent state $\left|
j,n\rag$ describes a vector in $\mathbb{R}^3$ with length $j$, its direction is determined by the
action of $n$ on a unit reference vector (the direction of $\sig_3$). From the expression $n
\sig_3n^{-1}$ we see that $n$ can be parameterized by the coset $\text{SU(2)}/\text{U(1)}=S^2$. In
addition, the integral in the resolution of identity is essentially over $\text{SU(2)}/\text{U(1)}=S^2$.
It is not hard to show that the (Euclidean) BF partition function can be expressed in terms of the
coherent states (we write $(g^+,g^-)$ for each SO(4) element, $(j^+,j^-)$ for an SO(4) unitary
irreducible representation)
\be
Z_{\text{BF}}(\ck)&=&\sum_{\{j_f^\pm\}_f}\prod_{f}\dim(j^+_f)\dim(j^-_f)\int\prod_{(\sig,t)}\rmd g^+_
{\sig t}\rmd g^-_{\sig t} \nonumber\\
&&\prod_{(t,f)}\dim(j^+_f)\dim(j^-_f)\int\rmd n^+_{tf}\rmd n^-_{tf}\prod_{(\sig f)}\lag j^+_f,n_{tf}^+
\right|g^+_{t\sig}g^+_{\sig t}\left|j^+_f,n^+_{t'f}\rag\lag j^-_f,n_{tf}^-\right|g^-_{t\sig}g^-_{\sig t}\left|j^-
_f,n^-_{t'f}\rag\label{BF1}
\ee
where $(g_{\sig t}^+,g_{sig t}^-)$ is a SO(4) holonomy along the edge from the center of 4-simplex
$\sig$ to the center of tetrahedron $t$. Then the strategy of imposing simplicity constraint in $\text
{FK}_\g$ Model is to use the interpretation (\ref{interpretation}) of the coherent state labels
$j_f^\pm n_{tf}^\pm\t_3(n_{tf}^\pm)^
{-1}$ as the self-dual/anti-self-dual part $X_{tf}^\pm$ of the so(4) variable $B_{tf}$ associated
with a triangle $f$ seen from a tetrahedron $t$. (More precisely, we know that the previously
defined $B_f$ can be decomposed into self-dual and anti-self-dual part $X_f^\pm$. The \emph
{interpretations} of $j_f^\pm n_{tf}^\pm\t_3(n_{tf}^\pm)^{-1}$, $X_{tf}^\pm$ are considered as the
parallel transport of $X_f^\pm$ from the center of triangle $f$ to the center of tetrahedron $t$, i.e.
$X^\pm_{tf}=g^\pm_{tf}X^\pm_fg^\pm_{ft}$, where $g^\pm_{tf}$ is the holonomy along the edge
from the center of triangle $f$ to the center of tetrahedron $t$). That is, the simplicity constraint is
imposed on the coherent state labels, which results in the following restrictions:
\be \label{restriction}
\frac{j^+}{j^-}=\lt|\frac{\g+1}{\g-1}\rt|,\ \ \ \ \text{and}\ \ \ \  \left\{
                                              \begin{array}{ll}
                                                (n^+_{tf},n^-_{tf})=(n_{tf}h_{\phi_{tf}},u_tn_{tf}h^{-1}_{\phi_{tf}}), & \hbox
{\text{for $-1<\g<1$};} \\
                                                (n^+_{tf},n^-_{tf})=(n_{tf}h_{\phi_{tf}},u_tn_{tf}h^{-1}_{\phi_{tf}}\eps), &
\hbox{\text{for $\g<-1$ or $\g>1$}.}
                                              \end{array}
                                            \right.
\ee
where $u_t$ is some normal to $t$, $h_{\phi_{tf}}$ takes values in the U(1) subgroup of SU(2) generated
by $\sigma_3$ and $\epsilon=i\sigma_2$. In more detail, the proposal is then to simply replace in
(\ref{BF1}) $n^\pm_{tf}$ by these
expressions and the Haar measure $dn^+_{tf}\;dn^-_{tf}$ by the Haar measure $dn_{tf}\;du_t\;dh_{\phi_{tf}}$.
We emphasize that this is an interesting but non standard procedure:
while the identification of the coherent state labels $j_f^\pm, n_{tf}^\pm$ with
the so(4) variables $B_{tf}$ is certainly well motivated, the resulting expression does not
arise by integrating out the $B$ fields in the presence of the delta distributions enforcing the
simplicit constraints. Rather, in (\ref{BF1}) the B fields have already been integrated out.
To restrict measure and integrand by hand afterwards according to (\ref{restriction}) is not
obviously equivalent with the standard procedure of solving the $\delta-$distributions.
One would hope that the resulting procedures coincide in the semiclassical or
the ``large-$j$'' limit \cite{pirsf}. Indeed, the ``large-$j$'' limit result in Section \ref{NSF} will
support this
expectation. Finally the spin-foam partition function of $\text{FK}_\g$ Model coincides (at least up
to a slight change of edge amplitude) with EPRL partition function when the Barbero-Immirzi
parameter $-1<\g<1$. However when $\g<-1$ or $\g>1$, $\text{FK}_\g$ partition function is rather
different from the EPRL partition function. Here we only show explicitly the 4-simplex/vertex
amplitude of $\text{FK}_\g$ model when $\g<-1$ or $\g>1$
\be
\sum_{i_t^\pm}\big\{15j\big\}_{\text{SO(4)}}\lt(j_f^\pm,i_t^\pm\rt)\prod_{\overrightarrow{(\sig,t)}}f^
{i_t}_{i^+_t,i^-_t}\lt(j^\pm_f,k_{tf}\rt)
\ee
Here although the relation between $j_f^\pm$
\be
\frac{j^+}{j^-}=\lt|\frac{\g+1}{\g-1}\rt|
\ee
is the same as in EPRL Model, in $\text{FK}_\g$ model for $\g<-1$ or $\g>1$,
there are some additional degrees of freedom associated with the label $k_{tf}$, which are the
values of spins from the coupling of $j^+_f$ and $j^-_f$, i.e. $k_{tf}$ could take values in $|j^+_f-j^-
_f|,\cdots,j^+_f+j^-_f$. The final partition function is obtained by summing over $j^-_f$, $i_t$, and
$k_{tf}$ with some measure factors (see \cite{FK} for details).

In the previous three paragraphs, we briefly revisited the main strategies of imposing simplicity
constraint in Barrett-Crane, EPRL and $\text{FK}_\g$ Models. We have seen that these in general
different spin-foam models came from two different ways of imposing simplicity constraint, i.e.
Barrett-Crane and EPRL Model quantize the simplicity constraint as operators and imposed them
(strongly or weakly) on the boundary spin-networks, while $\text{FK}_\g$ Model imposes the
constraint on the coherent state labels. However, as we have reviewed, none of the three models is
derived from the original path integral formula
Eq.(\ref{PL}) of the Plebanski action (or the discretized version of the path integral) without
using some non standard methods. Therefore a natural question arises:\\
Is any of those three spin-foam models consistent with the path
integral formula Eq.(\ref{PL}) and its discretized version? This question is non trivial
because in all three types of models one deals with non commutative B fields and simplicity constraints
as operators
on some Hilbert space while the original path integral is in terms of commutative c-number
variables so that anomalies cannot arise.
Because of this issue, it is interesting to investigate
what kind of spin-foam model we will obtain, if we start from the (discretization of) the path integral
formula Eq.(\ref{PL}) with commutative $B^{IJ}$ variables. It is also interesting to find some possible
bridges linking the (discretization of) the path integral formula Eq.(\ref{PL}) with commutative $B^{IJ}$
variables to the existing spin-foam models using non-commutative $B^{IJ}$ variables.

In this article, we consider the discretization of the path integral formula Eq.(\ref{PL}), which will be
Eq.(\ref{start}). As announced in \cite{BFT}, in contrast to the Barrett-Crane, EPRL, and $\text{FK}_\g$
Models, we always
consider the variables $B^{IJ}$ as commutative c-numbers. The simplicity constraint (and
closure constraint) is (are) imposed by the c-number delta functions inserted in the path integral
formula, which one gets by integrating over the Lagrange multiplier and which constrain the path integral
measure. In our concrete analysis in Section \ref
{NSF}, the most important difference between our derivation and the derivation in any of Barrett-
Crane, EPRL, and $\text{FK}_\g$ Models is the following: in any of Barrett-Crane, EPRL, and $\text
{FK}_\g$ Models, one always imposes the respective version of the simplicity constraint constraint
on the BF spin-foam partition
function Eq.(\ref{BF}) or (\ref{BF1}) {\it after} integration over $B^{IJ}$.
This feature is essentially the reason why it is difficult to find a relation between the
simplicity constraint imposed in any of Barrett-Crane, EPRL, and $\text{FK}_\g$ Models and the
simplicity constraint in the path integral formula Eq.(\ref{PL}). By contrast, our derivation in Section
\ref{NSF} will \emph{not} start from the spin-foam partition function of BF theory, but instead we
impose the delta function of the simplicity constraint (and closure constraint) before the integration
over $B^{IJ}$, and we will see that solving these constraints gives rise to a non trivial modification
of the path integral measure. There were early works analyzing the simplicity constraint toward this direction, see e.g. \cite{BL}.

As also announced in \cite{BFT}, regarding the $B^{IJ}$ variables as commutative c-numbers also makes the
treatment of closure
constraint different. We know that the closure constraint Eq.(\ref{closure}) is necessary in order that
the full set of simplicity constraint Eq.(\ref{3simple}) and (\ref{4simple}) is satisfied. In Barrett-Crane
Model the closure constraint is argued to be automatically satisfied by the SO(4) gauge invariance
of the vertex amplitude. However, as shown in \cite{BFT}, this is only true {\it after} performing the
Haar measure integrals which essentially project everything on the gauge invariant sector.
It is clear that the closure constraint must be imposed {\it before} performing the integral over the
connections.
In the EPRL Model, the argument is improved in that both simplicity constraint and
closure constraint vanishes \emph{weakly} on the EPRL boundary Hilbert space \cite{dingyou}.
Moreover, in \cite{CF}, it is shown that in both EPRL and $\text{FK}_\g$ Model, the closure
constraint can be implemented in terms of geometric quantization and by the commutativity of the
quantization and phase space reduction \cite{GS}. As defined, an additional
closure constraint would be redundant for both EPRL and $\text{FK}_\g$ Model, since they are
already on the constraint surface of closure constraint (if one interprets the coherent state labels to
be the $B^{IJ}$ variables), although the original definitions of both models didn't impose closure
constraint explicitly. We feel that this is again due to the fact that the Haar integrals have already
been performed.
%However we find that the above conclusions quite rely on the non-
%commutative nature of the $B^{IJ}$ variables in Barrett-Crane, EPRL, and $\text{FK}_\g$ Models.
In our analysis we find that the
implementation of closure constraint gives non-trivial restrictions on the measure.

In order to understand what happens when one ignores the clsoure constraint and to follow more closely
the procedure followed by existing spin foam models, in
section \ref{NSF}, we first consider a simplified partition function $Z_{\text{Simplified}}(\ck)$ in which
the delta functions of closure constraint is dropped (as it is discussed in \cite{BL}), and derive an expression of $Z_{\text
{Simplified}}(\ck)$ as a sum of all possible spin-foam amplitudes (constrained only by the simplicity
constraints). Then we also compute the true partition function $Z(\ck)$ with the closure constraint implemented. When we compare $Z_{\text{Simplified}}(\ck)$ with the true partition function $Z(\ck)$, we find
the closure constraint non-trivially affect the spin-foam expression of partition
function. But all the spin-foams (transition channels) admitted in the simplified
partition function $Z_{\text{Simplified}}(\ck)$ still contribute to the full partition function $Z(\ck)$
(with some changes for the triangle/face amplitude and tetrahedron/edge amplitude).

Another key feature of our derivation is a different discretization of the BF action. Here we first
break the faces dual to the triangles into wedges (see FIG.\ref{face}) and then write the discretized BF
action in terms of the holonomies along the boundary of the wedges.
Here, as usual, a wedge in the dual face $f$ is determined by a dual vertex or original 4-simplex $\sigma$
and thus denoted by $(\sigma,f)$. Its boundary consists of four segments defined as follows: The
original (piecewise linear) 4-simplex has a barycentre $\hat{\sigma}$ which is the dual vertex. The
dual edges connect
these barycentres. A pair of dual edges $e,e'$ adjacent to the same dual vertex defines a face. Conversely,
given a face and a dual vertex which is one of the corners of the face, we obtain two dual edges. These
are dual to two tetrahedra $t,t'$ of the original complex. The boundary of the wedge $(\sigma,f)$ is now
given by
$(\hat{\sigma},\hat{e})\circ (\hat{e},\hat{f})\circ (\hat{f},\hat{e}')\circ(\hat{e}',\hat{\sigma})$
where the hat denotes the respective barycentres. In an unfortunate abuse of notation which exploits the
duality one also writes this as
$(\sigma,t)\circ (t,f)\circ (f,t')\circ(t',\sigma)$. Using this notation we have (cf. FIG.\ref
{face})
\be
&&\int_M \lt[B+\frac{1}{\g}*B\rt]^{IJ}\wedge F_{IJ}\ =\ \int_M\lt(1+\frac{1}{\g}\rt)\tr\lt(X^+\wedge F^+\rt)
+\int_M\lt(1-\frac{1}{\g}\rt)\tr\lt(X^-\wedge F^-\rt)\nonumber\\
&=&\sum_f\lt(1+\frac{1}{\g}\rt)\tr\lt(X_f^+ F^+_f\rt)+\sum_f\lt(1-\frac{1}{\g}\rt)\tr\lt(X_f^- F_f^-\rt)
\nonumber\\
&=&\sum_{(\sig,f)}\lt(1+\frac{1}{\g}\rt)\tr\lt(X_f^+ F^+_{(\sig,f)}\rt)+\sum_{(\sig,f)}\lt(1-\frac{1}{\g}\rt)\tr\lt
(X_f^- F_{(\sig,f)}^-\rt)\nonumber\\
&\simeq&\sum_{(\sig,f)}\lt(1+\frac{1}{\g}\rt)\tr\lt(X^+_fg^+_{f t}g^+_{t\sig}g^+_{\sig t'}g^+_{t'f}\rt)+
\sum_{(\sig,f)}\lt(1-\frac{1}{\g}\rt)\tr\lt(X_f^- g^-_{f t}g^-_{t\sig}g^-_{\sig t'}g^-_{t'f}\rt)\label{discreteaction}
\ee
where $F_{(\sig,f)}$ is the curvature 2-form integrated on the wedge determined by $(\sig,f)$ and $t,t'$
respectively are the afore mentioned unique tetrahedra (or dual edges). This
starting point results in the following structures in the resulting spin-foam model $Z_{\text
{Simplified}}(\ck)$ (these structures turn out to be similar to the structure proposed in \cite{BL}):
\begin{itemize}
  \item In contrast to the existing spin-foam models, where the SO(4) representations $(j_f^+,j_f^-)$
were labeling the faces $f$, the new spin-foam model derived in Section \ref
{NSF} have SO(4) representations $(j_{\sig f}^+,j_{\sig f}^-)$ labeling the wedges, i.e. a dual face
$f$ having $n$ vertices (corners) in general has $n$ different pairs $(j_{\sig f}^+,j_{\sig f}^-)$, one for
each wedge determined by the vertex dual to $\sig$. However in the large-$j$ limit, the triangle/
face amplitude is concentrated on SO(4) representations $j_{\sig f}^\pm=j_{\sig' f}^\pm)$ for any
vertices $\sigma,\sigma'$ of the same face $f$.

  \item Two neighboring wedges $(\sig,f)$ and $(\sig',f)$ of a face $f$
share a segment $(t,f)$ (c.f. FIG.\ref{face}) whose end points are the center of the face $f$ and
the center of the edge dual to the tetrahedron $t=\sig\cap \sig'$. For each segment
$(t,f)$ there is an SU(2) representation $k_{tf}$ ``mediating'' the SO(4) representations on the two
neighboring wedges, $(j_{\sig f}^+,j_{\sig f}^-)$ and $(j_{\sig' f}^+,j_{\sig' f}^-)$, in the sense that
$k_{tf}$ has to lie in the range of the joint Clebsh - Gordan decomposition of
 $j_{\sig f}^+\otimes j_{\sig f}^-$ and $j_{\sig' f}^+ \otimes j_{\sig' f}^-$ (c.f.
FIG.\ref{face2}), thus
\be
k_{tf}\in \lt\{|j_{\sig f}^+-j_{\sig f}^-|,\cdots, j_{\sig f}^++j_{\sig f}^-\rt\}\cap\lt\{|j_{\sig' f}^+-j_{\sig' f}^-|,
\cdots, j_{\sig' f}^++j_{\sig' f}^-\rt\}.
\ee

\end{itemize}
Note that the idea for implementing c-number simplicity constraint strongly in the spin-foam model is not new, and has been employed in \cite{BL}. Some calculations, e.g. solving the simplicity constraint, toward $Z_{\text{Simplified}}(\ck)$ is similar to the derivation in \cite{BL} (especially in the first reference in \cite{BL}). However the discrete action Eq.(\ref{discreteaction}) here is different from the one used in \cite{BL}. The action here turns out to be important to understand the non-commutative deformation and the relation to Barrett-Crane Model in Appendix \ref{deformation}, which is one of the key points in this paper.

An interesting result from the analysis here is the relations between the new spin-foam model derived here and the
existing spin-foam models e.g. Barrett-Crane, EPRL, and $\text{FK}_\g$ Models. From the analysis
in Section \ref{NSF}, we find that, firstly, in the large-$j$ and large-area limit the spin-foams in our new model $Z_
{\text{Simplified}}(\ck)$ reduces to the spin-foams in $\text{FK}_\g$ Model (with identical 4-simplex/
vertex amplitude but different tetrahedron/edge and triangle/face amplitudes) at least
for $|\g|>1$.
Secondly, in Appendix \ref{deformation}, we study the non-commutative deformation of the partition
function Eq.(\ref{start}), in order to study how the non-commutative nature of the $B^{IJ}$ variables
in the existing spin-foam models emerges in our commutative context. The non-commutative
deformation we employ here comes from a generalized Fourier transformation on the compact
group \cite{Gfourier} (the deformed partition function will be denote by $Z_\star(\ck)$). With this
deformation, we find that the closure constraint really becomes redundant when we set the
deformation parameter $a=\ell_p^2$, while the redundancy is hard to be shown with a general
deformation parameter. With the setting of the deformation parameter $a=\ell_p^2$, we show that
the non-commutative deformation of our new spin-foam model leads to Barrett-Crane model when
the Barbero-Immirzi parameter $\g=\infty$. This result explains how the non-commutative nature of
the $B^{IJ}$ variables in Barrett-Crane model relates to the commutative context of our new spin-
foam model in Section \ref{NSF}, and also explains to some extent the reason why in the Barrett-Crane model the closure constraint is redundant (such an explanation also appears in the first
reference of \cite{AOL} from the group field theory perspective). On the other hand,
the relation with EPRL Model and $\text{FK}_\g$ ($|\g|<1$) is still veiled. What we know is
that the allowed spin-foams (transition channels) in EPRL Model form a subset of those allowed
in our new spin-foam model (with the same 4-simplex/vertex amplitude but different
different tetrahedron/edge and triangle/face amplitudes) and this fact also holds for $\text{FK}_\g$
Model for any $\gamma$. All above relations between various spin-foam models are summarized in the following
diagram, where the sets $\{Z_{\cdots}\}$ are the collections of spin-foams (transition channels) which respectively
contribute their partition functions $Z_{\cdots}(\ck)$:
\be
\text{\tiny inclusion} && \text{\tiny inclusion} \ \ \ \ \ \ \ \ \ \ \text{\tiny noncomm. deform.}\nonumber\\
\{Z_{\text{EPRL}}\},\{Z_{\text{FK}_\g}\}\ \ \ \subset\ \ &\{Z_{\text{Simplified}}\}&\ \
\subset\ \ \ \{Z\}\ \ \ \ \ \ \ \--\!\!\!\--\!\!\!\rightsquigarrow\ \ \ \ \ \ \ \{Z_{\text{BC}}\}\nonumber\\
&\downarrow& \!\!\!\!\!\!\!\!\!\! \text{\tiny large-$j$, large area, $|\g|>1$}\nonumber\\
&\{Z_{\text{FK}_{|\g|>1}}\}&\nonumber
\ee
where $\subset$ means the inclusion in terms of contributing spin-foam amplitudes. We will discuss the details in Section 4.2.

\section{Starting Point of the New Model}\label{model}

\subsection{The Partition Function}

In the last section we reviewed the approaches of simplicity constraint and closure constraint in
the existing spin-foam models, and summarized the approach and main results of the present
article. In this section, we present the detailed construction and analysis of our new spin-foam
model. We take a simplicial complex $\ck$ of the 4-dimensional manifold $M$\footnote{in most of the discussions of the present paper, the manifold $M$ is assumed to be without boundary, then the partition function $Z(\ck)$ is a number associated to the triangulation. But the discussion can be easily generalized to the case with a boundary. }, where we denote
the simplices by $\sig$, the tetrahedra by $t$ and the triangles by $f$. And we take the following
discretized partition function as the staring point for constructing the spin-foam model\footnote{Such a spin-foam partition function can be undertood as a sum over the histories of SO(4) spin-networks, as we will see in the following discussion.}:
\be \label{2.1}
Z(\ck)&:=&\int\;\prod_{f}\rmd^3X^+_f\rmd^3X^-_f\prod_{(\sig,t)}\rmd g^+_{\sig t}\rmd g^-_{\sig
t}\prod_{(t,f)}\rmd g^+_{tf}\rmd g^-_{tf}\prod_{t;f,f'\subset t}\delta\lt(X_{tf}^+\cdot X^+_{tf'}-X_{tf}^-
\cdot X^-_{tf'}\rt)\prod_{t}\delta\Big(\sum_{f\subset t}X_{tf}^+\Big)\nonumber\\
&&\times\prod_{(\sig,f)}e^{i(1+\frac{1}{\g})\tr\lt(X^+_fg^+_{f t}g^+_{t\sig}g^+_{\sig t'}g^+_{t'f}\rt)}
\prod_{(\sig,f)}e^{i(1-\frac{1}{\g})\tr\lt(X^-_fg^-_{f t}g^-_{t\sig}g^-_{\sig t'}g^-_{t'f}\rt)}\label{start}
\ee
We explain the meaning of the variables appearing in the above definition:
\begin{itemize}
  \item $X_f^+,X_f^-\in \fs\fu(2)$ are respectively the self-dual and anti-self-dual part of the $\fs\fo
(4)$ flux variable $B_f^{IJ}$, which is the $\fs\fo(4)$-valued 2-form field $B_{\a\b}^{IJ}$ smeared
on the triangle dual to $f$ while
      \be
      X^\pm_{tf}:={g}^\pm_{tf}X^\pm_f{g}^\pm_{f t}.
      \ee
      So given two tetrahedra $t,t'$ sharing a face $f$, the relation between $X_{tf}$ and $X_{t'f}$ is thus 
      \be
       X^\pm_{t'f}:={g}^\pm_{t't}X^\pm_{tf}{g}^\pm_{tt'}
      \ee
      where ${g}^\pm_{t't}={g}^\pm_{t'f}{g}^\pm_{ft}$ and ${g}^\pm_{t't}=({g}^\pm_{tt'})^{-1}$. Such a ``parallel-transportation condition'' for $X^\pm_{tf}$ means that each triangle $f$ associates a unique pair $X^\pm_f$, which ensures the right number of degrees of freedom as a discretization of Plebanski-Holst gravity. $X^\pm_{tf}$ are the auxiliary variables which are useful in the following derivation.

  \item $\rmd g$ is the Haar measure on SU(2). $g^+_{\sig t},g^-_{\sig t}\in\text{SU(2)}$ is the self-
dual and anti-self-dual part of the SO(4) holonomy along the half edge $\overrightarrow{(\sig,t)}$
outgoing from the vertex $\sigma$ while
$g^+_{tf},g^-_{tf}$ are respectively the self-dual and anti-self-dual part of the SO(4) holonomy
along the segments $\overrightarrow{(t,f)}$ (see FIG.\ref{face}).

  \begin{figure}[h]
  \begin{center}
  \includegraphics[width=6cm]{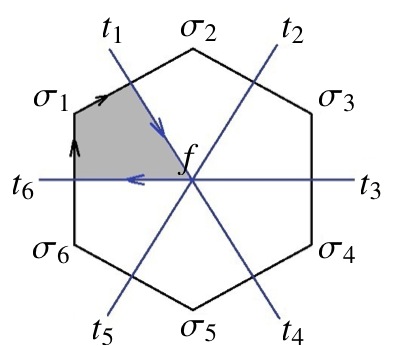}
  \caption{A face dual to the triangle $f$. The vertices of the face are dual to the 4-simplices $\sig$.
Each edge of the face is dual to a tetrahedron $t$. The fundamental region (in gray) in the
face determined by a 4-simplex and two tetrahedra is called a wedge. Each tetrahedron is shared
by two 4-simplices. A triangle is shared by $n$ simplices, where $n$ is the number of
vertices of the dual face.}
  \label{face}
  \end{center}
  \end{figure}

  \item The delta function $\delta\lt(X_{tf}^+\cdot X^+_{tf'}-X_{tf}^-\cdot X^-_{tf'}\rt)$ imposes the
simplicity constraint for each tetrahedron:
  \be
  \epsilon_{IJKL}B^{IJ}_{tf}B^{KL}_{tf'}=0\ \ \ \ \ \ \ \ f,f' \text{ belonging to the same tetrahedron}\label
{3simple1}
  \ee
  while the delta function $\delta\Big(\sum_{f\subset t}X_{tf}^+\Big)$ imposes the self-dual closure
constraint for each tetrahedron. Note that there is no closure constraint for $X^-_f$ because the
closure of $X^-_f$ is implied by the self-dual closure constraint and the simplicity constraint as we will
demonstrate shortly. So including it would be equivalent to multiplying the partition function
with a divergent constant which drops out in expectation values. In addition, the closure constraint and
simplicity constraint Eq.(\ref
{3simple}) imply the 4-simplex constraints ($i,j,k,l\in \{1,2,3,4,5\}$):
  \be
  &&\epsilon_{IJKL}B^{IJ}_{\sig f_{ij}}B^{KL}_{\sig f'_{kl}}=
  \epsilon_{IJKL}B^{IJ}_{\sig f_{ik}}B^{KL}_{\sig f'_{lj}}=
  \epsilon_{IJKL}B^{IJ}_{\sig f_{il}}B^{KL}_{\sig f'_{jk}}
\nonumber\\
  &&f_{ij}\text{face dual to the triangle $t_i\cap t_j$, where $t_i$ are the 5 tetrahedra of $\sigma$}
\label{4simple1}
  \ee
Here $X^\pm_{\sig f}=g^\pm_{\sig t}X_{tf}^\pm g_{t\sig}^\pm$ and $B_{\sig f}=X^+_{\sig f}+X^-_{\sig f}$.
In the continuum limit of
Eqs.(\ref{3simple1}) and (\ref{4simple1}), in which the holonomies can be replaced by the group unit,
we recover the Plebanski simplicity constraints (20 equations):
  \be
  \epsilon_{IJKL}B^{IJ}_{\a\b}B^{KL}_{\g\delta}=\cv\eps_{\a\b\g\delta}/4!\label{simplicity}
  \ee
  where $\cv:=\eps^{\a\b\g\delta}\epsilon_{IJKL}B^{IJ}_{\a\b}B^{KL}_{\g\delta}$ is the 4-dimensional volume element. Note that there are essentially 20 constraint equations while the
trace part of Eq.(\ref{simplicity}) is an identity. The solutions of the simplicity constraints is well-
known: given a non-degenerate co-tetrad $e_\a^I$, there are five sectors of solutions of the
simplicity constraints \cite{sfrevs}
  \be
I\pm:&&B^{IJ}=\pm e^I\wedge e^J\nonumber\\
II\pm:&&B^{IJ}=\pm \half\eps^{IJ}_{\ \ KL}e^K\wedge e^L\nonumber\\
\text{Deg}:&&B^{+}=B^-
\ee
where $B^\pm$ are the self-dual and anti-self-dual parts of $B^{IJ}$.

  \item The exponentials in $\prod_{(\sig,f)}e^{i(1+\frac{1}{\g})\tr\lt(X^+_fg^+_{f t}g^+_{t\sig}g^+_{\sig
t'}g^+_{t'f}\rt)}\prod_{(\sig,f)}e^{i(1-\frac{1}{\g})\tr\lt(X^-_fg^-_{f t}g^-_{t\sig}g^-_{\sig t'}g^-_{t'f}\rt)}$
come from the exponential of the BF action, discretized in terms of wedge holonomies $g^\pm_
{\sig t}g^\pm_{tt'}g^\pm_{t'\sig}$. In more detail,
\be
&&\int_M \lt[B+\frac{1}{\g}*B\rt]^{IJ}\wedge F_{IJ}\ =\ \int_M\lt(1+\frac{1}{\g}\rt)\tr\lt(X^+\wedge F^+\rt)
+\int_M\lt(1-\frac{1}{\g}\rt)\tr\lt(X^-\wedge F^-\rt)\nonumber\\
&=&\sum_f\lt(1+\frac{1}{\g}\rt)\tr\lt(X_f^+ F^+_f\rt)+\sum_f\lt(1-\frac{1}{\g}\rt)\tr\lt(X_f^- F_f^-\rt)
\nonumber\\
&=&\sum_{(\sig,f)}\lt(1+\frac{1}{\g}\rt)\tr\lt(X_f^+ F^+_{(\sig,f)}\rt)+\sum_{(\sig,f)}\lt(1-\frac{1}{\g}\rt)\tr\lt
(X_f^- F_{(\sig,f)}^-\rt)\nonumber\\
&\simeq&\sum_{(\sig,f)}\lt(1+\frac{1}{\g}\rt)\tr\lt(X^+_fg^+_{f t}g^+_{t\sig}g^+_{\sig t'}g^+_{t'f}\rt)+
\sum_{(\sig,f)}\lt(1-\frac{1}{\g}\rt)\tr\lt(X_f^- g^-_{f t}g^-_{t\sig}g^-_{\sig t'}g^-_{t'f}\rt)
\ee
where $F_{(\sig,f)}$ is the curvature 2-form integrated on the wedge determined by $(\sig,f)$

  \item Finally we note that under the SO(4) gauge transformations:
\be
g^\pm_{tf}\mapsto h^\pm_t g^\pm_{tf}(h^\pm_{f})^{-1}\ \ \ \ g^\pm_{\sig t}\mapsto h^\pm_\sig g^\pm_
{\sig t}(h^\pm_{t})^{-1}\ \ \ \ X^\pm_f\mapsto h^\pm_{f}X^\pm_f(h^\pm_{f})^{-1}\ \ \ \ X^\pm_{tf}
\mapsto h^\pm_{t}X^\pm_{tf}(h^\pm_{t})^{-1}
\ee
where $h:\Sigma\to SO(4);\;x\mapsto h(x)$ denotes a gauge transformation and $h_\sigma:=h(\hat{\sigma}),\;
h_t:=h(\hat{t}),\;h_f(\hat{f})$ with $\hat{\sigma}$ the barycenter of $\sigma$ etc.
% Notice that $B$ is really integrated over triangles T and that the face is composed out of the
%triangles $[\hat{T},\hat{t},\hat{\sigma}]$ where $T\cubset t\subset \sigma$. Therefore face $f$ and
%triangle $T$ always have the interior point $\hat{T}$ in common and $\hat{T}$ is the natural gauge
%transformation base point for $B_T=B_f$ and $g_{tf},g_{t\sigma}$ transform at $\hat{t},\hat{T}$
%and $\hat{t},hat{\sigma}$ respectively.

Hence the
traces of the exponentials
\be
\tr\lt(X_f^\pm g^\pm_{f t}g^\pm_{t\sig}g^\pm_{\sig t'}g^\pm_{t'f}\rt)
\ee
and the simplicity constraint
\be
X_{tf}^+\cdot X^+_{tf'}-X_{tf}^-\cdot X^-_{tf'}
\ee
are invariant quantities while the closure constraint transforms covariantly
\be
\sum_{f\subset t}X^+_{tf}\mapsto h_t\lt(\sum_{f\subset t}X^+_{tf}\rt) h_t^{-1}.
\ee
The desire to maintain gauge (co)invariance of action and constraints in the discretisation motivated
to introduce the quantities $X^\pm_{\sigma f}$ and $X^\pm_{\t f}$ which in the continuum limit reduce to
$X^\pm_f$ to leading order in the discretisation regulator.

\item One may wonder why we do not include $\delta$ functions enforcing the closure constraint
for the ``minus'' sector. As we will see, the measure is supported on configurations satisfying
$X^-_{tf}-u_t X_{tf}^+ u_t^{-1}$ for some $u_t\in$SU(2). Thus
\begin{equation}
\sum_{f\subset t} X^-_{tf}=-u_t[\sum_{f\subset t} X^+_{tf}]u_t^{-1}
\end{equation}
is already implied by the ``Plus'' sector. So we could include it but that would result
in an infinite constant $\delta(0)$ which drops out in correlators. We assume to have done this
already.

\end{itemize}
Remark: \\
It appears awkward, that here are more holonomies than B fields, suggesting a mismatch in the number of
$B$ and $A$ degrees of freedom in contrast to the classical theory.
Here we remark that the natural definition of the dual of a triangle really is the gluing of wedges
(see e.g. the second reference of \cite{book} in the notation used here and references therein).
The boundary $\partial f$ is naturally a composition of the half edges $[\hat{t},\hat{\sigma}]$
where the hat denotes the barycentre of tetrahedron and 4-simplex respectively. Thus, if we would
discretize the action using the holonomy around the $\partial f$ rather than around the wedges,
the discretized action only would depend on the
edges $e=[\hat{\sigma}\hat{\sigma\cap \sigma'}]\cap[\hat{\sigma\cap \sigma'},\hat{\sigma}']$
and the properties of the Haar measure ensure that the inegrals over $g^\pm_{\sigma f},g^\pm_{t f}$
reduce to the integrals over $g^\pm_e$. Thus, what we are doing here is to approximate
$\tr(B_f\cdot g_f)$ by $\sum_{\hat{\sigma}\in f} \tr(B_f\cdot g_{f,\sigma})$ where
$g_{f,\sigma}=g_{f t}\;g_{t\sigma}\; g_{\sigma t'}\;g_{t' f}$ is the corresponding wedge holonomy
{\it after} having introduced the redundant variables $g_{t\sigma},\; g_{t f}$. We are aware that this
presents a further modification of the model but it should be a mild one because both discretised actions
have the same continuum limit. In fact we will see that in the semiclassical (large-$j$) limit the
representations on the wedges essentially coincide so that effectively only the face holonomies are of
relevance. It is certainly possible to define the commutative B field model without this step, however,
it is very helpful to do so as it facilitates the solution to otherwise cumbersome bookkeeping problems.
We leave the definition of the model without a priori introduction of wedges for future work.

\subsection{Expansion of The Exponentials}

For the preparation of the integration of the holonomies $g^\pm_{\sig t}$ amd $g^{\pm}_{tf}$, we
would like to expand the factors $e^{i(1\pm\frac{1}{\g})\tr\lt(X^\pm_f g^\pm_{f t}g^\pm_{t\sig}g^\pm_
{\sig t'}g^\pm_{t'f}\rt)}$ in terms of the SU(2) unitary irreducible representation matrix elements $
\pi^j_{mn}(g)$. So we define the matrix $K_{mn}^{j}(Y)$, $Y\in\fs\fu(2)$, such that
\be
e^{i\tr\lt(Yg\rt)}=\sum_{j,m,n}K_{mn}^{j}(Y)\pi^j_{mn}(g)\label{decomp}
\ee
while the expression of $K_{mn}^{j}(Y)$ can be obtained by
\be
\frac{1}{\dim(j)}K_{mn}^{j}(Y)=\int\rmd g\ e^{i\tr\lt(Yg\rt)}\overline{\pi^j_{mn}(g)}=\int\rmd g\ e^{i\tr\lt
(Yg\rt)}{\pi^j_{nm}(g^{-1})}
\ee
Since $iY\equiv i\vec{y}\cdot\vec{\t}=\vec{y}\cdot\vec{\sig}$ ($\sig_j$ are Pauli matrices, $\t_j=-i
\sig_j$), we have the following relation
\be
iY=\frac{|\vec{y}|}{i}i\hat{y}\cdot\vec{\sig}=\frac{|\vec{y}|}{i}e^{i\frac{\pi}{2}\hat{y}\cdot\vec{\sig}}
\ee
Therefore
\be
\frac{1}{\dim(j)}K_{mn}^{j}(Y)=\int\rmd g\ e^{\frac{|\vec{y}|}{i}\tr\lt(ge^{i\frac{\pi}{2}\hat{y}\cdot\vec
{\sig}}\rt)}{\pi^j_{nm}(g^{-1})}=\int\rmd g\ e^{\frac{|\vec{y}|}{i}\tr\lt(g\rt)}{\pi^j_{nm}(e^{i\frac{\pi}{2}\hat
{y}\cdot\vec{\sig}}g^{-1})}
\ee
where in the last step we made a translation $g\to ge^{-i\frac{\pi}{2}\hat{y}\cdot\vec{\sig}}$.
Moreover we can expand the function $e^{\frac{|\vec{y}|}{i}\tr\lt(g\rt)}$ by the SU(2) characters
\be
e^{-i|\vec{y}|\tr\lt(g\rt)}=\sum_{k\in\mathbb{N}/2}\b_k(|\vec{y}|)\chi_k(g)\label{beta}
\ee
Then
\be
\frac{1}{\dim(j)}K_{mn}^{j}(Y)&=&\sum_{k\in\mathbb{N}/2}\b_k(|\vec{y}|)\sum_{l}\pi^j_{nl}(e^{i\frac
{\pi}{2}\hat{y}\cdot\vec{\sig}})\int\rmd g\ \pi^j_{lm}(g^{-1})\chi_k(g)\nonumber\\
&=&\sum_{k\in\mathbb{N}/2}\b_k(|\vec{y}|)\sum_{l}\pi^j_{nl}(e^{i\frac{\pi}{2}\hat{y}\cdot\vec{\sig}})\int
\rmd g\ \overline{\pi^j_{ml}(g)}\chi_k(g)\nonumber\\
&=&\sum_{k\in\mathbb{N}/2}\b_k(|\vec{y}|)\sum_{l}\pi^j_{nl}(e^{i\frac{\pi}{2}\hat{y}\cdot\vec{\sig}})
\frac{1}{\dim(j)}\delta_{jk}\delta_{ml}\nonumber\\
&=&\frac{1}{\dim(j)}\b_j(|\vec{y}|)\pi^j_{nm}(e^{i\frac{\pi}{2}\hat{y}\cdot\vec{\sig}})
\ee
Then plugging this result back into Eq.(\ref{decomp}) yields
\be
e^{i\tr\lt(Yg\rt)}=\sum_{j}\b_j(|\vec{y}|)\ \tr_j\lt(e^{i\frac{\pi}{2}\hat{y}\cdot\vec{\sig}}g\rt)=\sum_{j}\b_j(|
\vec{y}|)\ \tr_j\lt(i\hat{y}\cdot\vec{\sig}g\rt)
\ee
by using this identity, we have ($X^\pm\equiv \vec{X}^\pm\cdot\vec{\t}=\vec{X}^\pm\cdot(-i\vec
{\sig})$)
\be
e^{i(1\pm\frac{1}{\g})\tr\lt(X^\pm_f g^\pm_{f t}g^\pm_{t\sig}g^\pm_{\sig t'}g^\pm_{t'f}\rt)}=\sum_{j^
\pm_{\sig f}}\b_{j^\pm_{\sig f}}\lt(\lt|1\pm\frac{1}{\g}\rt|\lt|\vec{X}^\pm_{f}\rt|\rt)\ \tr_{j^\pm_{\sig f}}\lt(i
\hat{X}^\pm_f\cdot\vec{\sig} g^\pm_{f t}g^\pm_{t\sig}g^\pm_{\sig t'}g^\pm_{t'f}\rt)
\ee
Inserting this result into the expression of the partition function, we obtain
\be
Z(\ck)&=&\int_{II\pm}\prod_{f}\rmd^3X^+_f\rmd^3X^-_f\prod_{(\sig,t)}\rmd g^+_{\sig t}\rmd g^-_{\sig
t}\prod_{(t,f)}\rmd g^+_{tf}\rmd g^-_{tf}\prod_{t;f,f'\subset t}\delta\lt(X_{tf}^+\cdot X^+_{tf'}-X_{tf}^-
\cdot X^-_{tf'}\rt)\prod_{t}\delta\Big(\sum_{f\subset t}X_{tf}^+\Big)\nonumber\\
&&\times\sum_{\{j^+_{\sig f}\}}\prod_{(\sig,f)}\b_{j^+_{\sig f}}\lt(\lt|1+\frac{1}{\g}\rt|\lt|\vec{X}^+_{f}\rt|
\rt) \tr_{j^+_{\sig f}}\lt(i\hat{X}^+_f\cdot\vec{\sig} g^+_{f t}g^+_{t\sig}g^+_{\sig t'}g^+_{t'f}\rt)
\nonumber\\
&&\times\sum_{\{j^-_{\sig f}\}}\prod_{(\sig,f)}\b_{j^-_{\sig f}}\lt(\lt|1-\frac{1}{\g}\rt|\lt|\vec{X}^-_{f}\rt|\rt)
\tr_{j^-_{\sig f}}\lt(i\hat{X}^-_f\cdot\vec{\sig} g^-_{f t}g^-_{t\sig}g^-_{\sig t'}g^-_{t'f}\rt).\label{Z0}
\ee

%Furthermore for the convenience of the implementing simplicity constraint,
%we distinguish the
%triangles
%corresponding to different tetrahedra and introduce the variables
%$X^\pm_{tf}$, thus
%\be
%Z(\ck)&=&\int_{II\pm}\prod_{(t,f)}\rmd^3X^+_{tf}\rmd^3X^-_{tf}\prod_{(\sig,t)}
%\rmd g^+_{\sig t}\rmd
%g^-_{\sig t}\prod_{(t,t')}\rmd g^+_{tt'}\rmd g^-_{tt'}\prod_{(t,f),(t,f')}
%\dela\lt(X_{tf}^+\cdot X^+_{tf'}-X_{tf}^-\cdot X^-_{tf'}\rt)\nonumber\\
%&&\times\prod_{t}\delta\Big(\sum_{f\subset t}X_{tf}^+\Big)\prod_{(t,t'f)}
%\delta\lt(X^+_{tf}-X^+_{t'f}\rt)
%\delta\lt(X^-_{tf}-X^-_{t'f}\rt)\nonumber\\
%&&\times\sum_{\{j^+_{\sig f}\}}\prod_{(\sig,f)}\b_{j^+_{\sig f}}
%\lt([1+\frac{1{\g}]\lt|\vec{X}^+_{f}\rt|\rt)\ \tr_{j^+_{\sig f}}
%\lt(i\hat{X}^+_f\cdot\vec{\sig}g^+_{\sig t}g^+_{tt'}g^+_{t'\sig}\rt)
%\nonumber\\
%&&\times\sum_{\{j^-_{\sig f}\}}\prod_{(\sig,f)}\b_{j^-_{\sig f}}\lt([1-\frac{1}%{\g}]\lt|\vec{X}^-_{f}\rt|\rt)\tr_j^-_{\sig f}}\lt(i\hat{X}^-_f
%\cdot\vec{\sig}g^-_{\sig t}g^-_{tt'}g^-_{t'\sig}\rt)
%\ee
%where we introduce the delta functions $\delta\lt(X^\pm_{tf}-X^\pm_{t'f}\rt)$
%o identify the flux variables $X^\pm_{tf}$ and $X^\pm_{t'f}$ associated with
%the same triangle shared by two different
%tetrahedra.

\section{Implementation of Simplicity Constraint}\label{implementation}

\subsection{Linearizing the Simplicity Constraint}

In order to implement the simplicity constraints via the delta functions
$\delta\lt(X_{f}^+\cdot X^+_
{f'}-X_{f}^-\cdot X^-_{f'}\rt)$ for each
tetrahedron it proves convenient to pass from this quadratic expression
to an integral of linear expressions directly at the level of measures
(in the gravitational sector II$\pm$). In this subsection
we are dealing with a single tetrahedron $t$, thus we ignore the $t$ label of
$X^\pm_{tf}$. \\
\\
Consider the four flux variables $X_{f}^\pm$ $(f=1,\cdots,4)$ associated with a tetrahedron $t$.
Define the symmetric matrix $l_{ff'}^\pm:=X_{f}^\pm\cdot X_{f'}^\pm,\;1\le f,f'\le 4$. Then
$l^\pm_{ff'}$ determines the $X^\pm_f$ up to an O(3) matrix $O$. Denote by
L the range of the map $\{X^\pm_f\}_{f=1}^f\mapsto \{l^\pm_{ff'}\}_{1\le f\le f'\le 4}$ (as a subset
of $\mathbb{R}^{10}$, L is
constrained in particular by the Cauchy--Schwarz inequality). Then we can define a map
$Y:\; O(3)\otimes L\to \mathbb{R}^{12},\;(g,l)\mapsto (g X_1(l),g X_2(l),g X_3(l),g X_4(l))$
where $X_f(l)$ is any solution of $l_{ff'}=X_f\cdot X_{f'}$.

In the following result we drop the $\pm$ for convenience.
\begin{Lemma}\label{measure}
We have $\det((l_{ff'})=0$. Given $F:\;\mathbb{R}^{12}\to \mathbb{R}$ define
$\tilde{F}:\;O(3)\times L\to \mathbb{R}$ by $\tilde{F}:=F\circ Y$. Then
\begin{equation}
\int_{\mathbb{R}^{12}}\; \prod_{f=1}^4\mathrm{d}^3X_{f}\;F =
\int_{O(3)}\mathrm{d}g\int_{\mathbb{R}\times L}\;d^{10}l\;
\delta\lt(\det((l_{ff'}))\rt)\;\tilde{F}
\end{equation}
where $\mathrm{d}g$ is the SU(2) Haar measure (up to normalisation) and $\tilde{F}$ is trivially
extended off the surface $\det(l)=0$.
\end{Lemma}
\textbf{Proof:}
Up to measure zero sets, $X_1, X_2, X_3$ will be linearly independent and define a 3 metric
$l_{ab}=X_a\cdot X_b$. Accordingly (since $X_4$ is a linear combination of $X_1,X_2,X_3$)
\begin{equation}
X_4=l^{ab} (X_b\cdot X_4) X_a=l^{ab}\; l_{b4}\;X_a
\end{equation}
is a linear combination of these vectors and $l^{ac} l_{cb}=\delta^a_b$. We obtain the constraint
\begin{equation}
l_{44}=X_4\cdot X_4=l^{ab} l_{4a} l_{4b}\label{3.3}
\end{equation}
among the $l_{ff'}$. On the other hand
\begin{equation}
\det(l_{ff'})=\det\left(\begin{array}{cc}
l_{44} & l_{4b}\\
l_{a4} & l_{ab}
\end{array} \right)
=l_{44} \det(l_{ab}-l_{4a} l_{4b}/l_{44})
=l_{44}^{-2}\; [\det(X_a^i)]^2\;\det(l_{44} \delta_{ij}-l_{4i} l_{4j})
\end{equation}
with $l_{4i}=X^a_i l_{4a}$ and $X^a_i$ is the inverse of $X_a^i$. The computation of the remaining
determinant is elementary and yields
\begin{equation}
\det(l_{ff'})=\det(l_{ab})[l_{44}-l^{ab} l_{4a} l_{4b}]
\end{equation}
which is proportional to the constraint Eq.(\ref{3.3}), hence $\det(l_{ff'})=0$.

In order to write an integral over $X_1,..,X_4$ in terms of the independent coordinates
$l_{ab},l_{4a},\vec{\alpha}$ where $\alpha$ parametrises the rotation $g$, we must compute the Jacobian
\begin{equation}
J=|\det(\frac{\partial(X_1,X_2,X_3,X_4)}{\partial(l_{ab},\vec{\alpha},l_{4a})})|
\end{equation}
Since only $X_4$ depends on $l_{4a}$ this immediately simplifies to
\begin{equation}
J=\frac{1}{\sqrt{\det(l_{ab})}}\;
|\det(\frac{\partial(X_1,X_2,X_3)}{\partial(l_{ab},\vec{\alpha})})|
\end{equation}
To compute the remaining determinant we choose for instance the following parametrisation
\begin{eqnarray}
X_1 &=& \frac{l_{13}}{\sqrt{l_{33}}} b_3+\sqrt{l_{11}-l_{13}^2/l_{33}}(\cos(\gamma+\chi) b_1
+\sin(\gamma+\chi) b_2)
\nonumber\\
X_2 &=& \frac{l_{23}}{\sqrt{l_{33}}} b_3+\sqrt{l_{22}-l_{23}^2/l_{33}}(\cos(\chi) b_1
+\sin(\chi) b_2)
\nonumber\\
X_3 &=& \sqrt{l_{33}} b_3
\end{eqnarray}
with the Euler angles $\vec{\alpha}=(\phi,\theta,\chi),\;\phi,\chi\in [0,2\pi],\;\theta\in [0,\pi]$ and
the orthonormal right oriented basis
\begin{equation}
b_3=(\sin(\theta)\cos(\phi),\sin(\theta)\sin(\phi),\cos(\theta)),\;b_1=b_{3,\theta},\;b_2
=b_{3,\phi}/\sin(\theta)
\end{equation}
together with
\begin{equation}
\cos(\gamma)
=\frac{l_{12} l_{33}-l_{13} l_{23}}{\sqrt{(l_{11} l_{33}-l_{13}^2)(l_{22} l_{33}-l_{23}^2)}},\;\;
\sin(\gamma)
=\frac{\sqrt{\det(l_{ab}) l_{33}}}{\sqrt{(l_{11} l_{33}-l_{13}^2)(l_{22} l_{33}-l_{23}^2)}}
\end{equation}
This defines the map $Y$ above and the reader may check that the relations $X_a\cdot X_b=l_{ab}$
are satisfied for any $\vec{\alpha}$.
The computation of the Jacobian is much simplified by noticing that the matrix
\begin{equation}
\frac{\partial (X_3, X_2, X_1)}{\partial(l_{33},\theta,\phi,\chi,l_{22},l_{23},l_{11},l_{12},l_{13})}
\end{equation}
consists of 3x3 blocks and is upper block trigonal with non singular matrices as diagonal block
entries. Accordingly its determinant is the product of the determinants of the diagonal block matrices and yields
after a short comutation the value $\sin(\theta)/(8\sqrt{\det(l_{ab})})$. Due to the absolute value
the Jacobian is thus given by
\begin{equation}
J=\frac{\sin(\theta)}{8\det(l_{ab})}\;
\end{equation}
It is not difficult to check that for the Euler angle parametrisation we have up to a normalisation
constant the following expression for the Haar measure
\begin{equation}
dg=d\chi\;d\phi\;d\theta\;\sin(\theta)/8
\end{equation}
We can therefore finish the proof by
\begin{eqnarray}
&& \int_{\mathbb{R}^{12}}\; d^3X_1\; d^3X_2\; d^3X_3\; d^3X_4\; F
\nonumber\\
&=& \int_{O(3)}\; dg\;\int_{B}\; \prod_{a\le b\le 3} dl_{ab}\; \prod_{a=1}^3 \; dl_{4a}\;
\frac{\tilde{F}}{\det((l_{ab}))}
\nonumber\\
&=& \int_{O(3)}\; dg\;\int_{B\times\mathbb{R}}\; \prod_{a\le b\le 3} dl_{ab}\; \prod_{f=1}^4 \; dl_{4f}\;
\frac{\tilde{F}}{\det((l_{ab}))}\; \delta(l_{44}-l^{ab} l_{4a} l_{4b})
\nonumber\\
&=& \int_{O(3)}\; dg\;\int_{B\times\mathbb{R}}\; \prod_{f\le f'\le 4} dl_{ff'}\;
\delta(\det(l_{ff'}))\;\tilde{F}
\end{eqnarray}
$\Box$\\

As usual in path integrals we will not worry about normalisation constants as they drop out in
correlators. The preceding lemma is crucial for establishing the following result.

\begin{Lemma}
For each tetrahedron $t$ ($u_t\in \text{SO(3)}$\footnote{SO(3) is considered as the upper hemisphere of SU(2), while their Haar measure is different by a factor of 2.}). $u_t$ can be viewed as the parametrization of the normal for the tetrahedron $t$ (see Eq.(\ref{3.18})).
\begin{eqnarray}
\prod_{f,f'=1}^4\delta\lt(X_{f}^+\cdot X_{f'}^+-X^-_{f}\cdot X^-_{f'}\rt)=\delta\lt(\det(X_{f}^+\cdot X_{f'}^+)\rt)\int\mathrm{d}u_{t}\prod_{f=1}^4\delta\lt(X_{f}^++u_{t}X_{f}^-u_{t}^{-1}\rt)
\end{eqnarray}
in the solution sector II$\pm$ of the simplicity constraint \cite{FK}.
\end{Lemma}

\textbf{Proof:} Essentially we need to prove that for all continuous function $f(X^+_f,X^-_f)$ $f=1,\cdots,4$ vanishing in the topological solution sector I$\pm$ of the simplicity constraint
\be
&&\int\prod_{f=1}^4\mathrm{d}^3X_{f}^+\mathrm{d}^3X_{f}^-\prod_{f,f'=1}^4\delta\lt(X_{f}^+\cdot X_{f'}^+-X^-_{f}\cdot X^-_{f'}\rt)f\lt(X^+_f,X^-_f\rt)\nonumber\\
&=&\int\prod_{f=1}^4\mathrm{d}^3X_{f}^+\mathrm{d}^3X_{f}^-\delta\lt(\det(X_{f}^+\cdot X_{f'}^+)\rt)\int\mathrm{d}u_{t}\prod_{f=1}^4\delta\lt(X_{f}^++u_{t}X_{f}^-u_{t}^{-1}\rt)f\lt(X^+_f,X^-_f\rt)
\ee
From the left hand side, by using Lemma \ref{measure}, we transform the coordinates from $X^\pm_f$ to $u^\pm_t$ and $l_{ff'}^\pm$, constrained by $\det(l_{ff'}^\pm)=0$
\be
&&\int\prod_{f=1}^4\mathrm{d}^3X_{f}^+\mathrm{d}^3X_{f}^-\prod_{f,f'=1}^4\delta\lt(X_{f}^+\cdot X_{f'}^+-X^-_{f}\cdot X^-_{f'}\rt)f\lt(X^+_f,X^-_f\rt)\nonumber\\
&=&\int\mathrm{d}u_t^+\rmd u_t^-\prod_{f,f'=1}^4\mathrm{d}l_{ff'}^+\mathrm{d}l_{ff'}^-
\delta\lt(\det(l_{ff'}^+)\rt)\delta\lt(\det(l_{ff'}^-)\rt)\prod_{f,f'=1}^4\delta\lt(l_{ff'}^+-l_{ff'}^-\rt)
f\lt(l_{ff'}^+,l_{ff'}^-,u_t^+,u^-_t\rt)\nonumber\\
&=&\int\mathrm{d}u_t^+\rmd u_t^-\prod_{f,f'=1}^4\mathrm{d}l_{ff'}^+
\delta\lt(\det(l_{ff'}^+)\rt)\delta\lt(\det(l_{ff'}^+)\rt)f\lt(l_{ff'}^+,l_{ff'}^+,u_t^+,u^-_t\rt)\nonumber\\
&=&\int\prod_{f=1}^4\mathrm{d}^3X_{f}^+\int\rmd u_t^-\delta\lt(\det(X_{f}^+\cdot X_{f'}^+)\rt)f\lt(X^+_f,-u^-_tX^+_f(u^-_t)^{-1}\rt)\nonumber\\
&=&\int\prod_{f=1}^4\mathrm{d}^3X_{f}^+\mathrm{d}^3X_{f}^-\delta\lt(\det(X_{f}^+\cdot X_{f'}^+)\rt)\int\mathrm{d}u_{t}\prod_{f=1}^4\delta\lt(X_{f}^++u_{t}X_{f}^-u_{t}^{-1}\rt)f\lt(X^+_f,X^-_f\rt)
\ee
where we restrict ourself in the gravitational sector $II\pm$.
$\Box$

Notice that strictly speaking we should be using the Haar measure $\rmd u_t^{\pm}$ on O(3) rather than SO(3)
which is just the sum of two Haar measures on SO(3) twisted by a reflection so that we actually
get an integral over $SO(3)$
of a sum of $\delta$ distributions
$\delta(X-u_t X^+u_t^{-1})+\delta(X+u_t X^+u_t^{-1})$ with $u_t\in SO(3)$. This is expected because
the simplicity constraints do not select either of the two sectors (gravitational and topological).
As usual in spin-foam models, we consider a restriction of the model to the purely gravitational sector in the above lemma.

Here we note that the singular factor $\delta(\det(X_{tf}^+\cdot X_{tf'}^+))$ is essentially a $\delta(0)$ and can be divided out by an appropriate Faddeev-Popov procedure \cite{FK}. And the linearized simplicity constraint $\delta\lt(X_{f}^++u_{t}X_{f}^-u_{t}^{-1}\rt)$ has clear geometrical interpretation that for each tetrahedron $t$, there exists a unit 4-vector ${n}_t=(n_t^1,n_t^2,n_t^3,n_t^4)$ corresponding to the SU(2) element
\be
u_t=\left(
    \begin{array}{cc}
      n_t^1+in_t^2 & n_t^3+in_t^4 \\
      -(n_t^3-in_t^4) & n_t^1-in_t^2 \\
    \end{array}
  \right)\label{3.18}
\ee
such that $*B^{IJ}_{f}n_{t,I}=0$.

Thus the constrained measure of the flux variables in Eq.(\ref{Z0}) is written as (we denote
$g_t=u_t$ in what follows)
\be
&&\prod_{f}\rmd^3X^+_{f}\rmd^3X^-_{f}\prod_{t}\int\mathrm{d}u_{t}\prod_{f\subset t}\delta\lt(X_{tf}^
+ +u_{t}X_{tf}^-u_{t}^{-1}\rt)
\prod_{t}\delta\Big(\sum_{f\subset t}X_{tf}^+\Big)\label{1}
\ee
Note that the measure $\rmd^3X^\pm_f$ can be considered as the measure $\rmd^3X^\pm_{tf}$ constrained by the parallel-transportation condition $\delta\lt(X^\pm_{tf}-g^\pm_{tf} X_f^\pm (g^\pm_{tf})^{-1}\rt)$.

In particular we see, that it is possible to justify the passing between the quadratic simplicity
constraints employed by the BC model and the lineraised simplicity constraints of the EPRL and
FK models respectively, at the level of measures in terms of the commuting B variables.

\subsection{Imposing the Simplicity Constraint}

In what follows we make the ad hoc restriction to the gravitational sector as mentioned at the
end of the previous subsection.\\
\\
Performing a polar decomposition of the variables $X^{\pm}_{f}$ and $X^\pm_{tf}$, we
introduce the new variables $\rho_{f}^\pm\in \mathbb{R}^+$ and $N_{f}^\pm\in \text{SU(2)}$
\be
X^{\pm}_{f}=\rho_{f}^\pm N_{f}^\pm\t_3(N_{f}^\pm)^{-1}\ \ \ \ X^{\pm}_{tf}=\rho_{f}^\pm N_{tf}^\pm
\t_3(N_{tf}^\pm)^{-1}\ \ \ \ N_{tf}^\pm=g_{tf}^\pm N_f^\pm
\ee
where $\rho_{f}^\pm=||X^\pm_{f}||$, $\hat{X}^\pm_{f}\cdot\vec{\t}=N_{f}^\pm\t_3(N_{f}^\pm)^
{-1}$ and the same for $X^\pm_{tf}$. Note that given $X^\pm\in\fs\fu(2)$, $N^\pm\in\text{SU(2)}$ is
determined up to a U(1) rotation $h_\phi\in\text{U(1)}$, which leaves $\t_3$ invariant.
\be
h_\phi=\left(
         \begin{array}{cc}
           e^{i\phi} & 0 \\
           0 & e^{-i\phi} \\
         \end{array}
       \right)
\ee
The associated equivalence relation is called the Hopf fibration of $SU(2)=S^3$ as a $U(1)$ bundle
over the coset space $SU(2)/U(1)\cong S^2$. It is convenient for given unit vector
$\vec{n}(\theta,\phi)=(\sin(\theta)\cos(\phi),\sin(\theta)\sin(\phi),\cos(\theta)$ to fix the
representative $N=ie^j(\theta,\phi)\sigma_j$ with the unit vector
$\vec{e}(\theta,\phi)=(\sin(2\theta)\cos(\phi),\sin(2\theta)\sin(\phi),\cos(2\theta)$ parametrising
a point on $S^2$.  

The linearized simplicity constraint $X_{tf}^+=-u_{t}X_{tf}^-u_{t}^{-1}$ implies that there exists
a $h_{\phi_{tf}}\in \text{U(1)}$ for each pair of $(t,f)$ such that
\be
\rho_{f}^+=\rho_{f}^-=\rho_{f}\ \ \ \ \ \text{and} \ \ \ \ \lt(N^+_{tf},N^-_{tf}\rt)=\lt(N_{tf}h_{\phi_{tf}},u_tN_
{tf}h^{-1}_{\phi_{tf}}\eps\rt)\label{**}
\ee
where the diagonal U(1) invariance is absorbed into the definition of $N_{tf}$, we only take care of the anti-diagonal one by introducing $\phi_{tf}$, and
\be
\eps=\left(
       \begin{array}{cc}
         0 & 1 \\
         -1 & 0 \\
       \end{array}
     \right).
\ee

We now reexpress the constrained measure in terms of the new variables $\rho^\pm_{f}$
and $N^\pm_{f}$. The Lebesgue measure $\rmd^3X$ can be expressed in the spherical
coordinates (when one integrates any function $f$ of $X$ independent of the U(1) part)
\be \label{d3X}
\int f \rmd^3X=\int f\rho^2\rmd\rho\rmd^2\O=\int f\rho^2 \; d\rho\;\rmd N
\ee
where $\rmd^2\O$ is the round measure on $S^2$ and $\rmd N$ is the Haar measure on SU(2).

\begin{Lemma}

For any continuous function $f(N^+,N^-)$ on SU(2)$\times$SU(2), up to overall constant factor ($\delta_{S^2}(\cdots)$ is the delta function on $S^2$)
\be
&&\int_{\text{SU(2)}\times\text{SU(2)}}\rmd N^+\rmd N^-\ \delta_{S^2}\Big(N^-\t_3(N^-)^{-1}+uN^+\t_3(N^+)^{-1}u^{-1}\Big)\ f(N^+,N^-)\nonumber\\
&=&\int_{\text{SU(2)}\times\text{SU(2)}}\rmd N^+\rmd N^-\int_{\text{SU(2)}}\rmd N\int_0^{2\pi}\rmd\phi\ \delta_{\text{SU(2)}}\lt(N^+,Nh_{\phi}\rt)\ \delta_{\text{SU(2)}}\lt(N^-,u N h^{-1}_{\phi}\eps\rt)\ f(N^+,N^-)\label{lemma}
\ee
which gives
\be
\delta_{S^2}\Big(N^-\t_3(N^-)^{-1}+uN^+\t_3(N^+)^{-1}u^{-1}\Big)=\int_0^{2\pi}\rmd\phi\ \delta_{\text{SU(2)}}\lt(N^-,uN^+h^{-1}_{2\phi}\eps\rt)
\ee

\end{Lemma}

\textbf{Proof:} On the right hand side of Eq.(\ref{lemma}),
\be
&&\int_{\text{SU(2)}\times\text{SU(2)}}\rmd N^+\rmd N^-\int_{\text{SU(2)}}\rmd N\int_0^{2\pi}\rmd\phi\ \delta_{\text{SU(2)}}\lt(N^+,Nh_{\phi}\rt)\ \delta_{\text{SU(2)}}\lt(N^-,u N h^{-1}_{\phi}\eps\rt)\ f(N^+,N^-)\nonumber\\
&=&\int_{\text{SU(2)}\times\text{SU(2)}}\rmd N^+\rmd N^-\int_0^{2\pi}\rmd\phi\ \delta_{\text{SU(2)}}\lt(N^-,uN^+h^{-1}_{2\phi}\eps\rt)\ f(N^+,N^-)\nonumber\\
&=& \int_{\text{SU(2)}}\rmd N^+\int_0^{2\pi}\rmd\phi\ f\lt(N^+,uN^+h^{-1}_{2\phi}\eps\rt)\label{lemma1}
\ee

On the left hand side, we can express the Haar measure $\rmd N^-$ in terms of Euler angles
\be
\int_{\text{SU(2)}}\rmd N^-\ldots=\frac{1}{16\pi^2}\int_0^{2\pi}\rmd\phi_2\int_0^\pi\rmd\theta\ \sin\theta\int_0^{4\pi}\rmd\phi_1\ldots
\ee
And the delta function $\delta_{S^2}\Big(N^-\t_3(N^-)^{-1}+uN^+\t_3(N^+)^{-1}u^{-1}\Big)$ is the delta function on $S^2$, which is coordinatized by $\theta\in[0,\pi]$ and $\phi_2\in[0,2\pi]$. By explicit computation
\be
&&\int_0^{2\pi}\rmd\phi_2\int_0^\pi\rmd\theta\ \sin\theta\ \delta_{S^2}\Big(N^-\t_3(N^-)^{-1}+uN^+\t_3(N^+)^{-1}u^{-1}\Big)\ f(N^+,N^-)\nonumber\\
&=&f\lt(N^+,uN^-h^{-1}_{\phi_1}\eps\rt)
\ee
Therefore the left hand side of Eq.(\ref{lemma}) reduces to
\be
\int_{\text{SU(2)}}\rmd N^+\int_0^{4\pi}\rmd\phi_1\ f\lt(N^+,uN^-h^{-1}_{\phi_1}\eps\rt).
\ee
which is identical to the right hand side Eq.(\ref{lemma1}).\\
$\Box$

Using this we rewrite the constrained measure up to an
unimportant overall constant as
\be
&&\prod_{f}\rmd^3X^+_{f}\rmd^3X^-_{f}\prod_{t}\int\mathrm{d}u_{t}\prod_{f\subset t}\delta\lt(X_{tf}^
++u_{t}X_{tf}^-u_{t}^{-1}\rt)
\prod_{t}\delta\Big(\sum_{f\subset t}X_{tf}^+\Big)\nonumber\\
&=&\prod_{f}\rmd\rho^+_{f}\rmd N^+_{f}\lt(\rho^+_{f}\rt)^2\rmd\rho^-_{f}\rmd N^-_{f}\prod_{t}\int
\mathrm{d}u_{t}\prod_{f\subset t}\delta\lt(\rho_f^+-\rho^-_f\rt)\int_0^{2\pi}\rmd\phi_{tf} \delta\lt(N^-_
{tf}, u_{t}N_{tf}^+h^{-1}_{2\phi_{tf}}\eps\rt)\nonumber\\
&&\times\prod_{t}\delta\Big(\sum_{f\subset t}\rho_{f}^+N_{tf}^+\t_3(N_{tf}^+)^{-1}\Big)\label{2}
\ee

We insert this result into the partition function
\be
Z(\ck)&=&\int\prod_{(\sig,t)}\rmd g^+_{\sig t}\rmd g^-_{\sig t}\prod_{(t,f)}\rmd g^+_{tf}\rmd g^-_{tf}
\prod_{f}\rmd\rho^+_{f}\rmd N^+_{f}\lt(\rho^+_{f}\rt)^2\rmd\rho^-_{f}\rmd N^-_{f}\prod_{t}\mathrm{d}
u_{t}\prod_{f\subset t}\delta\lt(\rho_f^+-\rho^-_f\rt)\nonumber\\
&&\times\int_0^{2\pi}\rmd\phi_{tf} \delta\lt(N^-_{tf}, u_{t}N_{tf}^+h^{-1}_{2\phi_{tf}}\eps\rt)\prod_{t}
\delta\Big(\sum_{f\subset t}\rho_{f}^+N_{tf}^+\t_3(N_{tf}^+)^{-1}\Big)\nonumber\\
&&\times\sum_{\{j^+_{\sig f}\}}\prod_{(\sig,f)}\b_{j^+_{\sig f}}\lt(\lt|1+\frac{1}{\g}\rt|\rho^+_{f}\rt)
\tr_{j^+_{\sig f}}\lt(iN^+_f\sig_3\lt(N_f^+\rt)^{-1}g^+_{f t}g^+_{t\sig}g^+_{\sig t'}g^+_{t'f}\rt) \nonumber\\
&&\times\sum_{\{j^-_{\sig f}\}}\prod_{(\sig,f)}\b_{j^-_{\sig f}}\lt(\lt|1-
\frac{1}{\g}\rt|\rho^-_{f}\rt)
\tr_{j^-_{\sig f}}\lt(iN^-_f\sig_3\lt(N_f^-\rt)^{-1}g^-_{f t}g^-_{t\sig}
g^-_{\sig t'}g^-_{t'f}\rt)\label{*}
\ee
Performing a translation of the Haar measure $\rmd g^\pm_{tf}$
\be
\rmd g^\pm_{tf}\mapsto \rmd \lt(g_{tf}^\pm N^\pm_f\rt)=\rmd N_{tf}^\pm
\ee
(notice that $\rmd N_{tf}^\pm$ and $\rmd N_{f}^\pm$ are Haar measures on SU(2))
we see that the integrand depends on $N^\pm_{tf}$ only so that the
integrals over $\rmd N_f$ are trivial and give unity (upon proper normalisation). The partition function
therefore reduces to
\be
Z(\ck)&=&\int\prod_{(\sig,t)}\rmd g^+_{\sig t}\rmd g^-_{\sig t}\prod_{(t,f)}\rmd N^+_{tf}\rmd N^-_{tf}
\prod_{f}\rmd\rho_{f}\lt(\rho_{f}\rt)^2\prod_{t}\mathrm{d}u_{t}\prod_{f\subset t}\nonumber\\
&&\times\int_0^{2\pi}\rmd\phi_{tf} \delta\lt(N^-_{tf}, u_{t}N_{tf}^+h^{-1}_{2\phi_{tf}}\eps\rt)\prod_{t}
\delta\Big(\sum_{f\subset t}\rho_{f} N_{tf}^+\t_3(N_{tf}^+)^{-1}\Big)\nonumber\\
&&\times\sum_{\{j^+_{\sig f}\}}\prod_{(\sig,f)}\b_{j^+_{\sig f}}\lt(\lt|1+\frac{1}{\g}\rt|\rho_{f}\rt)
\tr_{j^+_{\sig f}}\lt(iN^+_{t'f}\sig_3\lt(N_{tf}^+\rt)^{-1}g^+_{t\sig}g^+_{\sig t'}\rt) \nonumber\\
&&\times\sum_{\{j^-_{\sig f}\}}\prod_{(\sig,f)}\b_{j^-_{\sig f}}\lt(\lt|1-\frac{1}{\g}\rt|\rho_{f}\rt)
\tr_{j^-_{\sig f}}\lt(iN^-_{t'f}\sig_3\lt(N_{tf}^-\rt)^{-1}g^-_{t\sig}g^-_{\sig t'}\rt)
\ee
where we also performed the integral over $\rho^-_f$.

Next we perform the integral over $\rmd N_{tf}^-$ to solve the simplicity constraint
(implementing Eq.(\ref{**}))
\be
Z(\ck)&=&\int\prod_{(\sig,t)}\rmd g^+_{\sig t}\rmd g^-_{\sig t}\prod_{(t,f)}\rmd N_{tf}\prod_{f}\rmd\rho_
{f}\lt(\rho_{f}\rt)^2\prod_{t}\mathrm{d}u_{t}\prod_{(t,f)}\rmd\phi_{tf}\prod_{t}\delta\Big(\sum_{f\subset
t}\rho_{f} N_{tf}\t_3N_{tf}^{-1}\Big)\nonumber\\
&&\times\sum_{\{j^+_{\sig f}\}}\prod_{(\sig,f)}\b_{j^+_{\sig f}}\lt(\lt|1+\frac{1}{\g}\rt|\rho_{f}\rt)
\tr_{j^+_{\sig f}}\lt(iN_{t'f}h_{\phi_{t'f}}\sig_3h_{\phi_{tf}}^{-1}\lt(N_{tf}\rt)^{-1}g^+_{t\sig}g^+_{\sig t'}\rt)
\nonumber\\
&&\times\sum_{\{j^-_{\sig f}\}}\prod_{(\sig,f)}\b_{j^-_{\sig f}}\lt(\lt|1-\frac{1}{\g}\rt|\rho_{f}\rt)
\tr_{j^-_{\sig f}}\lt(iu_{t'}N_{t'f}h^{-1}_{\phi_{t'f}}\eps\sig_3\eps^{-1}h_{\phi_{tf}}N_{tf}^{-1}u_{t}^{-1}
g^-_{t\sig}g^-_{\sig t'}\rt)
\ee
where we also have performed the translation $N^+_{tf}\to N^+_{tf} h_{\phi_{tf}}$.
Performing the translation $g^-_{\sig t}\to g^-_{\sig t}u_t^{-1}$, the integrand no longer
depends on $u_t$ and the $u_t$ integral gives unity, leaving us with
\be
Z(\ck)&=&\int\prod_{(\sig,t)}\rmd g^+_{\sig t}\rmd g^-_{\sig t}\prod_{(t,f)}\rmd N_{tf}\prod_{f}\rmd\rho_
{f}\lt(\rho_{f}\rt)^2\prod_{(t,f)}\rmd\phi_{tf}\prod_{t}\delta\Big(\sum_{f\subset t}\rho_{f}N_{tf}\t_3N_
{tf}^{-1}\Big)\nonumber\\
&&\times\sum_{\{j^+_{\sig f}\}}\prod_{(\sig,f)}\b_{j^+_{\sig f}}\lt(\lt|1+\frac{1}{\g}\rt|\rho_{f}\rt)
\tr_{j^+_{\sig f}}\lt(iN_{t'f}h_{\phi_{t'f}}\sig_3h_{\phi_{tf}}^{-1}\lt(N_{tf}\rt)^{-1}g^+_{t\sig}g^+_{\sig t'}\rt)
\nonumber\\
&&\times\sum_{\{j^-_{\sig f}\}}\prod_{(\sig,f)}\b_{j^-_{\sig f}}\lt(\lt|1-\frac{1}{\g}\rt|\rho_{f}\rt)
\tr_{j^-_{\sig f}}\lt(iN_{t'f}h^{-1}_{\phi_{t'f}}\eps\sig_3\eps^{-1}h_{\phi_{tf}}N_{tf}^{-1}g^-_{t\sig}g^-_
{\sig t'}\rt)\label{3.36}
\ee

%In order to further perform the integration of $N_{tf}$ and $g_{\sig t}$, we ex%press the $\tr_j$ in terms of the SU(2) representation matrix elements. For the%self-dual part (summing repeated indices):
%\be
%&&\tr_{j^+_{\sig f}}\lt(iN_{t'f}\sig_3\lt(N_{tf}\rt)^{-1}g^+_{t\sig}g^+_{\sig t%'}\rt)\nonumber\\
%&=&\pi^{j^+_{\sig f}}_{a_{\sig f}b_{\sig f}}\lt(N_{t'f}\rt) \pi^{j^+_{\sig f}}_%{b_{\sig f}c_{\sig f}}\lt(i
%\sig_3\rt) \pi^{j^+_{\sig f}}_{c_{\sig f}d_{\sig f}}\lt(N_{tf}^{-1}\rt) \pi^{j^%+_{\sig f}}_{d_{\sig f}a_{\sig f}}\lt g^+_{t\sig}g^+_{\sig t'}\rt)
%\ee

Recall that for any SL(2,$\bbc$) matrix $g$
\be
g=\left(
    \begin{array}{cc}
      a & b \\
      c & d \\
    \end{array}
  \right)
\ee
the representation matrix element $\pi^j_{mn}(g)$ reads
\be
\pi^j_{mn}(g)=\sum_l\frac{\sqrt{(j+m)!\ (j-m)!\ (j+n)!\ (j-n)!}}{(j+n-l)!\ (m-n+l)!\ (j-m-l)!\ l!}\ a^{j+n-l}\ b^
{m-n+l}\ c^l\ d^{j-m-l}
\ee
Applying this to $i\sig_3$
\be
i\sig_3=\left(
    \begin{array}{cc}
      i & 0 \\
      0 & -i \\
    \end{array}
  \right)
\ee
yields
\be
\pi^j_{mn}(i\sig_3)&=&\sum_l\frac{\sqrt{(j+m)!\ (j-m)!\ (j+n)!\ (j-n)!}}{(j+n-l)!\ (m-n+l)!\ (j-m-l)!\ l!}\ i^{j
+n-l}\ 0^{m-n+l}\ 0^l\ (-i)^{j-m-l}\nonumber\\
&=&i^{j+m} \ (-i)^{j-m}\ \delta_{mn}
\ee
Likewise for
\be
h_\phi=\left(
         \begin{array}{cc}
           e^{i\phi} & 0 \\
           0 & e^{-i\phi} \\
         \end{array}
       \right)
\ee
we obtain
\be
\pi^j_{mn}(h_\phi)&=&\sum_l\frac{\sqrt{(j+m)!\ (j-m)!\ (j+n)!\ (j-n)!}}{(j+n-l)!\ (m-n+l)!\ (j-m-l)!\ l!}\ (e^{i
\phi})^{j+n-l}\ 0^{m-n+l}\ 0^l\ (e^{-i\phi})^{j-m-l}\nonumber\\
&=&(e^{i\phi})^{j+m}(e^{-i\phi})^{j-m}\delta_{mn}\ =\ (e^{2im\phi})\delta_{mn}
\ee
We conclude (summing over repeated indices),
\be
&&\tr_{j^+_{\sig f}}\lt(iN_{t'f}h_{\phi_{t'f}}\sig_3h_{\phi_{tf}}^{-1}\lt(N_{tf}\rt)^{-1}g^+_{t\sig}g^+_{\sig
t'}\rt)\nonumber\\
&=&i^{2b_{\sig f}}e^{2ib_{\sig f}(\phi_{t'f}-\phi_{tf})}\pi^{j^+_{\sig f}}_{a_{\sig f}b_{\sig f}}\lt(N_{t'f}\rt)
\pi^{j^+_{\sig f}}_{b_{\sig f}c_{\sig f}}(N_{tf}^{-1}) \pi^{j^+_{\sig f}}_{c_{\sig f}a_{\sig f}}\lt(g^+_{t\sig}g^
+_{\sig t'}\rt)
\ee
Since $i\eps\sig_3\eps^{-1}=-i\sig_3$ we have similarly for the anti-self-dual part
\be
\pi^j_{mn}(-i\sig_3)&=&\sum_l\frac{\sqrt{(j+m)!\ (j-m)!\ (j+n)!\ (j-n)!}}{(j+n-l)!\ (m-n+l)!\ (j-m-l)!\ l!}\ (-i)^{j
+n-l}\ 0^{m-n+l}\ 0^l\ i^{j-m-l}\nonumber\\
&=&(-i)^{j+m} \ i^{j-m}\ \delta_{mn}
\ee
thus
\be
&&\tr_{j^-_{\sig f}}\lt(iN_{t'f}h^{-1}_{\phi_{t'f}}\eps\sig_3\eps^{-1}h_{\phi_{tf}}N_{tf}^{-1}g^-_{t\sig}g^-
_{\sig t'}\rt)\nonumber\\
&=&(-i)^{2b_{\sig f}} e^{-2ib_{\sig f}(\phi_{t'f}-\phi_{tf})}
\pi^{j^-_{\sig f}}_{a_{\sig f}b_{\sig f}}\lt(N_{t'f}\rt) \pi^{j^-_{\sig f}}_{b_{\sig f}c_{\sig f}}(N^{-1}_{tf}) \pi^
{j^-_{\sig f}}_{c_{\sig f}a_{\sig f}}\lt(g^-_{t\sig}g^-_{\sig t'}\rt)
\ee
We insert these formulae into the partition function Eq.(\ref{3.36})
\be
Z(\ck)
&=&\int\prod_{(\sig,t)}\rmd g^+_{\sig t}\rmd g^-_{\sig t}\prod_{f}\rmd\rho_{f}\lt(\rho_{f}\rt)^2\prod_
{(t,f)}\rmd N_{tf}\prod_{(t,f)}\rmd\phi_{tf}\prod_{t}\delta\Big(\sum_{f\subset t}\rho_{f}N_{tf}\t_3N_{tf}^
{-1}\Big)\nonumber\\
&&\times\sum_{\{j^\pm_{\sig f}\}}\prod_{(\sig,f)}\b_{j^+_{\sig f}}\lt(\lt|1+\frac{1}{\g}\rt|\rho_{f}\rt)\ \b_{j^-
_{\sig f}}\lt(\lt|1-\frac{1}{\g}\rt|\rho_{f}\rt)\nonumber\\
&&\times \sum_{a,b,c}i^{2b^+_{\sig f}}\ e^{2ib^+_{\sig f}(\phi_{t'f}-\phi_{tf})}\pi^{j^+_{\sig f}}_{a^+_
{\sig f}b^+_{\sig f}}\lt(N_{t'f}\rt)  \pi^{j^+_{\sig f}}_{b^+_{\sig f}c^+_{\sig f}}(N_{tf}^{-1}) \pi^{j^+_{\sig f}}
_{c^+_{\sig f}a^+_{\sig f}}\lt(g^+_{t\sig}g^+_{\sig t'}\rt)\nonumber\\
&&\times(-i)^{2b^-_{\sig f}} \ e^{-2ib^-_{\sig f}(\phi_{t'f}-\phi_{tf})}
\pi^{j^-_{\sig f}}_{a^-_{\sig f}b^-_{\sig f}}\lt(N_{t'f}\rt) \pi^{j^-_{\sig f}}_{b^-_{\sig f}c^-_{\sig f}}(N^{-1}_
{tf}) \pi^{j^-_{\sig f}}_{c^-_{\sig f}a^-_{\sig f}}\lt(g^-_{t\sig}g^-_{\sig t'}\rt).
\ee
and perform the integrals over $\rmd\phi_{tf}$ which enforce $b^-_{\sig f}=b^+_{\sig f}\equiv b_{\sig f}$, and
restrict the
range of the sum over $b_{\sig f}$ to the set $\{-j_{\sig f}^+,\cdots,j_{\sig f}^+\}\cap\{-j_{\sig f}^-,\cdots,j_{\sig
f}^-\}$. Accordingly,
\be
Z(\ck)
&=&\int\prod_{(\sig,t)}\rmd g^+_{\sig t}\rmd g^-_{\sig t}\prod_{f}\rmd\rho_{f}\lt(\rho_{f}\rt)^2\prod_
{(t,f)}\rmd N_{tf}\prod_{t}\delta\Big(\sum_{f\subset t}\rho_{f}N_{tf}\t_3N_{tf}^{-1}\Big)\nonumber\\
&&\times\sum_{\{j^\pm_{\sig f}\}}\prod_{(\sig,f)}\b_{j^+_{\sig f}}\lt(\lt|1+\frac{1}{\g}\rt|\rho_{f}\rt)\ \b_{j^-
_{\sig f}}\lt(\lt|1-\frac{1}{\g}\rt|\rho_{f}\rt)\nonumber\\
&&\times \sum_{a,b,c}\pi^{j^+_{\sig f}}_{a^+_{\sig f}b_{\sig f}}\lt(N_{t'f}\rt)  \pi^{j^+_{\sig f}}_{b_{\sig f}
c^+_{\sig f}}(N_{tf}^{-1}) \pi^{j^+_{\sig f}}_{c^+_{\sig f}a^+_{\sig f}}\lt(g^+_{t\sig}g^+_{\sig t'}\rt)
\nonumber\\
&&\times
\pi^{j^-_{\sig f}}_{a^-_{\sig f}b_{\sig f}}\lt(N_{t'f}\rt) \pi^{j^-_{\sig f}}_{b_{\sig f}c^-_{\sig f}}(N^{-1}_{tf})
\pi^{j^-_{\sig f}}_{c^-_{\sig f}a^-_{\sig f}}\lt(g^-_{t\sig}g^-_{\sig t'}\rt).\label{Z2}
\ee

\subsection{Topological/Gravitational Sector Duality, $\g$-Duality}

Before performing further computations, in this subsection we consider the topological sector I$\pm
$ of the simplicity constraint. Because we consider the model with finite Barbero-Immirzi
parameter, the sector I$\pm$ is actually also gravitational here in the following sense:
By definition, $\tr(F\wedge \ast(e\wedge e))$ is the Palatini (gravitational) term while
$\tr(F\wedge (e\wedge e))$ is the topological term. Since we are considering the Plebanski -- Holst
Lagrangian $\tr(F\wedge (B+\frac{1}{\gamma}\ast B))$, inserting the gravitational solution
$B=\pm \ast (e\wedge e)$ yields (due to $\ast^2=$id in Euclidian signature)
the Palatini -- Holst Lagrangian with Immirzi parameter $\gamma$, that is,
$\pm \tr(F\wedge (\ast(e\wedge e)+\frac{1}{\gamma} e\wedge e))$ while inserting the topological
solution $B=\pm e\wedge e$ yields Palatini -- Holst Lagrangian with Immirzi parameter
$1/\gamma$, that is
$\pm \frac{1}{\gamma} \tr(F\wedge (\ast(e\wedge e)+\gamma e\wedge e))$.
rescaled by $1/\gamma$. If we change coordinates from $X^\pm_f$ to $\pm X^\pm_f/\gamma$ in
the partition function $Z_\gamma$ (\ref{2.1}) then we obtain the relation
\begin{equation}
Z_\gamma({\cal K})=\gamma^{6F-21T} \; Z_{1/\gamma}({\cal K})
\end{equation}
where $F,T$ respectively denote the number of triangles and tetrahedra respectively in $\cal K$
(the powers arise from the Lebesgue measure and the $\delta$ functions respectively).
The appearing power of $\gamma$ drops out in correlators, hence up to the rescaling of the
n-point functions of involving $X^\pm_f$, $Z_\gamma,\; Z_{1/\gamma}$ yield the same
correlators. It follows that the model (\ref{2.1}) is a mixture of gravitational and
topological sectors as it should be.

This is before restriction to either the gravitational or topological sector respectively
and the manipulations (dropping infinite constants) that followed.
For comparison, the partition function for the topological (I) and gravitational sector
with Immirzi parameter $\gamma$ respectively read (before expanding the exponentials)
\begin{eqnarray}
Z_\gamma^{I/II}({\cal K}) &=& \int\;[\prod_f\; d^3X^+_f \;d^3X^-_f]\;
[\prod_{(\sigma,t)} \; dg^+_{\sigma t}\; dg^-_{\sigma t}] \;
[\prod_{(t,f)} \; dg^+_{t f}\; dg^-_{t f}]
\nonumber\\
&\times& [\prod_t\; \delta(\sum_{f\subset t}\; X^+_{tf})]\;\int\; [\prod_t \; du_t]
[\prod_{(t,f)} \delta(X^+_{tf}\mp u_t X^-_{tf} u_t^{-1})]
\nonumber\\
&\times& \exp(i[1+\gamma^{-1}]\sum_{(\sigma,f)}\; {\rm Tr}(X^+_f w^+_{\sigma f})
+i[1-\gamma^{-1}]\sum_{(\sigma,f)}\; {\rm Tr}(X^-_f w^-_{\sigma f}))
\end{eqnarray}
The only difference is the sign in the $\delta$ distribution enforcing the linearised
simplicity constraint. Now change variables $X_f^\pm\to \pm X_f^\pm/\gamma$ in the
model I (this induces also $X^\pm_{tf}\to \pm X^\pm_{tf}/\gamma$). This switches the
sign of the simplicity constraint to that of the model II, maps the $1/\gamma$ in the
exponent to $\gamma$ and rescales the Lebesgue measure and the $\delta$ distributions
according to
\begin{equation}
Z^I_\gamma({\cal K})=\gamma^{6F-5T} Z^{II}_{1/\gamma}({\cal K})
\end{equation}
The power of $\gamma$ again drops out in correlators and thus up to $\gamma$ powers
coming from n-point functions, ``topological'' correlators with respect to $\gamma$ are
essentially the same as ``gravitational'' correlators with respect to $1/\gamma$.
We coin this relation between the two sectors ``$\gamma$ duality''.
We will therefore not discuss model I any further in this article.

\section{The Spin-foam Model}\label{NSF}

\subsection{A Simplified Model without Closure Constraint}

In this subsection we discuss a simplified model by removing the closure constraint in the partition
function $Z(\ck)$ by hand as it is also done in existing spin foam models. We do this just for a better
comparison between our model and those models as far as the modifications are concerned that result
from commuting rather than non commuting B fields. The discussion of the full model and the additional
modifications that come from a proper treatment of the closure constraint will follow in the subsequent
subsection.

The simplified partition function reads (from Eq.(\ref{Z2}))
\be
&&Z_{\text{Simplified}}(\ck)
=\int\prod_{(\sig,t)}\rmd g^+_{\sig t}\rmd g^-_{\sig t}\prod_{f}\rmd\rho_{f}\lt(\rho_{f}\rt)^2\prod_{(t,f)}
\rmd N_{tf}\sum_{\{j^\pm_{\sig f}\}}\prod_{(\sig,f)}\b_{j^+_{\sig f}}\lt(\lt|1+\frac{1}{\g}\rt|\rho_{f}\rt)\ \b_
{j^-_{\sig f}}\lt(\lt|1-\frac{1}{\g}\rt|\rho_{f}\rt)\nonumber\\
&&\sum_{a,b,c}\lt[\pi^{j^+_{\sig f}}_{a^+_{\sig f}b_{\sig f}}\lt(N_{t'f}\rt)  \pi^{j^+_{\sig f}}_{b_{\sig f}c^
+_{\sig f}}(N_{tf}^{-1}) \pi^{j^+_{\sig f}}_{c^+_{\sig f}a^+_{\sig f}}\lt(g^+_{t\sig}g^+_{\sig t'}\rt)\rt]
\lt[\pi^{j^-_{\sig f}}_{a^-_{\sig f}b_{\sig f}}\lt(N_{t'f}\rt) \pi^{j^-_{\sig f}}_{b_{\sig f}c^-_{\sig f}}(N^{-1}_
{tf}) \pi^{j^-_{\sig f}}_{c^-_{\sig f}a^-_{\sig f}}\lt(g^-_{t\sig}g^-_{\sig t'}\rt)\rt]\label{Z3}
\ee

In order to explore the structure of the spin-foam amplitude (e.g. vertex amplitude) for this
partition function, we use the following recoupling relation ($N\in\text{SU(2)}$):
\be
&&\pi^{j^+}_{a^+b^+}\lt(N\rt)\pi^{j^-}_{a^-b^-}\lt(N\rt)\nonumber\\
&=&\langle j^+,b^+|\otimes\langle j^-,b^-|U(N)|j^+,a^+\rangle\otimes|j^-,a^-\rangle\nonumber\\
&=&\sum_{k=|j^+-j^-|}^{j^++j^-}\lag k,a^++a^-|j^+,a^+;j^-,a^-\rag\lag j^+,b^+;j^-,b^-|k,b^++b^-\rag
\pi^k_{a^++a^-,b^++b^-}(N)
\ee
We denote by $c(k,j^\pm)^{a^+a^-}_\a=c(k,j^\pm)_{a^+a^-}^\a$ the Clebsch-Gordan coefficients $
\lag k,\a|j^+,a^+;j^-,a^-\rag$, which are real and vanish unless $\a=a^++a^-$. Thus (summing
repeated indices)
\be
\pi^{j^+}_{a^+b^+}\lt(N\rt)\pi^{j^-}_{a^-b^-}\lt(N\rt)
&=&\sum_{k=|j^+-j^-|}^{j^++j^-}c(k,j^\pm)_{a^+a^-}^\a{c(k,j^\pm)_{b^+b^-}^{\b}}\pi^k_{\a\b}(N)
\nonumber\\
\pi^k_{\a\b}(N)
&=&{c(k,j^\pm)^{a^+a^-}_\a}c(k,j^\pm)^{b^+b^-}_{\b}\pi^{j^+}_{a^+b^+}\lt(N\rt)\pi^{j^-}_{a^-b^-}\lt(N
\rt)\nonumber\\
&&(k\in\{|j^+-j^-|,\cdots,j^++j^-\})
\ee
By using this recoupling relation we find
\be
&&\pi^{j^+}_{a^+b}\lt(N_{t'f}\rt)
\pi^{j^-}_{a^-b}\lt(N_{t'f}\rt)\pi^{j^+}_{bc^+}(N_{tf}^{-1}) \pi^{j^-}_{bc^-}(N_{tf}^{-1})\nonumber\\
&=&\sum_{k,k'=|j^+-j^-|}^{j^++j^-}c(k,j^\pm)_{a^+a^-}^{\a}\ \pi^{k}_{\a\b}(N_{t'f})\ {c(k,j^\pm)_{b,b}^
{\b}}\
c(k',j^\pm)_{b,b}^{\a'}\ \pi^{k'}_{\a'\b'}(N^{-1}_{tf})\ {c(k',j^\pm)_{c^+c^-}^{\b'}}
\ee
where $\b$ and $\a'$ are fixed to be $2b$. We note that $k$ and $k'$ are restricted to be
greater than or equal to $|2b|$ which we take care of by defining ${c(k,j^\pm)_{b,b}^{2b}}$
to be zero when $k<|2b|$. Inserting
this result back into the partition function $Z_{\text{Simplified}}(\ck)$ results in
\be
&&Z_{\text{Simplified}}(\ck)\nonumber\\
&=&\int\prod_{(\sig,t)}\rmd g^+_{\sig t}\rmd g^-_{\sig t}\prod_{f}\rmd\rho_{f}\lt(\rho_{f}\rt)^2\prod_
{(t,f)}\rmd N_{tf}\sum_{\{j^\pm_{\sig f}\}}\prod_{(\sig,f)}\b_{j^+_{\sig f}}\lt(\lt|1+\frac{1}{\g}\rt|\rho_{f}\rt)\
\b_{j^-_{\sig f}}\lt(\lt|1-\frac{1}{\g}\rt|\rho_{f}\rt)\sum_{k_{\sig f},k'_{\sig f}=|j_{\sig f}^+-j_{\sig f}^-|}^{j_
{\sig f}^++j_{\sig f}^-}\nonumber\\
&&\prod_{(\sig, f)}\sum_{b_{\sig f}}\pi^{k_{\sig f}}_{\a_{\sig f},2b_{\sig f}}(N_{t'f})\lt[{c(k_{\sig f},j_{\sig
f}^\pm)_{b_{\sig f}b_{\sig f}}^{2b_{\sig f}}}\
c(k'_{\sig f},j_{\sig f}^\pm)_{b_{\sig f}b_{\sig f}}^{2b_{\sig f}}\rt]\pi^{k'_{\sig f}}_{2b_{\sig f},\b'_{\sig f}}
(N^{-1}_{tf})\ \nonumber\\
&&\lt[{c(k'_{\sig f},j^\pm)_{c_{\sig f}^+c_{\sig f}^-}^{\b'_{\sig f}}}\pi^{j^+_{\sig f}}_{c^+_{\sig f}a^+_{\sig
f}}\lt(g^+_{t\sig}g^+_{\sig t'}\rt)\pi^{j^-_{\sig f}}_{c^-_{\sig f}a^-_{\sig f}}\lt(g^-_{t\sig}g^-_{\sig t'}\rt)c(k_
{\sig f},j_{\sig f}^\pm)_{a_{\sig f}^+a_{\sig f}^-}^{\a_{\sig f}}\rt]\label{structure}
\ee

  \begin{figure}[h]
  \begin{center}
  \includegraphics[width=6cm]{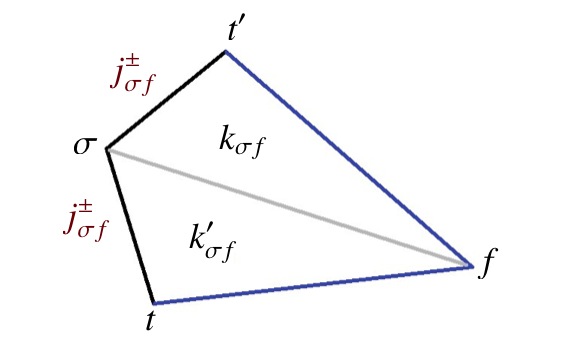}
  \caption{A wedge $(\sig,f)$ with a SO(4) representation $(j^+_{\sig f},j^-_{\sig f})$ and two SU(2)
representations $k_{\sig f}$ and $k'_{\sig f}$, where $k_{\sig f},k'_{\sig f}\in\lt\{|j_{\sig f}^+-j_{\sig f}
^-|,\cdots, j_{\sig f}^++j_{\sig f}^-\rt\}\cap\lt\{|j_{\sig' f}^+-j_{\sig' f}^-|,\cdots, j_{\sig' f}^++j_{\sig' f}^-\rt\}
$.}
  \label{wedge}
  \end{center}
  \end{figure}

\begin{figure}[h]
  \begin{center}
  \includegraphics[width=6cm]{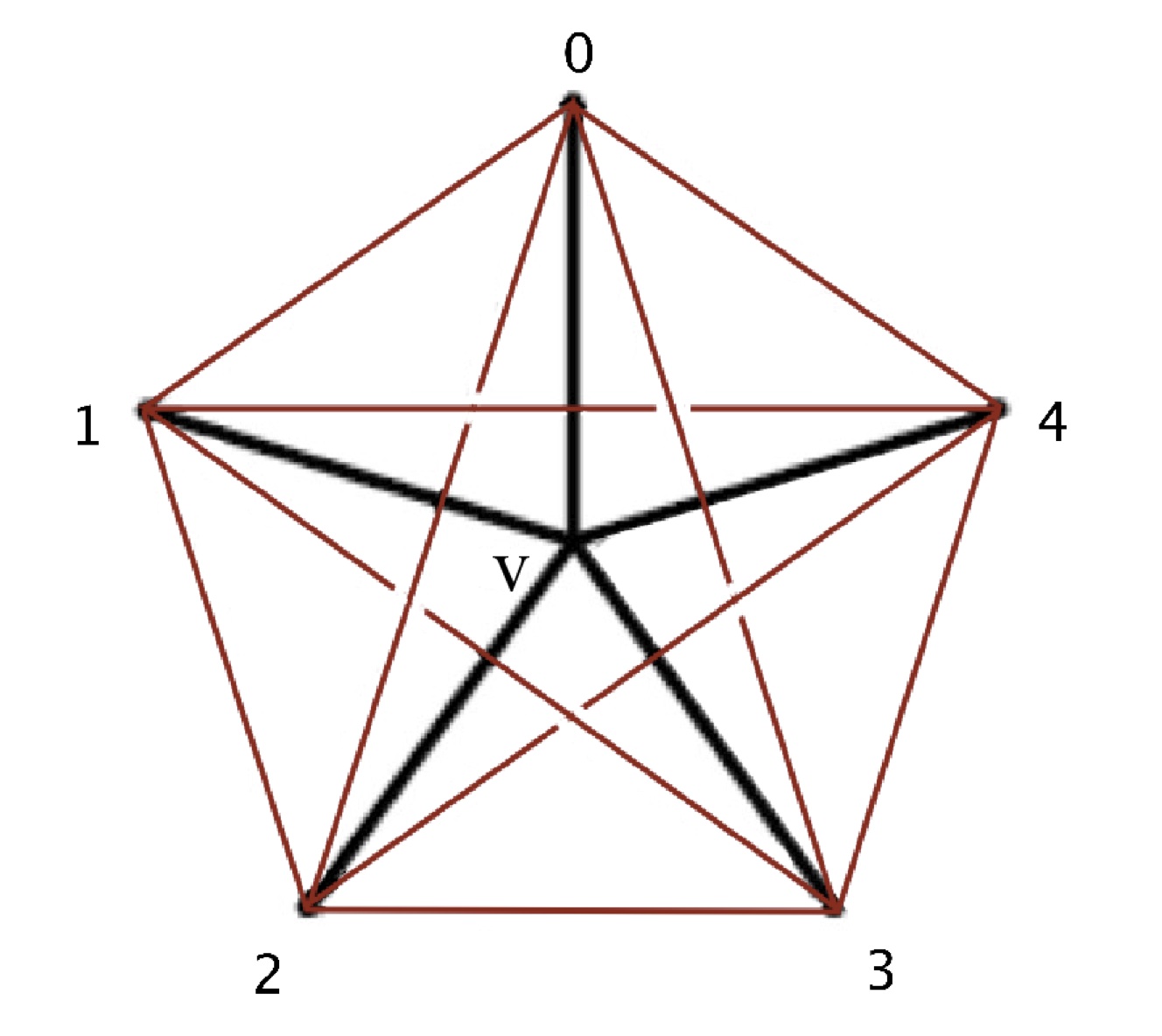}
  \caption{$v$ is the vertex dual to the 4-simplex $\sig$. $\mu=0,1,\cdots,4$ are the labels for the
tetrahedra $t_\mu$ forming the boundary of $\sig$. The edges $(v,\mu)$ $\mu=0,\cdots,4$ are the
edges dual to the tetrahedron $t_\mu$. The face determined by $v,\mu,\nu$ $\mu,\nu=0,\cdots,4$
is the wedge determined by $\sig,t_\mu,t_\nu$.}
  \label{4simplex}
  \end{center}
  \end{figure}

Now we focus on a vertex $v$ dual to a 4-simplex $\sig$. We fix the orientation of each dual
half edge $\overrightarrow{(\sig,t)}(=\overrightarrow{(v,\mu)}$ in the notation of FIG. \ref
{4simplex}) to be outgoing from the vertex and integrate the SU(2) holonomies $g_{\sig t}^\pm$.
The integration of $g_{\sig t}^
\pm$ leads to a result that depends on the orientations of the wedges bounded by $\overrightarrow
{(\sig,t)}$. We say a wedge $w$ bounded by $\overrightarrow{(\sig,t)}$ is incoming to the edge $
\overrightarrow{(\sig,t)}$, if the orientation along its boundary agrees with $\overrightarrow{(\sig,t)}$,
otherwise we call it to be outgoing from $\overrightarrow{(\sig,t)}$. The integrations of $g_{\sig t}^\pm$ in
Eq.(\ref{structure})
\be
\int\rmd g_{\sig t}\bigotimes_{w\ \text{incoming}\ \overrightarrow{(\sig,t)}}\pi^{j_w}\Big(g_{\sig t}\Big)
\bigotimes_{w\ \text{outgoing}\ \overrightarrow{(\sig,t)}}\pi^{j_w}\Big(g_{\sig t}^{-1}\Big)
\ee
equals a projection operator $\Fp_{\sig t}$ for each dual half edge $\overrightarrow{(\sig,t)}$
\be
\Fp_{\sig t}&:&\lt[\bigotimes_{w\ \text{incoming}\ \overrightarrow{(\sig,t)}}V_{j^+_w}\bigotimes_{w\
\text{outgoing}\ \overrightarrow{(\sig,t)}}V^*_{j^+_w}\rt]\bigotimes\lt[\bigotimes_{w\ \text{incoming}\
\overrightarrow{(\sig,t)}}V_{j^-_w}\bigotimes_{w\ \text{outgoing}\ \overrightarrow{(\sig,t)}}V^*_{j^-
_w}\rt]\nonumber\\
&\to&\text{Inv}\lt(\bigotimes_{w\ \text{incoming}\ \overrightarrow{(\sig,t)}}V_{j^+_w}\bigotimes_{w\
\text{outgoing}\ \overrightarrow{(\sig,t)}}V^*_{j^+_w}\rt)\bigotimes\text{Inv}\lt(\bigotimes_{w\ \text
{incoming}\ \overrightarrow{(\sig,t)}}V_{j^-_w}\bigotimes_{w\ \text{outgoing}\ \overrightarrow
{(\sig,t)}}V^*_{j^-_w}\rt)\nonumber\\
\Fp_{\sig t}&:=&\lt[\sum_{i^+_{\sig t}}C^4_{i^+_{\sig t}}\lt(j_{\sig f}^+\rt)\otimes C^4_{i^+_{\sig t}}\lt(j_
{\sig f}^+\rt)^\dagger\rt]\otimes
\lt[\sum_{i^-_{\sig t}}C^4_{i^-_{\sig t}}\lt(j_{\sig f}^-\rt)\otimes C^4_{i^-_{\sig t}}\lt(j_{\sig f}^-\rt)^
\dagger\rt]
\ee
where $V_j$ is the representation space for SU(2) unitary irreducible representation, and we keep
in mind that each pair $(\sig,f)$ determines a wedge $w$, and $C^4_{i^\pm_{\sig t}}\lt(j_{\sig f}^
\pm\rt)$ are the 4-valent SU(2) intertwiners forming an orthonormal basis in
\be
\text{Inv}\lt(\bigotimes_{w\ \text{incoming}\ \overrightarrow{(\sig,t)}}V_{j^\pm_w}\bigotimes_{w\ \text
{outgoing}\ \overrightarrow{(\sig,t)}}V^*_{j^\pm_w}\rt)
\ee

Thus the result of the integrations of $g_{\sig t}^\pm$ in Eq.(\ref{structure}) is a product of the
projection operators $\Fp_{\sig t}$ for all the dual half edges $\overrightarrow{(\sig,t)}$.
According to the index structure appearing in Eq.(\ref{structure}), we find that in each $\Fp_{\sig t}$ the
adjoint intertwiners $C^4_{i^\pm_{\sig t}}\lt(j_{\sig f}^\pm\rt)^\dagger$ are combined with the
indices $a^\pm_{\sig f},c^\pm_{\sig f}$, where $a^\pm_{\sig f}$ are for the incoming wedges and
$c^{\pm}_{\sig f}$ are for the outgoing wedges. However the intertwiners $C^4_{i^\pm_{\sig t}}\lt(j_
{\sig f}^\pm\rt)$ for each half edge $\overrightarrow{(\sig,t)}$ are contracted with other half edge
intertwiners of $\overrightarrow{(\sig,t')}$ at the vertex dual to $\sig$. Summing over the indices $a^\pm_
{\sig f},c^\pm_{\sig f}$, the integrations of $g_{\sig t}^\pm$ in Eq.(\ref{structure}) result in a product
of
\be
\lt[\sum_{i^+_{\sig t}}C^4_{i^+_{\sig t}}\lt(j_{\sig f}^+\rt)_{\cdots} \overline{C^4_{i^+_{\sig t}}\lt(j_{\sig
f}^+\rt)}_{\{a^+_{\sig f}\},\{c^+_{\sig f'}\}}\rt]\cdot
\lt[\sum_{i^-_{\sig t}}C^4_{i^-_{\sig t}}\lt(j_{\sig f}^-\rt)_{\cdots} \overline{C^4_{i^-_{\sig t}}\lt(j_{\sig f}
^-\rt)}_{\{a^-_{\sig f}\},\{c^-_{\sig f'}\}}\rt]\label{cc}
\ee
for all half edges $\overrightarrow{(\sig, t)}$, where $\cdots$ are the indices contracted with other
half edge intertwiners of $\overrightarrow{(\sig,t')}$ at the vertex dual to $\sig$. According to the
structure of Eq.(\ref{structure}), we assign the intertwiners
\be
\lt[C^4_{i^+_{\sig t}}\lt(j_{\sig f}^+\rt)_{\cdots}\rt]\cdot
\lt[C^4_{i^-_{\sig t}}\lt(j_{\sig f}^-\rt)_{\cdots}\rt]\label{intertwiner}
\ee
to the beginning point of $\overrightarrow{(\sig, t)}$, while we assign the adjoint intertwiners
\be
\lt[\overline{C^4_{i^+_{\sig t}}\lt(j_{\sig f}^+\rt)}_{\{a^+_{\sig f}\},\{c^+_{\sig f'}\}}\rt]\cdot
\lt[\overline{C^4_{i^-_{\sig t}}\lt(j_{\sig f}^-\rt)}_{\{a^-_{\sig f}\},\{c^-_{\sig f'}\}}\rt]\label
{adjointintertwiner}
\ee
to the end point of $\overrightarrow{(\sig, t)}$.

The contractions of the half edge intertwiners Eq.(\ref{intertwiner}) at each vertex dual to $\sig$
gives a SO(4) $15j$-symbol
\be
\Big\{15j\Big\}_{\text{SO(4)}}(j^\pm_{\sig f},i^\pm_{\sig t}):=\tr\lt[\bigotimes_{\overrightarrow{(\sig,t)}}
C^4_{i^+_{\sig t}}\lt(j_{\sig f}^+\rt)
\rt]\ \tr\lt[\bigotimes_{\overrightarrow{(\sig,t)}} C^4_{i^-_{\sig t}}\lt(j_{\sig f}^-\rt)\rt]
\ee
to each 4-simplex $\sig$ (to each vertex dual to $\sig$).

On the other hand, each of the adjoint intertwiners Eq.(\ref{adjointintertwiner}) at the end point of $
\overrightarrow{(\sig, t)}$ is contracted with the Clebsch-Gordan coefficients $c(k,j^\pm)_{a^+a^-}^
{\a}$ and $\overline{c(k',j^\pm)_{c^+c^-}^{\b'}}$. Thus we obtain a 4-valent SU(2) intertwiner $
\ci^4_{i^\pm_{\sig t}}(k_{\sig f},k'_{\sig f};j_{\sig f}^\pm)$ at the end point of each half edge $
\overrightarrow{(\sig, t)}$ (summing repeated indices)
\be
\ci^4_{i^\pm_{\sig t}}\lt(k_{\sig f},k'_{\sig f};j_{\sig f}^\pm\rt)_{\{\b'_{\sig f}\},\{\a_{\sig f'}\}}&:=&\lt
[\overline{C^4_{i^+_{\sig t}}\lt(j_{\sig f}^+\rt)}_{\{a^+_{\sig f}\},\{c^+_{\sig f'}\}}\rt]\cdot
\lt[\overline{C^4_{i^-_{\sig t}}\lt(j_{\sig f}^-\rt)}_{\{a^-_{\sig f}\},\{c^-_{\sig f'}\}}\rt]\nonumber\\
&&\times\prod_{(\sig,f)\ \text{incoming}}c(k_{\sig f},j_{\sig f}^\pm)_{a_{\sig f}^+a_{\sig f}^-}^{\a_{\sig
f}}\prod_{(\sig,f')\ \text{outgoing}}{c(k_{\sig f'}',j_{\sig f'}^\pm)_{c_{\sig f'}^+c_{\sig f'}^-}^{\b'_{\sig f'}}}
\ee
If we choose an orthonormal basis in the space of 4-valent SU(2) intertwiners (labeled by $l_{\sig
t}$)
\be
C^4_{l_{\sig t}}\lt(k_{\sig f},k'_{\sig f'}\rt)\in \text{Inv}\lt(\bigotimes_{f\ \text{incoming}\ \overrightarrow
{(\sig,t)}}V_{k_{\sig f}}\bigotimes_{f'\ \text{outgoing}\ \overrightarrow{(\sig,t)}}V^*_{k'_{\sig f'}}\rt)
\ee
we may expand $\ci^4_{i^\pm_{\sig t}}(k_{\sig f},k'_{\sig f};j_{\sig f}^\pm)$ in terms of this basis, explicitly
\be
&&\ci^4_{i^\pm_{\sig t}}\lt(k_{\sig f},k'_{\sig f};j_{\sig f}^\pm\rt)_{\{\b'_{\sig f}\},\{\a_{\sig f'}\}}\nonumber
\\
&=&\sum_{l_{\sig t}}\lt[\ci^4_{i^\pm_{\sig t}}\lt(k_{\sig f},k'_{\sig f};j_{\sig f}^\pm\rt)_{\{\rho_{\sig f}\},
\{\rho'_{\sig f'}\}}\ C^4_{l_{\sig t}}\Big(k_{\sig f},k'_{\sig f'}
\Big)^{\{\rho_{\sig f}\},\{\rho'_{\sig f'}\}}\rt]\
\overline{C^4_{l_{\sig t}}\lt(k_{\sig f},k'_{\sig f'}\rt)}_{\{\b'_{\sig f}\},
\{\a_{\sig f'}\}}\nonumber\\
&\equiv&\sum_{l_{\sig t}}f^{l_{\sig t}}_{i^\pm_{\sig t}}\lt(k_{\sig f},k'_{\sig f};j_{\sig f}^\pm\rt)\
\overline{C^4_{l_{\sig t}}\lt(k_{\sig f},k'_{\sig f'}\rt)}_{\{\b'_{\sig f}\},\{\a_{\sig f'}\}}
\ee
Insert these findings into the partition function $Z_{\text{Simplified}}(\ck)$ yields
\be
&&Z_{\text{Simplified}}(\ck)\nonumber\\
&=&\sum_{\{j^\pm_{\sig f}\}}\sum_{k_{\sig f},k'_{\sig f}=|j_{\sig f}^+-j_{\sig f}^-|}^{j_{\sig f}^++j_{\sig f}
^-}\sum_{\{l_{\sig t}\}}
\int\prod_{f}\rmd\rho_{f}\lt(\rho_{f}\rt)^2\prod_{(t,f)}\rmd N_{tf}\prod_{(\sig,f)}\b_{j^+_{\sig f}}\lt(\lt|1+
\frac{1}{\g}\rt|\rho_{f}\rt)\ \b_{j^-_{\sig f}}\lt(\lt|1-\frac{1}{\g}\rt|\rho_{f}\rt)\nonumber\\
&&\sum_{\{i_{\sig t}^\pm\}}\prod_{\sig}\Big\{15j\Big\}_{\text{SO(4)}}\lt(j^\pm_{\sig f},i^\pm_{\sig t}\rt)
\prod_{\overrightarrow{(\sig,t)}}f^{l_{\sig t}}_{i^\pm_{\sig t}}\lt(k_{\sig f},k'_{\sig f};j_{\sig f}^\pm\rt)\
\overline{C^4_{l_{\sig t}}\lt(k_{\sig f},k'_{\sig f'}\rt)}_{\{\b'_{\sig f}\},\{\a_{\sig f'}\}}\nonumber\\
&&\prod_{(\sig, f)}\pi^{k_{\sig f}}_{\a_{\sig f},2b_{\sig f}}(N_{t'f})\lt[{c(k_{\sig f},j_{\sig f}^\pm)_{b_{\sig
f}b_{\sig f}}^{2b_{\sig f}}}\
c(k'_{\sig f},j_{\sig f}^\pm)_{b_{\sig f}b_{\sig f}}^{2b_{\sig f}}\rt]\pi^{k'_{\sig f}}_{2b_{\sig f},\b'_{\sig f}}
(N^{-1}_{tf})\label{structure1}
\ee
from which we we read the vertex amplitude for each vertex dual to a 4-simplex $\sig$
\be
A_\sig\lt(j^\pm_{\sig f};k_{\sig f},k'_{\sig f};l_{\sig t}\rt):=\sum_{\{i_{\sig t}^\pm\}}\Big\{15j\Big\}_{\text
{SO(4)}}\lt(j^\pm_{\sig f},i^\pm_{\sig t}\rt)\prod_{\overrightarrow{(\sig,t)}}f^{l_{\sig t}}_{i^\pm_{\sig t}}\lt
(k_{\sig f},k'_{\sig f};j_{\sig f}^\pm\rt)
\ee

Next, we consider the integrations of $\rmd N_{tf}$. Since the closure constraint is removed, the
integrals over $\rmd N_{tf}$ can be done immediately. Consider a tetrahedron $t_i$ is shared
by two 4-simplex $\sig _i, \sig_{i+1}$ (c.f. FIG.\ref{face}), the integral of $\rmd N_{t_i,f}$ is
essentially
\be
\int\rmd N_{t_i f}\ \pi^{k_{\sig_i f}}_{\a_{\sig_i f}\b_{\sig_i f}}(N_{t_if})\ \pi^{k'_{\sig_{i+1} f}}_{\a'_{\sig_
{i+1} f}\b'_{\sig_{i+1} f}}(N^{-1}_{t_if})=\frac{1}{\dim(k_{\sig_i f})}\delta^{k_{\sig_i f}k'_{\sig_{i+1} f}}
\delta_{\a_{\sig_i f}\b'_{\sig_{i+1} f}}\delta_{\b_{\sig_i f}\a'_{\sig_{i+1} f}}\label{N}
\ee
There are three consequences from these integrals:
\begin{enumerate}
\item The SU(2) representations $k_{\sig_i f}$ is identified with $k'_{\sig_{i+1} f}$, thus we label
\be
k_{\sig_i f}=k'_{\sig_{i+1} f}\equiv k_{t_if}
\ee
where $t_i$ is the tetrahedron shared by the 4-simplices $\sig_i,\sig_{i+1}$ (see FIG.\ref{face2}).

  \begin{figure}[h]
  \begin{center}
  \includegraphics[width=7cm]{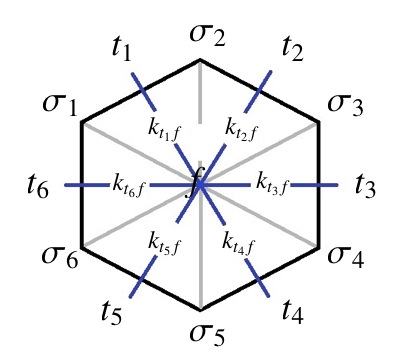}
  \caption{SU(2) representations $k_{tf}$ assigned to each pair of $(t,f)$. }
  \label{face2}
  \end{center}
  \end{figure}

\item For the SU(2) intertwiners on the half edges $\overrightarrow{(\sig_i,t_i)}$ and $
\overrightarrow{(\sig_{i+1},t_i)}$,
\be
&&\overline{C^4_{l_{\sig_i t_i}}\lt(k_{t_i f},k_{t_i f'}\rt)}_{\{\b'_{\sig_i f}\},\{\a_{\sig_i f'}\}}\overline
{C^4_{l_{\sig_{i+1} t_i}}\lt(k_{t_i f'},k_{t_i f}\rt)}_{\{\b'_{\sig_{i+1} f'}\},\{\a_{\sig_{i+1} f}\}}\nonumber\\
&&\prod_{f\ \text{incoming}\ \overrightarrow{(\sig_i,\sig_{i+1})}}\delta_{\a_{\sig_i f}\b'_{\sig_{i+1}f}}
\prod_{f'\ \text{outgoing}\ \overrightarrow{(\sig_i,\sig_{i+1})}}\delta_{\a_{\sig_i f'}\b'_{\sig_{i+1}f'}}
\nonumber\\
&=&\delta_{l_{\sig_i t_i},\ l_{\sig_{i+1} t_i}^\dagger}
\ee
which identify the half edge intertwiners into full edge intertwiners
\be
l_{\sig_i t_i}=l_{\sig_{i+1} t_i}^\dagger\equiv l_{\overrightarrow{(\sig_i,\sig_{i+1})}}\equiv l_{e_i}
\ee
where $e_i:=\overrightarrow{(\sig_i,\sig_{i+1})}$ is the edge dual to the tetrahedron $t_i$.

\item For each face dual to $f$, we have a factor
\be
&&\prod_{i=1}^{|\sig|_f}\lt[{c(k_{t_i f},j_{\sig_i f}^\pm)_{b_{\sig_i f}b_{\sig_i f}}^{2b_{\sig_i f}}}\
c(k_{t_{i-1} f},j_{\sig_i f}^\pm)_{b_{\sig_i f}b_{\sig_i f}}^{2b_{\sig_i f}}\rt]\prod_{i=1}^{|\sig|_f}\delta_
{2b_{\sig_i f},2b_{\sig_{i+1} f}}
\nonumber\\
&=&\sum_{b_f}\prod_{i=1}^{|\sig|_f}{c(k_{t_i f},j_{\sig_i f}^\pm)_{b_{ f}b_f}^{2b_f}}\ c(k_{t_{i} f},j_
{\sig_{i+1} f}^\pm)_{b_{f}b_f}^{2b_f}
\ee
where the indices $b_{\sig f}$ are identified for the different wedges belonging to the same dual
face and the range of $b_f$ is
\be
\bigcap_{i=1}^{|\sig|_f}\lt[\{-j_{\sig_i f}^+,\cdots,j_{\sig_i f}^+\}\cap\{-j_{\sig_i f}^-,\cdots,j_{\sig_i f}^-\}
\rt]
\ee
and $|\sig|_f$ is the number of vertices around a face dual to $f$.

\end{enumerate}

Finally we consider the integrals of $\rmd \rho_f$. We define a triangle/face amplitude
\be
A_f\lt(j_{\sig f}^\pm,k_{tf}\rt)&:=&%\prod_{i=1}^{|\sig|_f}\frac{1}{\dim(k_{t_i f})}
\sum_{b_f}\prod_{i=1}^{|\sig|_f}{c(k_{t_i f},j_{\sig_i f}^\pm)_{b_{ f}b_f}^{2b_f}}\ c(k_{t_{i} f},j_{\sig_{i
+1} f}^\pm)_{b_{f}b_f}^{2b_f}\nonumber\\
&&\times\int_0^\infty\rmd\rho_f(\rho_f)^2\prod_{i=1}^{|\sig|_f}\lt[\b_{j^+_{\sig_if}}\lt(\lt|1+\frac{1}{\g}
\rt|\rho_{f}\rt)\ \b_{j^-_{\sig_if}}\lt(\lt|1-\frac{1}{\g}\rt|\rho_{f}\rt)\rt]
\ee
By Eq.(\ref{beta}), we can directly compute the expression of the function $\beta_j$
\be
\b_j(r)=\int\rmd g\ e^{-ir\tr\lt(g\rt)}\chi_j(g)=i^{-2j}(2j+1)\frac{J_{2j+1}(2r)}{r}
\ee
where $J_n(x)$ is the Bessel function of the first kind. The proof of this relation uses the
recurrence relation:
\be
J_{2j+2}(2r)+J_{2j}(2r)=(2j+1)\frac{J_{2j+1}(2r)}{r}
\ee
Let's consider the integrand of the integration over areas $\rho_f$ (or considering an integral in large area regime). In the uniform limit of $j\to\infty,\rho\to\infty$ (or $r\to\infty$), the asymptotic behavior of the function $\b_j$ is (see e.g. \cite{bessel}, uniform limit can be made by the scaling $j\to \l j,r\to\l r$ and sending $\l\to\infty$)
%\footnote{$\lim_{j\to\infty}\delta(2r-2j)=\lim_{j\to\infty}\frac{1}{2j}\delta(\frac{r}{j}-1)=\lim_{j\to\infty}\lim_{\nu\to\infty}\frac{\nu}{2j}J_\nu(\frac{\nu}{2j}2r)$ \cite{bessel}}
\be
\text{Large-}(j,r):\ \ \ \ \b_j(r)\sim i^{-2j}\frac{2j+1}{r}\delta\big(2r-2j\big)
\ee
It follows that in the uniform limit $j^\pm_{\sig f}\to\infty,\rho_f\to\infty$, the asymptotic behaviour of the Bessel
functions constrains the SO(4) representations on the wedges by
\be
j^\pm_{\sig f}=j^\pm_{\sig' f}=j^\pm_f\label{wedgerep}
\ee
and also impose the well-known constraint on the self-dual and anti-self-dual representations
\be
\lt|1-\frac{1}{\g}\rt|j^+_{f}=\lt|1+\frac{1}{\g}\rt|j^-_{f}\label{simplerep}
\ee
which gives the quantization condition for the Barbero-Immirzi parameter
\be
\text{If}\ |\g|>1:&& \g=\frac{j^+_{f}+j_{f}^-}{j^+_{f}-j_{f}^-}\nonumber\\
\text{If}\ |\g|<1:&& \g=\frac{j^+_{f}-j_{f}^-}{j^+_{f}+j_{f}^-}
\ee
While it is nice to see that we obtain certain points of contact with the
EPRL and FK models respectively, one should keep in mind that
these constraints hold only in the sense of large-$j$. In general, the representations
which do not satisfy Eqs.(\ref{wedgerep}) and (\ref{simplerep}) still have nontrivial contributions to
the spin-foam amplitude and it is not clear whether these ``non EPRL/FK configurations'' have large
or low measure.

Let us summarize the structure of the partition function $Z_{\text{Simplified}}(\ck)$ in terms of the
4-simplex/vertex amplitude, tetrahedron/edge amplitude and triangle/face amplitude
\be
Z_{\text{Simplified}}(\ck)=\sum_{\{j_{\sig f}^\pm\}}\sum_{\{k_{tf}\}}\sum_{\{l_{e}\}}\prod_fA_f\lt(j_{\sig
f}^\pm,k_{tf}\rt)\prod_tA_t(k_{tf})\prod_{\sig}A_\sig\lt(j^\pm_{\sig f},k_{t f},l_{e}\rt)
\ee
where $k_{tf}$ is constrained by the condition that for a tetrahedron $t$ shared by both 4-simplices $\sig,\sig'$ we have
\be
k_{tf}\in \lt\{|j_{\sig f}^+-j_{\sig f}^-|,\cdots, j_{\sig f}^++j_{\sig f}^-\rt\}\cap\lt\{|j_{\sig' f}^+-j_{\sig' f}^-|,
\cdots, j_{\sig' f}^++j_{\sig' f}^-\rt\}\label{simplifiedmodel}
\ee
and the 4-simplex/vertex amplitudes, tetrahedron/edge amplitudes and triangle/face
amplitudes are respectively identified as
\be
A_\sig\lt(j^\pm_{\sig f},k_{t f},l_{e}\rt)&:=&\sum_{\{i_{\sig t}^\pm\}}\Big\{15j\Big\}_{\text{SO(4)}}\lt(j^
\pm_{\sig f},i^\pm_{\sig t}\rt)\prod_{\overrightarrow{(\sig,t)}}f^{l_{e}}_{i^\pm_{\sig t}}\lt(k_{t f};j_{\sig f}
^\pm\rt)\nonumber\\
A_t(k_{tf})&:=&\prod_{f\subset t}\frac{1}{\dim(k_{tf})}\nonumber\\
A_f\lt(j_{\sig f}^\pm,k_{tf}\rt)&=&
\sum_{b_f}\prod_{i=1}^{|\sig|_f}{c(k_{t_i f},j_{\sig_i f}^\pm)_{b_{ f}b_f}^{2b_f}}\ c(k_{t_{i} f},j_{\sig_{i
+1} f}^\pm)_{b_{f}b_f}^{2b_f}\nonumber\\
&&\times\int_0^\infty\rmd\rho_f(\rho_f)^2\prod_{i=1}^{|\sig|_f}\lt[\b_{j^+_{\sig_if}}\lt(\lt|1+\frac{1}{\g}
\rt|\rho_{f}\rt)\ \b_{j^-_{\sig_if}}\lt(\lt|1-\frac{1}{\g}\rt|\rho_{f}\rt)\rt]
\ee
When we take the uniform limit: $j^\pm_{\sig f},\rho_f\to\infty$ for the integrand, by the previous discussion, we obtain the
constraints:
\be
j^\pm_{\sig f}=j^\pm_{\sig' f}=j^\pm_f\ \ \ \ \text{and}\ \ \ \ \lt|1-\frac{1}{\g}\rt|j^+_{f}=\lt|1+\frac{1}{\g}\rt|
j^-_{f}\label{jj}
\ee
Thus the spins $j^\pm_{\sig f}$ for different wedges are identical on the same
face dual to $f$, and $j^+_f$ and $j^-_f$ satisfies the ``$\g$-simple'' relation in this limit. Then the
vertex amplitude reduces to
\be
A_\sig&\sim&\sum_{\{i_{t}^\pm\}}\Big\{15j\Big\}_{\text{SO(4)}}\lt(j^\pm_{f},i^\pm_{t}\rt)\prod_
{\overrightarrow{(\sig,t)}}f^{l_{e}}_{i^\pm_{ t}}\lt(k_{t f};j_{f}^\pm\rt)\label
{FK}
\ee
where $j^+_{f}$ and $j^-_f$ subject the relation in Eq.(\ref{jj}). We notice that in this
limit Eq.(\ref{FK}) is nothing but the vertex amplitude of the $\text{FK}_{\g}$ Model (when $|
\g|>1$) \cite{FK}. And in the large-$j$ limit the integral over area $\rho_f$ in the large area regime can be approximated
by a discrete sum over $j^-_f$ or $j^+_f$ in the path integral Eq.(\ref{start}). In the usual context of spinfoam formulation, the large-j limit is understood as a semiclassical limit in a certain sense \cite{semiclassical, HZ}.

\subsection{On the Implementation of Closure Constraint}\label{icc}

In this subsection we properly keep the closure constraint in the partition function:
\be
Z(\ck)
&=&\sum_{\{j^\pm_{\sig f}\}}\sum_{k_{\sig f},k'_{\sig f}=|j_{\sig f}^+-j_{\sig f}^-|}^{j_{\sig f}^++j_{\sig f}
^-}\sum_{\{l_{\sig t}\}}
\int\prod_{f}\rmd\rho_{f}\lt(\rho_{f}\rt)^2\prod_{(t,f)}\rmd N_{tf}\prod_{(\sig,f)}\b_{j^+_{\sig f}}\lt(\lt|1+
\frac{1}{\g}\rt|\rho_{f}\rt)\ \b_{j^-_{\sig f}}\lt(\lt|1-\frac{1}{\g}\rt|\rho_{f}\rt)\nonumber\\
&&\sum_{\{i_{\sig t}^\pm\}}\prod_{\sig}\ca_\sig\lt(j^\pm_{\sig f};k_{\sig f},k'_{\sig f};l_{\sig t}\rt)\prod_
{\overrightarrow{(\sig,t)}}\overline{C^4_{l_{\sig t}}\lt(k_{\sig f},k'_{\sig f'}\rt)}_{\{\b'_{\sig f}\},\{\a_{\sig f'}
\}}\prod_{t}\delta\Big(\sum_{f\subset t}\rho_{f}N_{tf}\t_3N_{tf}^{-1}\Big)\nonumber\\
&&\prod_{(\sig, f)}\pi^{k_{\sig f}}_{\a_{\sig f},2b_{\sig f}}(N_{t'f})\lt[{c(k_{\sig f},j_{\sig f}^\pm)_{b_{\sig
f}b_{\sig f}}^{2b_{\sig f}}}\
c(k'_{\sig f},j_{\sig f}^\pm)_{b_{\sig f}b_{\sig f}}^{2b_{\sig f}}\rt]\pi^{k'_{\sig f}}_{2b_{\sig f},\b'_{\sig f}}
(N^{-1}_{tf})\label{structure2}
\ee
Here we can also extract the vertex/4-simplex amplitude $\ca_\sig$, the edge/tetrahedron
amplitude $\ca_t$, and the face/triangle amplitude $\ca_f$
\be
\ca_\sig\lt(j^\pm_{\sig f};k_{\sig f},k'_{\sig f};l_{\sig t}\rt)
&:=&\sum_{\{i_{\sig t}^\pm\}}\Big\{15j\Big\}_{\text{SO(4)}}\lt(j^\pm_{\sig f},i^\pm_{\sig t}\rt)\prod_
{\overrightarrow{(\sig,t)}}f^{l_{\sig t}}_{i^\pm_{\sig t}}\lt(k_{\sig f},k'_{\sig f};j_{\sig f}^\pm\rt)\nonumber
\\
\ca_t\Big(\rho_f;k_{\sig f},k'_{\sig f};l_{\sig t};b_{\sig f}\Big)
&:=&\int\prod_{f\subset t}\rmd N_{tf}\ \delta\Big(\sum_{f\subset t}\rho_{f}N_{tf}\t_3N_{tf}^{-1}\Big)
\nonumber\\
&&\lt[\overline{C^4_{l_{\sig t}}\lt(k_{\sig f},k'_{\sig f'}\rt)}_{\{\a_{\sig f}\},\{\a_{\sig f'}\}}\rt]\lt[\overline
{C^4_{l_{\sig' t}}\lt(k_{\sig' f},k'_{\sig' f'}\rt)}_{\{\b_{\sig' f'}\},\{\b_{\sig' f}\}}\rt]\nonumber\\
&&\prod_{f\ \text{outgoing}}\pi^{k'_{\sig f}}_{2b_{\sig f},\a_{\sig f}}(N^{-1}_{tf})\ \pi^{k_{\sig' f}}_{\b_
{\sig' f},2b_{\sig' f}}(N_{tf})\prod_{f'\ \text{incoming}}\pi^{k_{\sig f'}}_{\a_{\sig f'},2b_{\sig f'}}(N_{tf'})\pi^
{k'_{\sig' f'}}_{2b_{\sig' f'},\b_{\sig' f'}}(N^{-1}_{tf'})\nonumber\\
\ca_f\lt(\rho_f;j^\pm_{\sig f};k_{\sig f},k'_{\sig f};b_{\sig f}\rt)
&:=&(\rho_f)^2\prod_{(\sig,f)}\b_{j^+_{\sig f}}\lt(\lt|1+\frac{1}{\g}\rt|\rho_{f}\rt)\ \b_{j^-_{\sig f}}\lt(\lt|1-
\frac{1}{\g}\rt|\rho_{f}\rt)\nonumber\\
&&\prod_{(\sig,f)}{c(k_{\sig f},j_{\sig f}^\pm)_{b_{\sig f}b_{\sig f}}^{2b_{\sig f}}}\
c(k'_{\sig f},j_{\sig f}^\pm)_{b_{\sig f}b_{\sig f}}^{2b_{\sig f}}
\ee
Then the partition function can be written in terms of these amplitudes as:
\be
Z(\ck)
&=&\sum_{\{j^\pm_{\sig f}\}}\sum_{\{k_{\sig f},k'_{\sig f}\}}\sum_{\{l_{\sig t}\}}\sum_{\{b_{\sig f}\}}
\int\prod_{f}\rmd\rho_{f}\times\nonumber\\
&&\ca_f\lt(\rho_f;j^\pm_{\sig f};k_{\sig f},k'_{\sig f};b_{\sig f}\rt)\
\ca_t\Big(\rho_f;k_{\sig f},k'_{\sig f};l_{\sig t};b_{\sig f}\Big)\
\ca_\sig\lt(j^\pm_{\sig f};k_{\sig f},k'_{\sig f};l_{\sig t}\rt)
\ee
Note that in the large-$(j,\rho)$ limit,
\be
\text{Large-}(j,\rho):\ \ \ \ \b_{j^\pm_{\sig f}}\lt(\lt|1\pm\frac{1}{\g}\rt|\rho_{f}\rt)\sim i^{-2j^\pm_{\sig f}}\frac{2j^
\pm_{\sig f}+1}{\lt|1\pm\frac{1}{\g}\rt|\rho_{f}}\ \delta\lt(2\lt|1\pm\frac{1}{\g}\rt|\rho_{f}-2j^\pm_{\sig f}\rt)
\ee
In this limit, the integral of $\rho_f$ in the large area regime is completely constrained by the delta functions. Thus, as
in the previous section, the delta functions impose the constraints:
\be
j^\pm_{\sig f}=j^\pm_{\sig' f}=j^\pm_f\ \ \ \ \text{and}\ \ \ \ \lt|1-\frac{1}{\g}\rt|j^+_{f}=\lt|1+\frac{1}{\g}\rt|
j^-_{f}\label{jj}
\ee
This shows that in the large-$(j,\rho)$ limit the spins $j^\pm_{\sig f}$ for different wedges are identical
on the same face dual to $f$, and $j^+_f$ and $j^-_f$ satisfy the ``$\g$-simple'' relation in this
limit. However, since at the current stage the constraint
\be
k_{\sig_i f}=k'_{\sig_{i+1} f}\equiv k_{t_if}
\ee
are not obviously imposed by the integral of $N_{tf}$ (because of the present of closure constraint
in $\ca_t$), the vertex amplitude $\ca_\sig$, even in the large-$j$ limit, does not approximate the $
\text{FK}_\g$ vertex amplitude in general.

To explore the structure of this amplitude, we
consider the integral of $N_{tf}$ in the expression of $\ca_t$, for a tetrahedron $t$ shared
by $\sig, \sig'$
\be
&&\ca_t\Big(\rho_f;k_{\sig f},k'_{\sig f};l_{\sig t};b_{\sig f}\Big)\nonumber\\
&:=&\int\prod_{f\subset t}\rmd N_{tf}\ \delta\Big(\sum_{f\subset t}\rho_{f}N_{tf}\t_3N_{tf}^{-1}\Big)\ \lt
[\overline{C^4_{l_{\sig t}}\lt(k_{\sig f},k'_{\sig f'}\rt)}_{\{\a_{\sig f}\},\{\a_{\sig f'}\}}\rt]\lt[\overline{C^4_{l_
{\sig' t}}\lt(k_{\sig' f},k'_{\sig' f'}\rt)}_{\{\b_{\sig' f'}\},\{\b_{\sig' f}\}}\rt]\nonumber\\
&&\prod_{f\ \text{outgoing}}\pi^{k'_{\sig f}}_{2b_{\sig f},\a_{\sig f}}(N^{-1}_{tf})\ \pi^{k_{\sig' f}}_{\b_
{\sig' f},2b_{\sig' f}}(N_{tf})\prod_{f'\ \text{incoming}}\pi^{k_{\sig f'}}_{\a_{\sig f'},2b_{\sig f'}}(N_{tf'})\pi^
{k'_{\sig' f'}}_{2b_{\sig' f'},\b_{\sig' f'}}(N^{-1}_{tf'})\label{Nint}
\ee

Recall that $N_{tf}=g_{tf}N_{f}$ ($N_{tf}=N^+_{tf}=g^+_{tf}N^+_{f}$) while the integrand of the
partition function only depends on the combination $N_f\t_3N_f^{-1}$ (recall that the integrand
depends on $X_f^\pm$), i.e. the integrand is invariant under $N_f\mapsto N_fh_\phi$ where $h_
\phi\in\text{U(1)}$ thus it only depends on $\text{SU(2)}/\text{U(1)}$. Let us parameterize $N_f$ in
terms of the spherical coordinates: In terms of the complex coordinates ($z_f,\bar{z}_f$) on the unit
sphere we have
\be
N(z_f)&=&\frac{1}{\sqrt{1+|z_f|^2}}\left(
                                                    \begin{array}{cc}
                                                      1 & z_f \\
                                                      -\bar{z}_f & 1 \\
                                                    \end{array}
                                                  \right)
\ee
where the complex coordinates $z,\bar{z}$ are defined by the stereographic projection, and the
unit vector $\vec{\O}$ on $S^2$ is expressed in terms of the complex coordinates
\be
\vec{\O}(z)&=&-{i}\lt(-\frac{z+\bar{z}}{1+|z|^2}\ \sig_1+\frac{1}{i}\frac{z-\bar{z}}{1+|z|^2}\ \sig_2+\frac
{1-|z|^2}{1+|z|^2}\ \sig_3\rt)=N(z)\t_3N(z)^{-1}
\ee
Under the action of SU(2) group
\be
gN(z)=N(z^g)\left(
              \begin{array}{cc}
                \frac{\bar{a}-\bar{b}z}{|a-b\bar{z}|} & 0 \\
                0 & \frac{{a}-{b}\bar{z}}{|a-b\bar{z}|} \\
              \end{array}
            \right)^{-1}\ \ \ \ \text{where}\ \ \ \ \left(
              \begin{array}{cc}
                \frac{\bar{a}-\bar{b}z}{|a-b\bar{z}|} & 0 \\
                0 & \frac{{a}-{b}\bar{z}}{|a-b\bar{z}|} \\
              \end{array}
            \right)\in\text{U(1)}
\ee
where
\be
z^g=\frac{az+b}{\bar{a}-\bar{b}z}\ \ \ \ \text{with}\ \ \ \ g=\left(
                                                    \begin{array}{cc}
                                                      a & b \\
                                                      -\bar{b} & \bar{a} \\
                                                    \end{array}
                                                  \right)
\ee
Therefore
\be
N_{tf}=g_{tf}N(z_f)=N(z_{tf})h_{\phi_{tf}}^{-1}\ \ \ \ \ h_{\phi_{tf}}\in \text{U(1)}
\ee
where $z_{tf}=z_f^{g_{tf}}$. Note that the above decomposition may also be understood by writing
the SU(2) matrix in terms of Euler coordinates, i.e. $u=u(\phi_2)\ u(\theta)\ u(\phi_1)$ for all $u\in
\text{SU(2)}$
\be
u(\phi_1)=\pm\left(
            \begin{array}{cc}
              e^{i\phi_1/2} & 0 \\
              0 & e^{i\phi_1/2} \\
            \end{array}
          \right)\ \ \ \
u(\theta)=\pm\left(
            \begin{array}{cc}
              \cos\frac{\theta}{2} & i\sin\frac{\theta}{2} \\
              i\sin\frac{\theta}{2} & \cos\frac{\theta}{2} \\
            \end{array}
          \right)\ \ \ \
u(\phi_2)=\pm\left(
            \begin{array}{cc}
              e^{i\phi_2/2} & 0 \\
              0 & e^{i\phi_2/2} \\
            \end{array}
          \right)\ \ \ \
\ee
where $u(\phi_1),u(\phi_2)\in\text{U(1)}$, and $0\leq\phi_1,\phi_2\leq2\pi$, $0\leq\theta\leq\pi$,
while the SU(2) Haar measure can also be written as
\be
\rmd g=\frac{1}{16\pi^2}\sin\theta\rmd\phi_1\rmd\theta\rmd\phi_2
\ee

Hence the integral Eq.(\ref{Nint}) can be written as
\be
&&\ca_t\Big(\rho_f;k_{\sig f},k'_{\sig f};l_{\sig t};b_{\sig f}\Big)\nonumber\\
&=&\int\prod_{f\subset t}\rmd^2 \O_{tf}\rmd\phi_{tf}\ \delta\Big(\sum_{f\subset t}\rho_{f}\vec{\O}_{tf}
\Big)\ \lt[\overline{C^4_{l_{\sig t}}\lt(k_{\sig f},k'_{\sig f'}\rt)}_{\{\a_{\sig f}\},\{\a_{\sig f'}\}}\rt]\lt[\overline
{C^4_{l_{\sig' t}}\lt(k_{\sig' f},k'_{\sig' f'}\rt)}_{\{\b_{\sig' f'}\},\{\b_{\sig' f}\}}\rt]\nonumber\\
&&\prod_{f\ \text{outgoing}}\pi^{k'_{\sig f}}_{2b_{\sig f},\a_{\sig f}}\Big(h_{\phi_{tf}}N(z_{tf})^{-1}\Big)\
\pi^{k_{\sig' f}}_{\b_{\sig' f},2b_{\sig' f}}\Big(N(z_{tf})h_{\phi_{tf}}^{-1}\Big)\nonumber\\
&&\prod_{f'\ \text{incoming}}\pi^{k_{\sig f'}}_{\a_{\sig f'},2b_{\sig f'}}\Big(N(z_{tf'})h_{\phi_{tf'}}^
{-1}\Big)\pi^{k'_{\sig' f'}}_{2b_{\sig' f'},\b_{\sig' f'}}\Big(h_{\phi_{tf'}}N(z_{tf'})^{-1}\Big)\label{Nint1}
\ee
where $\rmd\O_{tf}$ is the standard spherical measure on $S^2=\text{SU(2)}/\text{U(1)}$. Since $
\pi^j_{mn}(h_\phi)=(e^{2im\phi})\delta_{mn}$ it follows
\be
&&\ca_t\Big(\rho_f;k_{\sig f},k'_{\sig f};l_{\sig t};b_{\sig f}\Big)\nonumber\\
&=&\int\prod_{f\subset t}\rmd^2 \O_{tf}\rmd\phi_{tf}\ \delta\Big(\sum_{f\subset t}\rho_{f}\vec{\O}_{tf}
\Big)\ \lt[\overline{C^4_{l_{\sig t}}\lt(k_{\sig f},k'_{\sig f'}\rt)}_{\{\a_{\sig f}\},\{\a_{\sig f'}\}}\rt]\lt[\overline
{C^4_{l_{\sig' t}}\lt(k_{\sig' f},k'_{\sig' f'}\rt)}_{\{\b_{\sig' f'}\},\{\b_{\sig' f}\}}\rt]\nonumber\\
&&\prod_{f\ \text{outgoing}}e^{4i(b_{\sig f}-b_{\sig' f})\phi_{tf}}\ \pi^{k'_{\sig f}}_{2b_{\sig f},\a_{\sig f}}
\Big(N(z_{tf})^{-1}\Big)\ \pi^{k_{\sig' f}}_{\b_{\sig' f},2b_{\sig' f}}\Big(N(z_{tf})\Big)\nonumber\\
&&\prod_{f'\ \text{incoming}}e^{4i(b_{\sig' f'}-b_{\sig f'})\phi_{tf'}}
\pi^{k_{\sig f'}}_{\a_{\sig f'},2b_{\sig f'}}\Big(N(z_{tf'})\Big)\pi^{k'_{\sig' f'}}_{2b_{\sig' f'},\b_{\sig' f'}}\Big
(N(z_{tf'})^{-1}\Big)\label{Nint2}
\ee
The integrals $\int_0^{2\pi}\rmd\phi_{tf}$ impose the constraint that $b_{\sig f}=b_{\sig' f}\equiv b_f
$ for all $f\subset t$, hence
\be
&&\ca_t\Big(\rho_f;k_{\sig f},k'_{\sig f};l_{\sig t};b_{\sig f}\Big)\nonumber\\
&=&\int\prod_{f\subset t}\rmd^2 \O_{tf}\ \delta\Big(\sum_{f\subset t}\rho_{f}\vec{\O}_{tf}\Big)\ \lt
[\overline{C^4_{l_{\sig t}}\lt(k_{\sig f},k'_{\sig f'}\rt)}_{\{\a_{\sig f}\},\{\a_{\sig f'}\}}\rt]\lt[\overline{C^4_{l_
{\sig' t}}\lt(k_{\sig' f},k'_{\sig' f'}\rt)}_{\{\b_{\sig' f'}\},\{\b_{\sig' f}\}}\rt]\nonumber\\
&&\prod_{f\ \text{outgoing}}
\pi^{k'_{\sig f}}_{2b_{f},\a_{\sig f}}\Big(N(z_{tf})^{-1}\Big)\ \pi^{k_{\sig' f}}_{\b_{\sig' f},2b_{f}}\Big(N(z_
{tf})\Big)
\prod_{f'\ \text{incoming}}
\pi^{k_{\sig f'}}_{\a_{\sig f'},2b_{f'}}\Big(N(z_{tf'})\Big)\ \pi^{k'_{\sig' f'}}_{2b_{f'},\b_{\sig' f'}}\Big(N(z_
{tf'})^{-1}\Big)\label{close}
\ee
Moreover for the outgoing dual face $f$, we have the relation
\be
&&\pi^{k'_{\sig f}}_{2b_{f},\a_{\sig f}}\Big(N^{-1}\Big)\ \pi^{k_{\sig' f}}_{\b_{\sig' f},2b_{f}}\Big(N\Big)
\ =\ \pi^{k'_{\sig f}}_{2b_{f},\a_{\sig f}}\Big(\eps N^{T}\eps^{-1}\Big)\ \pi^{k_{\sig' f}}_{\b_{\sig' f},2b_{f}}
\Big(N\Big)\nonumber\\
&=&(-1)^{2k'_{\sig f}-2b_{f}-\a_{\sig f}}\pi^{k'_{\sig f}}_{-\a_{\sig f},-2b_{f}}\Big(N\Big)\ \pi^{k_{\sig' f}}_
{\b_{\sig' f},2b_{f}}\Big(N\Big)\nonumber\\
&=&(-1)^{2k'_{\sig f}-2b_{f}-\a_{\sig f}}\sum_{l'_{tf}=|k'_{\sig f}-k_{\sig' f}|}^{k'_{\sig f}+k_{\sig' f}}c\lt(l'_
{tf};k'_{\sig f},k_{\sig' f}\rt)_{-\a_{\sig f},\b_{\sig' f}}^{\rho'_{tf}}{c\big(l'_{tf};k'_{\sig f},k_{\sig' f}\big)_
{-2b_f,2b_f}^{0}}\pi^{l'_{tf}}_{\rho'_{tf},0}\big(N\big)
\ee
while for the incoming dual face $f'$ we have similarly
\be
&&\pi^{k_{\sig f'}}_{\a_{\sig f'},2b_{f'}}\Big(N\Big)\ \pi^{k'_{\sig' f'}}_{2b_{f'},\b_{\sig' f'}}\Big(N^{-1}\Big)
\ =\ \pi^{k_{\sig f'}}_{\a_{\sig f'},2b_{f'}}\Big(N\Big)\ \pi^{k'_{\sig' f'}}_{2b_{f'},\b_{\sig' f'}}\Big(\eps N^{T}
\eps^{-1}\Big)\nonumber\\
&=&(-1)^{2k'_{\sig' f'}-2b_{f'}-\b_{\sig' f'}}\pi^{k_{\sig f'}}_{\a_{\sig f'},2b_{f'}}\Big(N\Big)\ \pi^{k'_{\sig'
f'}}_{-\b_{\sig' f'},-2b_{f'}}\Big(N\Big)\nonumber\\
&=&(-1)^{2k'_{\sig' f'}-2b_{f'}-\b_{\sig' f'}}\sum_{l_{tf'}=|k_{\sig f'}-k'_{\sig' f'}|}^{k_{\sig f'}+k'_{\sig' f'}}c\lt
(l_{tf'};k_{\sig f'},k'_{\sig' f'}\rt)_{\a_{\sig f'},-\b_{\sig' f'}}^{\rho_{tf'}}{c\big(l_{tf'};k_{\sig f'},k'_{\sig' f'}\big)
_{2b_{f'},-2b_{f'}}^{0}}\pi^{l_{tf'}}_{\rho_{tf'},0}\big(N\big)
\ee
Thus the integral reduces to
\be
\ca_t\Big(\rho_f;k_{\sig f},k'_{\sig f};l_{\sig t};b_{\sig f}\Big)=\int\prod_{f\subset t}\rmd^2 \O_{tf}\rmd
\phi_{tf}\ \delta\Big(\sum_{f\subset t}\rho_{f}\vec{\O}_{tf}\Big)\ \Theta_t\lt(l_{\sig t},l_{\sig' t},k'_{\sig
f},k_{\sig' f},k_{\sig f'},k'_{\sig' f'},b_f;z_{tf}\rt)\label{At}
\ee
with the integrand ($t$ is the tetrahedron shared by $\sig, \sig'$)
\be
&&\Theta_t\lt(l_{\sig t},l_{\sig' t},k'_{\sig f},k_{\sig' f},k_{\sig f'},k'_{\sig' f'},b_f;z_{tf}\rt):=\nonumber\\
&&\prod_{f\ \text{outgoing}}
(-1)^{2k'_{\sig f}-2b_{f}-\a_{\sig f}}\sum_{l'_{tf}=|k'_{\sig f}-k_{\sig' f}|}^{k'_{\sig f}+k_{\sig' f}}c\lt(l'_
{tf};k'_{\sig f},k_{\sig' f}\rt)_{-\a_{\sig f},\b_{\sig' f}}^{\rho'_{tf}}{c\big(l'_{tf};k'_{\sig f},k_{\sig' f}\big)_
{-2b_f,2b_f}^{0}}\pi^{l'_{tf}}_{\rho'_{tf},0}\big(N(z_{tf})\big)\nonumber\\
&&\prod_{f'\ \text{incoming}}
(-1)^{2k'_{\sig' f'}-2b_{f'}-\b_{\sig' f'}}\sum_{l_{tf'}=|k_{\sig f'}-k'_{\sig' f'}|}^{k_{\sig f'}+k'_{\sig' f'}}c\lt(l_
{tf'};k_{\sig f'},k'_{\sig' f'}\rt)_{\a_{\sig f'},-\b_{\sig' f'}}^{\rho_{tf'}}{c\big(l_{tf'};k_{\sig f'},k'_{\sig' f'}\big)_
{2b_{f'},-2b_{f'}}^{0}}\pi^{l_{tf'}}_{\rho_{tf'},0}\big(N(z_{tf'})\big)\nonumber\\
&&\lt[\overline{C^4_{l_{\sig t}}\lt(k_{\sig f},k'_{\sig f'}\rt)}_{\{\a_{\sig f}\},\{\a_{\sig f'}\}}\rt]\lt[\overline
{C^4_{l_{\sig' t}}\lt(k_{\sig' f},k'_{\sig' f'}\rt)}_{\{\b_{\sig' f'}\},\{\b_{\sig' f}\}}\rt]\label{theta}
\ee
%on the constraint surface defined by the closure constraint $\sum_{f\subset t}\%rho_{f}\vec{\O}_{tf}$.

The complete integration of Eq.(\ref{At}) turns out to be rather involved, thus is left as a future
research. What one can say is the following: From the expression Eq.(\ref{theta}), it is not hard to see
that the set of
amplitudes contributing to the simplified model $Z_{\text{Simplified}}(\ck)$ are part of those of the
full model $Z(\ck)$. To see this, notice  that the simplest nontrivial
contribution of the integral in $\ca_t$ comes from the term with $l_{tf'}=l'_{tf}=0$. With the
constraints $l_{tf'}=l'_{tf}=0$ (and thus $\rho'_{tf}=\rho_{tf'}=0$), we obtain the same set of
constraints as it was in the previous subsection for the simplified model $Z_{\text{Simplified}}(\ck)
$.
\be
k'_{\sig f}=k_{\sig' f}\equiv k_{tf}\ \ \ \ k_{\sig f'}=k'_{\sig' f'}\equiv k_{tf'}\ \ \ \ \a_{\sig f'}=\b_{\sig' f'}\ \ \ \
\a_{\sig f}=\b_{\sig' f}
\ee
Since
\be
c\lt(0;k,k'\rt)_{\a,\b}^{0}=\delta_{k,k'}\delta_{\a,-\b}\frac{(-1)^{k-\a}}{\sqrt{\dim(k)}}
\ee
we obtain, by extracting the term with $l_{tf'}=l'_{tf}=0$ and dropping the contribution from the
other terms
\be
&&\Theta_t\lt(l_{\sig t},l_{\sig' t},k'_{\sig f},k_{\sig' f},k_{\sig f'},k'_{\sig' f'},b_f;z_{tf}\rt)\nonumber\\
&\to&\prod_{f\ \text{outgoing}}
(-1)^{2k'_{\sig f}-2b_{f}-\a_{\sig f}}
\delta_{k'_{\sig f},k_{\sig' f}}\delta_{\a_{\sig f},\b_{\sig' f}}\frac{(-1)^{k'_{\sig f}+\a_{\sig f}}(-1)^{k'_{\sig
f}+2b_f}}{\dim(k'_{\sig f})}
\nonumber\\
&&\prod_{f'\ \text{incoming}}
(-1)^{2k'_{\sig' f'}-2b_{f'}-\b_{\sig' f'}}
\delta_{k_{\sig f'},k'_{\sig' f'}}\delta_{\a_{\sig f'},\b_{\sig' f'}}\frac{(-1)^{k_{\sig f'}-\b_{\sig' f'}}(-1)^{k_
{\sig f'}-2b_{f'}}}{\dim(k_{\sig f'})}\nonumber\\
&&\lt[\overline{C^4_{l_{\sig t}}\lt(k_{\sig f},k'_{\sig f'}\rt)}_{\{\a_{\sig f}\},\{\a_{\sig f'}\}}\rt]\lt[\overline
{C^4_{l_{\sig' t}}\lt(k_{\sig' f},k'_{\sig' f'}\rt)}_{\{\b_{\sig' f'}\},\{\b_{\sig' f}\}}\rt]\nonumber\\
&=&\prod_{f\subset t}\frac{1}{\dim(k_{tf})}\delta_{k'_{\sig f},k_{\sig' f}}\delta_{k_{\sig f'},k'_{\sig' f'}}
\delta_{l_{\sig t},l_{\sig' t}^\dagger}
\ee
For this subset of amplitude the edge/tetrahedron amplitude reduces to
\be
\ca_t\Big(\rho_f;k_{\sig f},k'_{\sig f};l_{\sig t};b_{\sig f}\Big)&\to& \prod_{f\subset t}\frac{1}{\dim(k_{tf})}
\delta_{k'_{\sig f},k_{\sig' f}}\delta_{k_{\sig f'},k'_{\sig' f'}}\delta_{l_{\sig t},l_{\sig' t}^\dagger}\int\prod_
{f\subset t}\rmd^2 \O_{tf}\rmd\phi_{tf}\ \delta\Big(\sum_{f\subset t}\rho_{f}\vec{\O}_{tf}\Big)\nonumber\\
&\equiv&\ca_t'\Big(\rho_f;k_{\sig f},k'_{\sig f};l_{\sig t};b_{\sig f}\Big)\label{add}
\ee
Then we can define a spin-foam model by pick out a subset of amplitudes in the full partition function $Z(\ck)$:
\be
Z'(\ck)
&=&\sum_{\{j^\pm_{\sig f}\}}\sum_{\{k_{\sig f},k'_{\sig f}\}}\sum_{\{l_{\sig t}\}}\sum_{\{b_{\sig f}\}}
\int\prod_{f}\rmd\rho_{f}\times\nonumber\\
&&\ca_f\lt(\rho_f;j^\pm_{\sig f};k_{\sig f},k'_{\sig f};b_{\sig f}\rt)\
\ca'_t\Big(\rho_f;k_{\sig f},k'_{\sig f};l_{\sig t};b_{\sig f}\Big)\
\ca_\sig\lt(j^\pm_{\sig f};k_{\sig f},k'_{\sig f};l_{\sig t}\rt)
\ee
The amplitudes in $Z'(\ck)$ are contributions with the closure constraint implemented, however unfortunately they may not exhaust all the contributions.

In Eq.(\ref{add}) the Kronecker deltas $\delta_{k'_{\sig f},k_{\sig' f}}\delta_{k_{\sig f'},k'_{\sig' f'}}\delta_{l_{\sig
t},l_{\sig' t}^\dagger}$ imply that there is an one-to-one correspondence between the transition channels in the simplified model $Z_
{\text{Simplified}}(\ck)$ and the transition channels in the model $Z'(\ck)$, which form a subset of the transition channels in $Z(\ck)$.
Consider the sets $\{Z_{\text
{Simplified}}\}$ and $\{Z\}$ respectively, which are the collections of spin-foams that contribute
to their respective
partition functions $Z_{\text{Simplified}}(\ck)$ and $Z(\ck)$. Our above analysis
then reveals
\be
\{Z_{\text{Simplified}}\}\subset \{Z\}
\ee
At this point this is all we can say about the relation between the models with the closure constraint
in place or not. The additional weights and contributions in the full model may severely change
the correlators (physical inner product) and it is by no means obvious that the simplified model
is a good approximation.

As a final remark, the above inclusion is in terms of spin-foam amplitude, in the sense that we write the partition functions as a sum of amplitude over possible spins and intertwiners. Moreover such an inclusion is natural from the path integral point of view. We consider a simple example: Consider a function $f(x,y)$ on $\mathbb{R}^2$ which has a Fourier transform $\tilde{f}(k,q)$ and that we have a ``closure constraint'' $y=0$. Then
the Z integral (with closure) corresponds to (dropping factors of $2\pi$)
\be
Z&=&\int dx dy \delta(y,0) f(x,y)
\ =\ \int dx dy \delta(y,0)\int dk dq \tilde{f}(k,q)\exp(i(kx+qy))\nonumber\\
&=&\int dx \int dk dq \tilde{f}(k,q)\exp(ikx)
\ =\ \int dk dq \tilde{f}(k,q) \delta(k,0)\nonumber\\
&=& \int dq \tilde{f}(0,q)
\ee
On the other hand the $Z_{\text{Simplied}}$ integral without closure is
\be
Z_{\text{Simplied}}&=&\int dx dy f(x,y)
\ =\ \int dx dy \int dk dq \tilde{f}(k,q)\exp(i(kx+qy))
\ =\ \int dk dq \tilde{f}(k,q) \delta(k,0) \delta(q,0)\nonumber\\
&=&\tilde{f}(0,0)
\ee
Hence the $Z$ amplitudes are more in Fourier space ($k,p$) corresponding to spin-foam representation, and less in real space ($x,y$).

\section{Outlook}

In section \ref{NSF} we first carried out the analysis for the simplified partition function without
closure constraint and obtained the spin-foam model $Z_{\text{Simplified}}(\ck)$, then we discussed
the complete partition function $Z(\ck)$ with closure constraint implemented, however we did not compute
yet explicitly the full set of possible spin-foam amplitudes. We were only able to show that all the spin-foam
amplituded contributing to $Z_{\text{Simplified}}(\ck)$ are contained in those contributing to
the full model $Z(\ck)$. Therefore, in addition to present spin foam models,
our commutative $B$ field model variable sums over additional amplitudes having non-
trivial contributions to the partition function $Z(\ck)$. While we have shown that in the large-$j$
limit the 4-simplex/vertex amplitude of $Z_{\text{Simplified}}(\ck)$ can be related to the 4-simplex/
vertex amplitude of $\text{FK}_{\g}$ Model ($|\g|>1$), for the full model $Z(\ck)$, even in the
large-$j$ limit, there exist additional, non-trivial spin-foam amplitudes.
It would be important to further specify those unknown spin-foams
contributing to $\{Z\}$ but not to $\{Z_{\text{Simplified}}\}$, at least for their large-$j$ asymptotics.

Unfortunately, the relation between our new model and EPRL model is almost untouched in the present article.
Although we have seen that all the EPRL spin-foams (with possibly different triangle/face and
tetrahedron/edge amplitudes) are included in $\{Z_{\text{Simplified}}\}$ (thus in $\{Z\}$), it seems to
us that, however, they are not quite special among the spin-foam amplitudes contributing $Z_{\text
{Simplified}}(\ck)$ or $Z(\ck)$. We expected that the relation between our model $Z(\ck)$ and
EPRL Model could be realized by the non-commutative deformation, like in the case of Barrett-Crane
Model. The reason for our expectation was that (1) both models are defined via the non-commutative operator constraint technique, and (2) when the Barbero-Immirzi parameter $\g\to
\infty$, EPRL Model reduces to Barrett-Crane Model. However it turns out that our expectation is
difficult to realize, since the non-commutative deformation via the group Fourier transformation
hardly works for the case of finite $\g$. It seems to us that if our model $Z(\ck)$ and the EPRL Model
could be related via any non-commutative deformation, we should rather choose a different
deformation scheme.

The present article starts from a purely path-integral/spin-foam point of view. If we also consider
the relation between the path integral and canonical quantization, then the partition function Eq.
(\ref{start}) should probably be modified. It is pointed in \cite{links} that a quantum gravity path
integral formula consistent with canonical physical inner product should not only be an naive path
integral Eq.(\ref{PL}) of Plebanski-Holst action, but also include a suitable local measure factor in
the path integral formula. The local measure factor is a product of a certain power of spacetime
volume elements and a certain power of spatial volume elements at all the spacetime points. The
implementation of such local measure factor in the partition function will modify both the 4-simplex/
vertex and tetrahedron/edge amplitudes. A detailed analysis of this issue will be postponed to future
research.

It is interesting to look for relations with other new approaches on the implementation of simplicity constraint in spinfoam models or GFTs. In the appendix, we show that a non-commutative deformation of the above model, as a noncommutative simplicial path integral, relates to the GFT model defined in \cite{AOL}. One may also compare the approach here with the ``holomorphic simplicity constraint'' in \cite{DFLS}, where the new version of simplicity constraints using spinor/twistor variables are commutative. However this approach closely relates to the operator-constraint approach reviewed in the introduction. The commutative holomorphic simplicity constraints come from the noncommutative algebra of flux variables. It may also be interesting to see the relation with quantum Regge calculus. As far as we have shown, the spinfoam model constructed here comes from a path integral of simplicial Plebanski-Holst action, where the discretization procedure is different from Regge calculus (in 1st or 2nd order formulations). So the resulting spinfoam model doesn't coincide with the quantum Regge calculus in general. But it is possible that they may be related in certain limit. Such a possibility should be studied in the future.

\section*{Acknowledgements}

MH is grateful to Aristide Baratin for many enlightening discussions, and thanks You Ding for
letting him know about the details of her recent work. MH also acknowledges the support by
International Max Planck Research School (IMPRS) and the partial support by NSFC Nos.
10675019 and 10975017. A part of the research has received funding from the People Programme (Marie Curie Actions) of the European Union's 7th Framework Programme (FP7/2007-2013) under REA grant agreement No. 298786.

\begin{appendix}

\section{Noncommutative Deformation and Barrett-Crane Model}\label{deformation} 

\subsection{Noncommutative Deformation}

In order to further investigate the question, in which sense the closure constraint is
redundant when working with non commutative B fields (as is common practice in existent
spin foam models), in this section, we explore a non-commutative deformation of our starting point,
the partition function
$Z(\ck)$ in Eq.(\ref{start}). The non-commutative deformation we will employ here comes from a
generalized Fourier transformation defined on a compact group \cite{Gfourier}. The
deformation replaces the normal c-number product in the expression of $Z(\ck)$ by
a non-commutative ``$\star$-product'' (we will briefly review the definition below). Interestingly,
this non-commutative deformation establishes a relation between the new spin-
foam model $Z(\ck)$ we analyzed in the previous section and the Barrett-Crane spin-foam model
\cite{BC}. In some sense it relates the recent approach of using noncommutative product in the simplicial path integral representation of the Group Field Theory (GFT) \cite{AOL}.

First of all, we recall the partition function $Z(\ck)$ in the commutative context (after the
linearization of simplicity constraint):
\be
Z(\ck)&:=&\int_{II\pm}\prod_{f}\rmd^3X^+_f\rmd^3X^-_f\prod_{(\sig,t)}\rmd g^+_{\sig t}\rmd g^-_{\sig
t}\prod_{(t,f)}\rmd g^+_{tf}\rmd g^-_{tf}\prod_t\rmd u_t\prod_{t,f}\delta\lt(X_{tf}^-+\b u_{t}X_{tf}^+u_{t}
^{-1}\rt)\prod_{t}\delta\Big(\sum_{f\subset t}X_{tf}^+\Big)\nonumber\\
&&\times\prod_{(\sig,f)}e^{i\tr\lt(X^+_fg^+_{f t}g^+_{t\sig}g^+_{\sig t'}g^+_{t'f}\rt)}\prod_{(\sig,f)}e^{i\tr
\lt(X^-_fg^-_{f t}g^-_{t\sig}g^-_{\sig t'}g^-_{t'f}\rt)}\label{start1}
\ee
where $\b=\frac{1-1/\g}{1+1/\g}$ and for the convenience of the following analysis, we have made
a change of variables
\be
X_f^\pm\mapsto\lt(1\pm \frac{1}{\g}\rt)^{-1}X_f^\pm.
\ee
and dropped a constant $\gamma$ dependent factor. Here we assume that our structure group is SO(3)$\times$SO(3) instead of SO(4)$\simeq$SU(2)$\times$SU(2)$/\mathbb{Z}_2$. The reason for this replacement is to be compatible with the group fourier transformation, which will be seen shortly.

We now replace (by hand) the commutative c-number product in Eq.(\ref{start}) by the non-
commutative $\star$-product on $\fs\fu(2)\simeq\mathbb{R}^3$ defined in \cite{Gfourier}, that is
\be
e^{\frac{i}{2a}\tr(X|g_1|)}\star e^{\frac{i}{2a}\tr(X|g_2|)}:=e^{\frac{i}{2a}\tr(X|g_1g_2|)}\label
{noncommutative product}
\ee
where $a$ is the deformation parameter, $X=X^j\t_j$ and $\t_j=-i\sig_j$ with $\sig_j$ the Pauli
matrices $\sig_i\sig_j=\delta_{ij}+i\eps_{ijk}\sig_k$, $g\in\text{SU}(2)$ represented by a $2\times 2$
matrix and $|g|=\text{sgn}(\tr g)g$ so that $|-g|=|g|$. We can write $g\in\text{SU}(2)$ as
\be \label{a}
g=P_0+ia\vec{P}\cdot\vec{\sig},\ \ \ \ \ P_0^2+a^2||\vec{P}||^2=1
\ee
Thus $|g|$ is the projection of $g$ on the upper ``hemisphere'' of SU(2) with $P_0\geq0$. Therefore
the ``plane wave'' in Eq.(\ref{noncommutative product}) can be written
\be
e_g(X):=e^{\frac{i}{2a}\tr(X|g|)}=e^{i\vec{P}\cdot\vec{X}{\rm sgn}(\tr g)}\label{PW}
\ee
depends on SO(3) only (its character expansio depnds on integral representations only
because it is an even function under reflection $g\to -g$). With these ``plane waves'' we can define
an invertible
``Group Fourier Transformation'' from the functions $f(g)$ on SO(3) ($f(g)=f(-g)$ for $g\in\text{SU
(2)}$) to the functions $\tilde{f}(X)$ on the Lie algebra $\fs\fu(2)$
\be
\tilde{f}(X)&=&\int\rmd g\ f(g)\ e_g(X)\nonumber\\
f(g)&=&\frac{1}{8\pi a^3}\int\rmd^3X\ \tilde{f}(X)\star e_{g^{-1}}(X)\ =\ \frac{\sqrt{1-a^2|\vec{P}(g)|^2}}
{8\pi a^3}\int\rmd^3X\ \tilde{f}(X)\ e_{g^{-1}}(X)
\ee
Given two functions $\tilde{f}_1(X)$ and $\tilde{f}_2(X)$ in the image of the group Fourier
transformation, their $\star$-product is defined as
\be
\tilde{f}_1(X)\star\tilde{f}_2(X)=\int\rmd g_1\rmd g_2\ f_1(g_1)\ f_2(g_2)\ e_{g_1}(X)\star e_{g_2}(X)
\ee
and when the deformation parameter turns to $a\to0$, the $\star$-product reproduces the normal
commutative product (if we keep $P_0,\vec{P}$ fixed, see (\ref{a})).

 We also have two identities for delta functions
\be
\delta_{\text{SO(3)}}(g)&=&\frac{1}{8\pi a^3}\int\rmd^3X\ e_g(X)\nonumber\\
\delta_X(X')&=&\int\rmd g\ e_{g^{-1}}(X)\ e_g(X')
\ee
where the second delta function is the Dirac distribution in the noncommutative sense, that is
\be
\int\rmd^3X'\ \lt(\delta_X\star f\rt)(X')=\int\rmd^3X'\ \lt(f\star\delta_X \rt)(X')=f(X)
\ee

With the above definitions, we can make a noncommutative deformation of the integrand in Eq.(\ref
{start1}). In the following we fix the deformation parameter to
\be
a=\ell_p^2=1
\ee
The reason for this choice is that only in this case the closure constraint turns out to be
redundant and can be
removed from Eq.(\ref{start1}), which is necessary in order to derive the Barrett-Crane model. We
will show this immediately in the next paragraph. On the other hand, fixing $a=\ell_p^2$
makes it impossible to study the commutative limit $a\to 0$ of the non commutative model
$Z_a(\ck)$ which we denote by and thus we cannot
compare with the commutative model $Z(\ck)$.

We first define the noncommutative deformation of
\be
\prod_{t}\delta\Big(\sum_{f\subset t}X_{tf}^+\Big)\prod_{(\sig,f)}e^{i\tr\lt(X^+_fg^+_{f t}g^+_{t\sig}g^
+_{\sig t'}g^+_{t'f}\rt)}=\prod_{t}\delta\Big(\sum_{f\subset t}g^+_{tf}X_{f}^+g^+_{ft}\Big)\prod_{(\sig,f)}
e^{i\tr\lt(X^+_fg^+_{f t}g^+_{t\sig}g^+_{\sig t'}g^+_{t'f}\rt)}\label{closure and E}
\ee
Given a face dual to the triangle $f$ with $n$ vertices dual to the 4-simplices $\sig_1,\cdots,\sig_n
$ (cf. FIG.\ref{face}), we define the quantity
\be
\cg^+_f(X_f^+,g_{\sig t}^+,g_{tf}^+,h_t)&:=&\Big[e_{g^+_{ft_n}h_{t_n}g_{t_n f}^+}\star e_{g^+_{f t_n}
g^+_{t_n\sig_1}g^+_{\sig_1 t_1}g^+_{t_1f}}\star e_{g_{ft_1}^+h_{t_{1}}g^+_{t_1f}}\star e_{g^+_{f
t_1}g^+_{t_1\sig_2}g^+_{\sig_2 t_2}g^+_{t_2f}}\star\nonumber\\
&&\star\cdots\star e_{g_{ft_{n-1}}^+h_{t_{n-1}}g^+_{t_{n-1}f}}\star e_{g^+_{f t_{n-1}}g^+_{t_
{n-1}\sig_n}g^+_{\sig_n t_n}g^+_{t_nf}}\Big](X^+_f)
\ee
A possible noncommutative deformation of Eq.(\ref{closure and E}) is
\be
\int\prod_t\rmd h_t\prod_f\cg^+_f(X_f^+,g_{\sig t}^+,g_{tf}^+,h_t)
\ee
because the noncommutative Dirac distribution for the closure constraint is
\be
\delta\Big(\sum_{f\subset t}g^+_{tf}X_{f}^+g^+_{ft}\Big)=\int\rmd h_t\prod_{f\subset t}e_{g^+_{ft}
h_tg^+_{tf}}(X_f^+).\label{14}
\ee
It is here where the choice $a=\ell_P^2$ was important because we have implicitly set $\ell_P^2=1$
in the exponential so far (it comes from the fact that the flux field has dimension cm$^2$ and
the Plebanski action is multiplied by $1/\kappa$ where $\kappa \hbar=\ell_P^2$) so restoring it
we can combine the ordinary product of exponentials into star products only if the deformation parameter
is given by $a=\ell_P^2$\footnote{the point here is that one should make the exponential of action to look like a ``plane-wave'' Eq.(\ref{PW}) of group Fourier transformation. However it is $\ell_p^{-2}$ in front of the action but not $a^{-1}$ (The plane-wave in Eq.(\ref{14}) is with $a^{-1}$ not $\ell_P^2$). So we have to set $a=\ell_p^2$ to resolve the mismatch, in order to remove the closure condition from the ($\star$-deformed) path integral. }.

However, since
\be
\cg^+_f(X_f^+,g_{\sig t}^+,g_{tf}^+,h_t)=\Big[e_{g^+_{f t_n}h_{t_n}g^+_{t_n\sig_1}g^+_{\sig_1 t_1}
g^+_{t_1f}}\star e_{g^+_{f t_1}h_{t_{1}}g^+_{t_1\sig_2}g^+_{\sig_2 t_2}g^+_{t_2f}}\star\cdots\star
e_{g^+_{f t_{n-1}}h_{t_{n-1}}g^+_{t_{n-1}\sig_n}g^+_{\sig_n t_n}g^+_{t_nf}}\Big](X^+_f)
\ee
we can absorb $h_t$ into $g_{t\sig}$ by a change of variables
\be
g^+_{t_i\sig_{i+1}}\mapsto h_{t_i}^{-1}g^+_{t_i\sig_{i+1}}
\ee
while $\rmd g_{\sig t}^+$ does not change since it is Haar measure. Therefore finally the integral of
$h_t$ gives unity, which shows the redundancy of the closure constraint for this particular non commutative
deformation!

Next we consider the simplicity constraint
\be
\delta(X_{tf}^-+\b u_{t}X^+_{tf}u_{t}^{-1})=\delta(g^-_{tf}X_{f}^-g^-_{ft}+\b u_{t}g^+_{tf}X^+_{f}g^+_
{ft}u_{t}^{-1})
\ee
whose noncommutative version is
\be
\delta(g^-_{tf}X_{f}^-g^-_{ft}+\b u_{t}g^+_{tf}X^+_{f}g^+_{ft}u_{t}^{-1})&=&\int\rmd v_{tf}\ e_{v_{tf}}
(g^-_{tf}X_{f}^-g^-_{ft}+\b u_{t}g^+_{tf}X^+_{f}g^+_{ft}u_{t}^{-1})\nonumber\\
&=&\int\rmd v_{tf}\ e_{v_{tf}}(g^-_{tf}X_{f}^-g^-_{ft})\ e_{v_{tf}}(\b u_{t}g^+_{tf}X^+_{f}g^+_{ft}u_{t}^
{-1})\nonumber\\
&=&\int\rmd v_{tf}\ e_{g^-_{ft}v_{tf}g^-_{tf}}(X_{f}^-)\ e_{g^+_{ft}u_{t}^{-1}v_{tf}u_{t}g^+_{tf}}(\b X^+_
{f})
\ee
For the above factor related to $\b$ in the above integrand, we can write
\be
e^\b_g(X):=e_g(\b X)
\ee
Thus for each face dual to the triangle $f$ with $n$ vertices dual to the 4-simplices $\sig_1,\cdots,
\sig_n$, we define
\be
\cf^+_f(X_f^+,g_{\sig t}^+,g_{tf}^+,h_t,u_t,v_{tf},\b)&:=&\Big[e_{g_{ft_n}^+h_{t_n}g_{t_n f}^+}\star e_
{g^+_{f t_n}g^+_{t_n\sig_1}g^+_{\sig_1 t_1}g^+_{t_1f}}\star e^\b_{g^+_{ft_1}u_{t_1}^{-1}v_{t_1f}u_
{t_1}g^+_{t_1f}}\star\nonumber\\
&&\star\ e_{g_{ft_1}^+h_{t_{1}}g^+_{t_1f}}\star e_{g^+_{f t_1}g^+_{t_1\sig_2}g^+_{\sig_2 t_2}g^+_
{t_2f}}\star e^\b_{g^+_{ft_2}u_{t_2}^{-1}v_{t_2f}u_{t_2}g^+_{t_2f}}\star\nonumber\\
&&\star\cdots\star\nonumber\\
&&\star\ e_{g_{ft_{n-1}}^+h_{t_{n-1}}g^+_{t_{n-1}f}}\star e_{g^+_{f t_{n-1}}g^+_{t_{n-1}\sig_n}g^+_
{\sig_n t_n}g^+_{t_nf}}\star e^\b_{g^+_{ft_n}u_{t_n}^{-1}v_{t_nf}u_{t_n}g^+_{t_nf}}\Big](X^+_f)
\ee
and
\be
\cf^-_f(X_f^-,g_{\sig t}^-,g_{tf}^-,v_{tf})&:=&\Big[e_{g^-_{f t_n}g^-_{t_n\sig_1}g^-_{\sig_1 t_1}g^-_
{t_1f}}\star e_{g^-_{ft_1}v_{t_1f}g^-_{t_1f}}\star e_{g^-_{f t_1}g^-_{t_1\sig_2}g^-_{\sig_2 t_2}g^-_
{t_2f}}\star e_{g^-_{ft_2}v_{t_2f}g^-_{t_2f}}\star\nonumber\\
&&\star\cdots\star e_{g^-_{f t_{n-1}}g^-_{t_{n-1}\sig_n}g^-_{\sig_n t_n}g^-_{t_nf}}\star e_{g^-_{ft_n}
v_{t_nf}g^-_{t_nf}}\Big](X^-_f)
\ee

Then the deformed partition function is defined by
\be
Z_{\star}(\ck)&:=&\int\prod_{f}\rmd^3X^+_f\rmd^3X^-_f\prod_{(\sig,t)}\rmd g^+_{\sig t}\rmd g^-_{\sig
t}\prod_{(t,f)}\rmd g^+_{tf}\rmd g^-_{tf}\prod_t\rmd h_t\rmd u_t\prod_{(t,f)}\rmd v_{tf}\nonumber\\
&&\prod_f\cf^+_f(X_f^+,g_{\sig t}^+,g_{tf}^+,h_t,u_t,v_{tf},\b)\ \cf^-_f(X_f^-,g_{\sig t}^-,g_{tf}^-,v_{tf})
\ee
which is the noncommutative deformation of Eq.(\ref{start1}). However since we have shown the
redundancy of the closure constraint in $Z_\star(\ck)$, we can equivalently write
\be
Z_{\star}(\ck)&:=&\int\prod_{f}\rmd^3X^+_f\rmd^3X^-_f\prod_{(\sig,t)}\rmd g^+_{\sig t}\rmd g^-_{\sig
t}\prod_{(t,f)}\rmd g^+_{tf}\rmd g^-_{tf}\prod_t\rmd u_t\prod_{(t,f)}\rmd v_{tf}\nonumber\\
&&\prod_f\cf^+_f(X_f^+,g_{\sig t}^+,g_{tf}^+,u_t,v_{tf},\b)\ \cf^-_f(X_f^-,g_{\sig t}^-,g_{tf}^-,v_{tf})
\ee
where $\cf^+_f$ is replaced by
\be
\cf^+_f(X_f^+,g_{\sig t}^+,g_{tf}^+,u_t,v_{tf},\b)&:=&\Big[e_{g^+_{f t_n}g^+_{t_n\sig_1}g^+_{\sig_1
t_1}g^+_{t_1f}}\star e^\b_{g^+_{ft_1}u_{t_1}^{-1}v_{t_1f}u_{t_1}g^+_{t_1f}}\star\nonumber\\
&&\star\ e_{g^+_{f t_1}g^+_{t_1\sig_2}g^+_{\sig_2 t_2}g^+_{t_2f}}\star e^\b_{g^+_{ft_2}u_{t_2}^
{-1}v_{t_2f}u_{t_2}g^+_{t_2f}}\star\nonumber\\
&&\star\cdots\star\nonumber\\
&&\star\ e_{g^+_{f t_{n-1}}g^+_{t_{n-1}\sig_n}g^+_{\sig_n t_n}g^+_{t_nf}}\star e^\b_{g^+_{ft_n}u_
{t_n}^{-1}v_{t_nf}u_{t_n}g^+_{t_nf}}\Big](X^+_f).
\ee

\subsection{$\g=\infty$ and Barrett-Crane Model}

The computation with general $\beta$ is difficult, because it involves the $\star$-product between
two different types of plane waves $e_g$ and $e^\b_g$, which is even not well-defined in general
(since they could consider having different deformation parameter). Therefore here we only
consider the simplified case that $\g=\infty$. Then $\b=1$ and in this case we can directly compute $\cf^+_f
$ to be
\be
&&\cf^+_f(X_f^+,g_{\sig t}^+,g_{tf}^+,u_t,v_{tf},\b=1)\nonumber\\
&=&e_{g^+_{f t_n}g^+_{t_n\sig_1}g^+_{\sig_1 t_1}u_{t_1}^{-1}v_{t_1f}u_{t_1}g^+_{t_1\sig_2}g^+_
{\sig_2 t_2}u_{t_2}^{-1}v_{t_2f}u_{t_2}\cdots g^+_{t_{n-1}\sig_n}g^+_{\sig_n t_n}u_{t_n}^{-1}v_
{t_nf}u_{t_n}g^+_{t_nf}}(X^+_f)\nonumber\\
&=&e_{g^+_{t_n\sig_1}g^+_{\sig_1 t_1}u_{t_1}^{-1}v_{t_1f}u_{t_1}g^+_{t_1\sig_2}g^+_{\sig_2 t_2}
u_{t_2}^{-1}v_{t_2f}u_{t_2}\cdots g^+_{t_{n-1}\sig_n}g^+_{\sig_n t_n}u_{t_n}^{-1}v_{t_nf}u_{t_n}}
(g^+_{t_nf}X^+_fg^+_{f t_n})
\ee
Similarly for the anti-self-dual part
\be
&&\cf^-_f(X_f^-,g_{\sig t}^-,g_{tf}^-,v_{tf})\nonumber\\
&=&e_{g^-_{f t_n}g^-_{t_n\sig_1}g^-_{\sig_1 t_1}v_{t_1f}g^-_{t_1\sig_2}g^-_{\sig_2 t_2}v_{t_2f}
\cdots g^-_{t_{n-1}\sig_n}g^-_{\sig_n t_n}v_{t_nf}g^-_{t_nf}}(X^-_f)\nonumber\\
&=&e_{g^-_{t_n\sig_1}g^-_{\sig_1 t_1}v_{t_1f}g^-_{t_1\sig_2}g^-_{\sig_2 t_2}v_{t_2f}\cdots g^-_{t_
{n-1}\sig_n}g^-_{\sig_n t_n}v_{t_nf}}(g^-_{t_nf}X^-_fg^-_{f t_n})
\ee

We define the following changes of the variables
\be
g^+_{\sig_i t_i}\mapsto g^+_{\sig_i t_i}u_{t_i}\ \ \ \ g^+_{t_i\sig_{i+1}}\mapsto u^{-1}_{t_i}g^+_{t_i
\sig_{i+1}}\ \ \ \ X^+_f\mapsto g^+_{f t_n}u_{t_n}^{-1}X^+_fu_{t_n}g^+_{t_n f}\ \ \ \ X^-_f\mapsto g^-_
{ft_n}X^-_fg^-_{t_nf},
\ee
where for each face dual to $f$, a unique $t_n(f)$ is chosen as the base point of the dual face.
Thus the partition function can be written as
\be
Z_{\star}(\ck)&:=&\int\prod_{f}\rmd^3X^+_f\rmd^3X^-_f\prod_{(\sig,t)}\rmd g^+_{\sig t}\rmd g^-_{\sig
t}\prod_{(t,f)}\rmd v_{tf}\nonumber\\
&&\prod_fe_{g^+_{t_n\sig_1}g^+_{\sig_1 t_1}v_{t_1f}g^+_{t_1\sig_2}g^+_{\sig_2 t_2}v_{t_2f}
\cdots g^+_{t_{n-1}\sig_n}g^+_{\sig_n t_n}v_{t_nf}}(X^+_f)\nonumber\\
&&\prod_fe_{g^-_{t_n\sig_1}g^-_{\sig_1 t_1}v_{t_1f}g^-_{t_1\sig_2}g^-_{\sig_2 t_2}v_{t_2f}\cdots
g^-_{t_{n-1}\sig_n}g^-_{\sig_n t_n}v_{t_nf}}(X^-_f)
\ee
We perform the integrals over $X^+_f,\;X^-_f$ and obtain
\be
Z_{\star}(\ck)&:=&\int\prod_{(\sig,t)}\rmd g^+_{\sig t}\rmd g^-_{\sig t}\prod_{(t,f)}\rmd v_{tf}\nonumber
\\
&&\prod_f\delta\lt({g^+_{t_n\sig_1}g^+_{\sig_1 t_1}v_{t_1f}g^+_{t_1\sig_2}g^+_{\sig_2 t_2}v_{t_2f}
\cdots g^+_{t_{n-1}\sig_n}g^+_{\sig_n t_n}v_{t_nf}}\rt)\nonumber\\
&&\prod_f\delta\lt({g^-_{t_n\sig_1}g^-_{\sig_1 t_1}v_{t_1f}g^-_{t_1\sig_2}g^-_{\sig_2 t_2}v_{t_2f}
\cdots g^-_{t_{n-1}\sig_n}g^-_{\sig_n t_n}v_{t_nf}}\rt)
\ee
which gives Barrett-Crane vertex amplitude \cite{BC}. This result is consistent
with the work done by colleagues \cite{AOL,BL}. On a given triangulation, the GFT model constructed by Baratin and Oriti in \cite{AOL} reproduce the $\star$-deformed simplicial path integral considered in this appendix, which is the $\star$-deformation of the c-number simplicial path integral Eq.(\ref{start}). Thus the spinfoam model constructed in the main part of the paper may be viewed as the commutative limit of the model in \cite{AOL} as the triangulation is fixed.

\end{appendix}

\end{document}